\documentclass[12pt]{article}

\usepackage{a4wide,epsfig,feynmp}

\newcommand{\eqcm}{\: ,}   % punctuation in equations
\newcommand{\eqpt}{\: .}
\def\nn{\nonumber}
\newcommand{\lsim}{\raisebox{-3pt}{$\,\stackrel{\textstyle <}{\sim}\,$}}
\newcommand{\gsim}{\raisebox{-3pt}{$\,\stackrel{\textstyle >}{\sim}\,$}} 
\newcommand{\LQCD}{\Lambda_{\rm{QCD}}}
\newcommand{\had}{\Lambda}        % hadronic scale
\newcommand{\lcvec}[3]{\left[\;#1\;,\;#2\;,\;#3\;\right]}
\def\bj{{\it Bj}}
\def\na{{\it N}}
\def\gev{\,{\rm GeV}}
\def\vk{{\bf k}_\perp}
\def\vp{{\bf p}_\perp}
\def\vd{{\bf \Delta}_\perp}
\newcommand{\g}{{\rm g}}
\newcommand{\q}{{\rm q}}
\renewcommand{\d}{{\rm d}}
\renewcommand{\u}{{\rm u}}
\newcommand{\s}{{\rm s}}

\newcommand{\qbar}{\overline{\rm q}}
\newcommand{\dbar}{\overline{\rm d}}
\newcommand{\ubar}{\overline{\rm u}}
\newcommand{\sbar}{\overline{\rm s}}

\newcommand{\ibid}[1]{{\it ibid.}~#1}
\newcommand{\da}{distribution amplitude}

\begin{document}
\begin{fmffile}{comppic}

\begin{flushright}
DESY-98-172 \\
WUB 98-37 \\
FNT/T-98/10 \\
hep-ph/9811253 \\
\end{flushright}

\begin{center}
\vskip 3.5\baselineskip
\textbf{\Large Linking Parton Distributions to \\[0.3\baselineskip]
  Form Factors and Compton Scattering}
\vskip 2.5\baselineskip
M.~Diehl$^{1}$, Th.~Feldmann$^{2}$, R.~Jakob$^{3}$ and P.~Kroll$^{2}$
\vskip \baselineskip
1. Deutsches Elektronen-Synchroton DESY, D-22603 Hamburg, Germany \\
2. Fachbereich Physik, Universit\"at Wuppertal, D-42097 Wuppertal,
   Germany \\ 
3. Universit\`{a} di Pavia and INFN, Sezione di Pavia, 
   I-27100 Pavia, Italy
\vskip 3\baselineskip
({\it to be published in Eur.~Phys.~J.~C}\/)
\vskip 3\baselineskip
\textbf{Abstract} \\[0.5\baselineskip]
\parbox{0.9\textwidth}{We relate ordinary and skewed parton
  distributions to soft overlap contributions to elastic form factors
  and large angle Compton scattering, using light-cone wave functions
  in a Fock state expansion of the nucleon. With a simple ansatz for
  the wave functions of the three lowest Fock states we achieve a good
  description of unpolarised and polarised parton distributions at
  large $x$, and of the data for the Dirac form factor and for Compton
  scattering, both of which can be saturated with soft contributions
  only. Large angle Compton scattering appears as a good case to
  investigate the relative importance of soft and hard contributions
  in exclusive processes which are sensitive to the end point regions
  of the nucleon wave function.}
\vskip 1.5\baselineskip
\end{center}

% end of title page

%%%%%%%%%%%%%%%%%%%%%%%%%%%%%%%%%%%%%%%%%%%%%%%%%%%%%%%%%%%%%%%%%%%%%%%%%
\section{Introduction}

The recent theoretical developments for real and virtual Compton
scattering, which have lead to the introduction of skewed parton
distributions\footnote{The name {\em skewed parton distributions} has
  been proposed to amalgamate the different terms (nonforward,
  off-forward, nondiagonal, off-diagonal) used in the literature for
  closely related quantities.} (SPDs)~\cite{leipzig,ji,rad97}, have renewed
the interest in the interplay between hard inclusive and exclusive
reactions. In the light-cone approach the link between these classes
of reactions is mediated by light-cone wave functions (LCWFs).
Although this connection has been known for quite some
time~\cite{bro80,BHL} it has not yet been exploited practically.

An important question in this context is the size of perturbative QCD
contributions to exclusive reactions. There is general agreement that
the conventional hard scattering approach (see~\cite{bro80} and
references therein), in which the collinear approximation is used,
gives the correct description of electromagnetic form factors and
perhaps other exclusive processes in the limit of asymptotically large
momentum transfer. The onset of that asymptotic behaviour is however
subject to controversy. It has turned out that for the electromagnetic
form factors of the pion and the nucleon or for Compton scattering
agreement between data and the perturbative contributions is only
obtained if distribution amplitudes are employed that are strongly
concentrated in the end point regions, where one of the parton
momentum fractions tends to zero. Such distribution amplitudes have
been proposed by Chernyak et al.~\cite{COZ} on the basis of QCD sum
rules, but their derivation has been severely criticised, cf.\ for
instance~\cite{rad91}. At least for form factors but likely also for
Compton scattering they lead to perturbative contributions which are
dominated by contributions from the end point regions where the use of
perturbative QCD is not justified~\cite{isg89}. In the case of the
pion distribution amplitudes concentrated in the end point region are
now excluded by the CLEO data~\cite{CLEO} on the $\pi\gamma$
transition form factor, where they lead to perturbative contributions
much too large in comparison with experiment~\cite{JKR,RM}. In the
case of the nucleon form factor it has been shown in Ref.~\cite{ber}
that the inclusion of transverse momentum effects as well as Sudakov
suppressions~\cite{bot89} in the perturbative analysis leads to a
substantial reduction of the perturbative contribution which then is
much smaller than experiment.
 
There is another difficulty with distribution amplitudes concentrated
in the end point regions: if they are combined with a plausible
Gaussian transverse momentum dependence in a wave function and if from
that LCWF the soft overlap contribution~\cite{DY} to the nucleon form
factor is evaluated one obtains a result that exceeds the form factor
data dramatically~\cite{isg89}. Such wave functions also lead to
valence quark distributions that are much larger at large $x$ than
those derived from deeply inelastic lepton-nucleon
scattering~\cite{sch}.  Starting from all these observations and from
the assumption of soft physics dominance, the authors of
Ref.~\cite{bol96} derived a LCWF for the nucleon's valence Fock state
by fitting its free parameters to the valence quark distribution
functions and the form factors in the momentum transfer region from
about 5 to 30 GeV$^2$. The LCWF obtained in~\cite{bol96} is close to
the asymptotic form and very different from the end point concentrated
ones. Recently Radyushkin~\cite{rad98a} generalised the overlap
approach to large angle Compton scattering and showed that soft
physics, evaluated from LCWFs similar to the one used in~\cite{bol96},
can account for high energy Compton scattering in the experimentally
accessible kinematical region as well. It goes without saying that the
soft contributions to form factors and Compton scattering are
suppressed by inverse powers of the hard scales compared with the
perturbative contributions, which will always dominate at \emph{very}
large energy and momentum transfer.

The purpose of the present paper is firstly to extend the analysis of
\cite{bol96} to higher Fock states in order to explore their
importance relative to the lowest one, and secondly to include Compton
scattering in the analysis, following Radyushkin's work~\cite{rad98a}.
In Sect.~\ref{kinematics} we present some kinematics of the elastic
form factor and of Compton scattering. We then give a general
discussion concerning soft contributions and the essential conditions
for a representation of form factors and other processes as an overlap
of LCWFs (Sect.~\ref{theory}). Soft contributions to real and virtual
Compton scattering arising form the handbag diagrams will be discussed
in Sect.~\ref{theory-compt}. In the next section, Sect.~\ref{wave}, we
introduce our parametrisations of LCWFs for the lowest Fock states.
In Sects.~\ref{distributions}, \ref{form}, \ref{compt} we respectively
evaluate parton distributions, form factors and large angle Compton
scattering. As an extension of evaluating parton distributions in the
Fock state approach we also calculate skewed parton distributions
(Sect.~\ref{skewed}). Since our LCWFs describe quite well the
quantities mentioned before, our results for the skewed distributions
may convey an impression how these functions look like. The paper ends
with our summary (Sect.~\ref{sum}).

%%%%%%%%%%%%%%%%%%%%%%%%%%%%%%%%%%%%%%%%%%%%%%%%%%%%%%%%%%%%%%%%%%%%%%%%
\section{Kinematics}
\label{kinematics}
%%%%%%%%%%%%%%%%%%%%%%%%%%%%%%%%%%%%%%%%%%%%%%%%%%%%%%%%%%%%%%%%%%%%%%%%
To begin we give our notation for the elastic form factor and for
Compton scattering and introduce several reference frames we will need
later.

\subsection{The elastic form factor and Compton scattering}
\label{general-kinematics}

\begin{figure}[hbtp]
\parbox{\textwidth}{\begin{center}
(a)
\unitlength0.8cm
\fmfframe(0.5,1.0)(3.0,1.0){
\begin{fmfgraph*}(4.5,2.5)
\fmfpen{thick}
\fmfleft{Q1,dummy1,p1}\fmfright{Q2,dummy3,p2} \fmftop{p3}
\fmf{fermion,tension=2.0}{Q1,v1} \fmf{double,tension=0.5}{v1,v2}
\fmf{fermion,tension=2.0}{v2,Q2} %\fmffreeze
\fmf{plain,label=$k$,label.side=left}{v1,v3}
\fmf{plain,label=$k'=k+\Delta$,label.side=left}{v3,v2} 
\fmf{photon,tension=2}{p3,v3} 
\fmfv{label=$p$}{Q1}\fmfv{label=$p'=p+\Delta$}{Q2}
\fmfv{label=$\Delta$}{p3}
\fmfblob{.1w}{v1}\fmfblob{.1w}{v2}\fmfblob{.05w}{v3}
\end{fmfgraph*}}
%\ 
(b)
\fmfframe(0.5,1.0)(3.0,1.0){
\begin{fmfgraph*}(4.5,2.5)
\fmfpen{thick}
\fmfleft{Q1,dummy1,p1}\fmfright{Q2,dummy3,p2}
\fmf{fermion,tension=2.0}{Q1,v1} \fmf{double,tension=0.5}{v1,v2}
\fmf{fermion,tension=2.0}{v2,Q2} %\fmffreeze
\fmf{plain,label=$k$,label.side=left}{v1,v3}
\fmf{plain,label=$k'=k+\Delta$,label.side=left}{v3,v2} 
\fmf{photon}{p1,v3} \fmf{photon}{p2,v3}
\fmfv{label=$p$}{Q1}\fmfv{label=$p'=p+\Delta$}{Q2}
\fmfv{label=$q$}{p1}\fmfv{label=$q'=q-\Delta$}{p2}
\fmfblob{.1w}{v1}\fmfblob{.1w}{v2}\fmfblob{.05w}{v3}
\end{fmfgraph*}}
\end{center}}
\caption{\label{fig1} Overlap diagrams for (a) the elastic form factor
  and (b) Compton scattering. Lines $p$ and $p'$ denote protons, $k$
  and $k'$ quarks or antiquarks, and the horizontal lines represent
  any number of spectator partons. The small blob attached to the
  photon lines stands for the pointlike quark-photon coupling in~(a)
  and for the two diagrams of Fig.~\protect\ref{handbagfig} in~(b).} 
\end{figure}
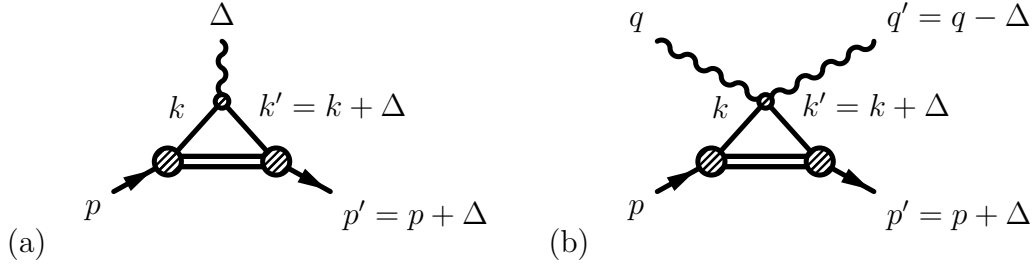

The external momenta of the one- and two-photon processes $\gamma^\ast
p \to p$ and $\gamma^\ast p \to \gamma p$ are denoted as shown in
Fig.~\ref{fig1} (a) and (b). We use the Mandelstam variables
$s=(p+q)^2$, $t= \Delta^2$, $u=(p-q')^2$, and write $Q^2= -q^2$ for
the incoming photon virtuality in Compton scattering and
$m^2=p^2=p'^2$ for the squared proton mass. Note that we write
$\Delta$ (and not $q$) for the momentum transfer to the proton in the
elastic form factor, reserving $q$ for the incoming photon in the
Compton process; this will be useful to display the similarities of
the one- and two-photon processes. We denote the momenta of the active
partons, i.e.\ those that couple to the photons by $k$ and $k'$, and
for the parton-photon subprocess in Fig.~\ref{fig1} (b) we use
Mandelstam variables $\hat{s}= (k+q)^2$, $\hat{t}=t$ and
$\hat{u}=(k-q')^2$. Whenever it is necessary to distinguish the
momenta of active and spectator partons we will label the active one
with an index $j$ and the spectators with an index $i$ ($i\ne j$);
outgoing momenta will always be indicated by a prime.

In the various reference frames described below we introduce light
cone variables $v^\pm = (v^0 \pm v^3) /\sqrt{2}$ and the transverse
part ${\bf v}_\perp = (v^1, v^2)$ for any four-vector $v$ and use
component notation $v= [v^+,v^-,{\bf v}_\perp]$. We finally define the
ratios
\begin{equation}
x = \frac{k^+}{p^+} \eqcm
\hspace{3em} 
\zeta = - \frac{\Delta^+}{p^+} = 1 - \frac{p'^+}{p^+} \eqcm
\hspace{3em} 
\eta  = \frac{q'^+}{p^+}
\end{equation}
of plus-components; positivity of the energy of the final state proton
and photon implies $\zeta < 1$ and $\eta \ge 0$.

Let us take a closer look at the physical region of the variables $t$
and $\zeta$. In any reference frame we can write
\begin{equation}  \label{generic-frame}
p = \lcvec{p^+}{\frac{m^2 + \vp^2}{2 p^+}}{\vp}  \eqcm \hspace{2em}
p'= \lcvec{(1-\zeta) p^+}
      {\frac{m^2 + \vp'^2}{2 (1-\zeta) p^+}}{\vp'}
\end{equation}
using our definition of $\zeta$ and the on-shell conditions for the
proton momenta. With (\ref{generic-frame}) and $\zeta < 1$ we have
\begin{equation}  \label{t-min}
-t = \frac{\zeta^2 m^2}{1-\zeta}
     + \frac{1}{1-\zeta} \Big( (1-\zeta)\, \vp - \vp' \Big)^2
 \ge \frac{\zeta^2 m^2}{1-\zeta}  \eqcm
\end{equation}
which imposes a minimum value on $-t$ at given $\zeta$. Note that this
is independent of the process considered.

\subsection{A symmetric frame}
\label{symmetric-frame}

For reasons that will become apparent in Sect.~\ref{theory} frames
where $\Delta^+ = 0$, i.e.\ $\zeta = 0$, play a special role in the
context of overlap contributions. We shall use a frame where
\begin{equation} \label{symm-frame-1}
p = \lcvec{p^+}{\frac{m^2 - t/4}{2 p^+}}
          {-\frac{1}{2}\,\vd} \eqcm \hspace{2em}
p'= \lcvec{p^+}{\frac{m^2 - t/4}{2 p^+}}
          {\frac{1}{2}\,\vd} \eqcm 
\end{equation}
which treats the transverse momenta of incoming and outgoing hadron in
a symmetric way and presents the further simplification that
$\Delta^-= 0$. Note that $t = -\vd^2$ here. Condition
(\ref{symm-frame-1}) fixes the frame up to a boost along the 3-axis.
For the elastic form factor one may take any frame satisfying
(\ref{symm-frame-1}); in the case of Compton scattering a symmetric
choice is to further impose $p^3 + q^3 = 0$.  Note that for real
Compton scattering this is just the c.m.\ frame with the 3-axis along
${\bf p} + {\bf p}'$, while with a virtual initial photon it does not
coincide with the centre of mass. For the photon momenta we have
\begin{eqnarray} \label{symm-frame-2}
q &=& \lcvec{\eta p^+}{\frac{(t + Q^2)^2}{-4t}\frac{1}{2\eta p^+}}
          {\frac{t - Q^2}{2 t}\,\vd} \eqcm \hspace{1em}  \nn \\
q'&=& \lcvec{\eta p^+}{\frac{(t + Q^2)^2}{-4t}\frac{1}{2\eta p^+}}
          {- \frac{t + Q^2}{2 t}\,\vd}
\end{eqnarray}
with
\begin{equation} \label{eta}
\eta = \frac{t + Q^2}{t} \, \frac{s + u - 2 m^2}{s - u + 2 
       \sqrt{s\, (u - u_1)\, (t_0 - t) / t}} \eqcm
\end{equation}
where $t=t_0$ corresponds to forward and $u=u_1$ to backward
scattering in the photon-proton c.m. We shall in the following refer
to this frame as the ``symmetric frame''.

\subsection{The photon-proton c.m.}
\label{cm-frame}

As we will see in Sect.~\ref{sub-compton} the symmetric frame just
described is not suitable for our discussion of deeply virtual Compton
scattering (DVCS). In that case we will use the c.m.\ frame with the
3-axis pointing in the incoming proton direction, where
\begin{eqnarray} 
p&=& \lcvec{p^+}{\frac{m^2}{2p^+}}{{\bf 0}_\perp} \eqcm
\hspace{2em} 
p'=  \lcvec{(1-\zeta)\,p^+}{\frac{m^2+{\bf\Delta}_\perp^2}{2
           (1-\zeta)p^+}}
         {{\bf\Delta}_\perp} \eqcm 
  \label{proton-momenta} \\
q&=& \lcvec{(\eta-\zeta)\,p^+}{\frac{- Q^2}{2(\eta-\zeta)p^+}}
         {{\bf 0}_\perp} \eqcm
\end{eqnarray}
and again $p^3 + q^3 = 0$. The non-vanishing plus component of the
momentum transfer is characterised by the skewedness parameter
$\zeta$; the total momentum transfer to the proton reads
\begin{equation} 
\Delta=\lcvec{-\zeta\,p^+}{\frac{\zeta m^2+\vd^2}{2p^+(1-\zeta)}}{\vd}
\end{equation} 
and its square is
\begin{equation} \label{t-through-zeta}
t= - \frac{\zeta^2m^2+\vd^2}{1-\zeta}  \eqpt
\end{equation}
Notice that the relation (\ref{t-through-zeta}) follows from
(\ref{proton-momenta}) alone and thus holds in any frame where $\vp =
0$.
In the photon-proton c.m.\ frame we have
\begin{equation} \label{zeta-eta}
  \zeta = x_\na\, \frac{Q^2 - t (1-x_\na)}{Q^2 + x_\na^2\, m^2}
  \eqcm  \hspace{3em}  
  \eta = \zeta - x_\na
\end{equation}
where
\begin{equation}
  x_\na = \frac{2 x_\bj}{1 + \sqrt{1+ 4x_\bj^2\, m^2/Q^2}} 
  \eqcm  \hspace{3em}  
  x_\bj = \frac{Q^2}{2 p\cdot q}
\end{equation}
respectively denote Nachtmann's and Bjorken's variable. In the
kinematical region of DVCS, i.e. when $-t$ is small, and $Q^2$ and $s$
are large (\ref{zeta-eta}) simplifies to $\zeta \approx x_\bj$ and
$\eta \approx 0$.

\subsection{Frames for the hadron LCWFs}
\label{hadron-frames}

The arguments of LCWFs are given as the plus-momentum fractions $x_i$
and the transverse parts ${\bf k}_{\perp i}$ of parton momenta in a
frame where the transverse momentum of the corresponding hadron is
zero. We will call those systems ``hadron frames'' and refer to
transverse parton momenta in an appropriate hadron frame as
``intrinsic'' transverse momenta.

The transformation from a given frame to a hadron frame can be
achieved by a ``transverse boost'' (cf.\ e.g.~\cite{bro89}) which
leaves the plus component of \emph{any} momentum vector $a$ unchanged,
and which involves a parameter $b^+$ and a transverse vector ${\bf
  b}_\perp$:
\begin{equation} 
\label{eq:plustrafo} 
\lcvec{a^+}{a^-}{{\bf a}_\perp}
\qquad\longrightarrow\qquad
\lcvec{a^{+}}{a^{-}-\frac{{\bf a}_\perp\cdot{\bf b}_\perp}{b^+}
+\frac{a^+\,{\bf b}_\perp^{\,2}}{2\,(b^+)^2}}
      {{\bf a}_\perp-\frac{a^+}{b^+}\,{\bf b}_\perp} \eqpt
\end{equation}
Starting for instance from the symmetric frame of
Sect.~\ref{symmetric-frame} the choice $b^+=p^+$, ${\bf
  b}_\perp=-\vd/2$ transforms the momenta of the incoming hadron and
its partons as
\begin{equation} 
p \longrightarrow
\tilde p=\lcvec{p^+}{\frac{m^2}{2 p^+}}{{\bf 0}_\perp} \eqcm
 \hspace{2em}
k_i \longrightarrow
\tilde k_i=\lcvec{x_i p^+}{\dots}
           {{\bf k}_{\perp i}+x_i \frac{{\bf\Delta}_\perp}{2}} \eqcm
\end{equation} 
where we suppressed the minus components of the parton momenta, whose
expression we will not need. This is an appropriate frame to read off
the arguments of the LCWF of the incoming hadron as $x_i$ and
$\tilde{\bf k}_{\perp i}={\bf k}_{\perp i}+x_i\,{\bf\Delta}_\perp/2$.
The analogous boost with the choice $b^+=p^+$, ${\bf b}_\perp =
+{\bf\Delta}_\perp/2$ relates the symmetric frame with a frame
appropriate for identifying the arguments of the LCWF for the
scattered hadron as $x'_i$ and $\hat{\bf k}'_{\perp i}={\bf k}'_{\perp
  i} - x'_i\,{\bf\Delta}^{\phantom{.}}_\perp/2$.

Incoming and outgoing parton momenta in the overlap contributions
Fig.~\ref{fig1} are related by $k'_i=k^{\phantom{.}}_i$ ($i\ne j$) for
the spectator partons and $k'_j=k^{\phantom{.}}_j+\Delta$ for the
active parton which takes the momentum transfer in the scattering.
Using the transformations between the symmetric frame and the
in/out-hadron frames established above we can directly express the
LCWF arguments for the outgoing hadron (denoted by a hat) in terms of
the ones for the incoming hadron (denoted by a tilde):
\begin{eqnarray} 
\hat x'_i=\tilde x^{\phantom{.}}_i \eqcm &\qquad& 
\hat{\bf k}'_{\perp i}=\tilde{\bf k}^{\phantom{.}}_{\perp i}
                       -\tilde x^{\phantom{.}}_i\,
                        {\bf\Delta}^{\phantom{.}}_\perp
\qquad \qquad \mbox{for }i\neq j \eqcm \nn\\
\hat x'_j=\tilde x^{\phantom{.}}_j \eqcm &\qquad& 
\hat{\bf k}'_{\perp j}=\tilde{\bf k}^{\phantom{.}}_{\perp j}
                       +(1-\tilde{x}^{\phantom{.}}_j)\,
                        {\bf\Delta}^{\phantom{.}}_\perp \eqcm
\label{tilde-args}
\end{eqnarray} 
where we could have dropped the hat/tilde notation for the momentum
fractions which are not changed by the boost (\ref{eq:plustrafo}).
   
For the case of deeply virtual Compton scattering the photon-proton
c.m.~frame introduced in Sect.~\ref{cm-frame} is already the
appropriate hadron frame to identify the arguments of the LCWF of the
incoming proton. By the boost (\ref{eq:plustrafo}) with the parameter
values $b^+=(1-\zeta)\,p^+$, ${\bf b}_\perp={\bf\Delta}_\perp$ one
obtains the momenta in the corresponding frame where the outgoing
proton has zero transverse momentum. LCWF arguments for the outgoing
proton (denoted by a breve) are related to the ones of the LCWF of the
incoming proton as
\begin{eqnarray} \label{breve-args}
\breve x'_i= \frac{x_i}{1-\zeta} \eqcm &\qquad& 
\breve{\bf k}'_{\perp i}={\bf k}^{\phantom{.}}_{\perp i}
                        -\frac{x_i}{1-\zeta}\,
                         {\bf\Delta}^{\phantom{.}}_\perp
\qquad \qquad \mbox{for }i\neq j \eqcm \nn\\
\breve x'_j=\frac{x_j-\zeta}{1-\zeta} \eqcm &\qquad& 
\breve{\bf k}'_{\perp j}={\bf k}^{\phantom{.}}_{\perp j}
                        +\frac{1-x_j}{1-\zeta}\,
                         {\bf\Delta}^{\phantom{.}}_\perp
                        \eqcm
\end{eqnarray} 
where according to its definition the plus momentum fraction in the
LCWF of the scattered proton is taken with respect to
$p'^+=(1-\zeta)\,p^+$ and not to $p^+$. We notice that for $\zeta=0$
Eq.~(\ref{breve-args}) takes the same form as (\ref{tilde-args}).

%%%%%%%%%%%%%%%%%%%%%%%%%%%%%%%%%%%%%%%%%%%%%%%%%%%%%%%%%%%%%%%%%%%%%%%%
\section{The theory of soft contributions}
\label{theory}
%%%%%%%%%%%%%%%%%%%%%%%%%%%%%%%%%%%%%%%%%%%%%%%%%%%%%%%%%%%%%%%%%%%%%%%%
In this section we are concerned with soft overlap contributions to
hard exclusive processes. They are contributions where only some of
the partons in the external hadrons are active, i.e.\ participate in a
hard scattering, while the other partons remain spectators.

\subsection{Bethe-Salpeter and light cone wave functions}

The evaluation of overlap contributions in terms of light cone wave
functions requires some care. An example is the Drell-Yan overlap
formula~\cite{DY} of the elastic form factor, for which Isgur and
Llewellyn Smith~\cite{isg89} observed that different results are
obtained in different reference frames. Sawicki~\cite{saw92} has shown
the origin of this discrepancy: in certain reference frames there are
overlap contributions which are not contained in the Drell-Yan
formula; when they are taken into account Lorentz invariance is
restored. We shall first review Sawicki's arguments~\cite{saw92,saw91}
for the form factor and then investigate the case of Compton
scattering.

\subsubsection{The elastic form factor}
\label{sub-form}

Our starting point to obtain the overlap formula for the form factor
is the diagram of Fig.~\ref{fig1} (a) in the framework of equal-time
quantisation and covariant perturbation theory. The hadron-parton
vertices, represented by the large blobs in the diagram, are described
by Bethe-Salpeter wave functions $\Psi_{BS}$.  For simplicity we
consider a scalar hadron coupling to two scalar partons, so that there
is only one spectator line in the diagram Fig.~\ref{fig1} (a). We
further work in a toy theory where the hadron has a pointlike coupling
to the two partons; to leading order in the coupling constant the wave
function $\Psi_{BS}(k)$ of the hadron with momentum $p$ is then given
by the coupling times the free propagators for the partons with
momenta $k$ and $p-k$. In general (and in particular for QCD)
$\Psi_{BS}(k)$ will have a more complicated analytic structure in the
virtualities $k^2$ and $(p-k)^2$ involving branch cuts in these
variables. Their discussion is beyond the scope of this paper and we
only retain the propagator poles in these variables. This will be
sufficient to exhibit the points we want to make.

The aim is now to perform the loop integration over $k^-$ in
Fig.~\ref{fig1} (a) so as to reduce $\Psi_{BS}(k)$ and $\Psi_{BS}(k')$
to LCWFs. For this we use that up to a normalisation factor a LCWF is
obtained from the corresponding Bethe-Salpeter wave function, say
$\Psi_{BS}(k)$, by the integral $\int \d k^-\, \Psi_{BS}(k)$ at fixed
$k^+$ and ${\bf k}_\perp$. Note that this relation does not only hold
in frames where the hadron has zero transverse momentum; with
(\ref{eq:plustrafo}) we see in particular that this integral is
invariant under transverse boosts. The $k^-$-integration in
Fig.~\ref{fig1} (a) is readily performed using Cauchy's theorem since
the analytic structure of the diagram is given by the propagator poles
in our model. Writing
\begin{eqnarray}
k^- &=& \frac{k^2 + \vk^2}{2 p^+ x}   \nonumber \\
    &=& \frac{(k-p)^2 + (\vk-\vp)^2}{2 p^+ (x - 1)} + p^- \nonumber \\
    &=& \frac{(k+\Delta)^2 + (\vk+\vd)^2}{2 p^+ (x - \zeta)} - \Delta^-
  \label{k-minus}
%
% please DO NOT change the arrangement of this equation as three lines
%
\end{eqnarray}
and using that the poles in $k^2$, $(k-p)^2$, $(k+\Delta)^2$ are
situated just below the real axes in these variables we see that the
propagator poles are below or above the real $k^-$-axis, depending on
the value of $x$. For definiteness we now take $\zeta \ge 0$, where we
have the following cases:
\begin{enumerate}
\item For $x>1$ and for $x<0$ all poles are on same side. Closing the
  integration contour in the half plane where there are no
  singularities one obtains a zero integral.
\item For $1>x>\zeta$ we pick up the pole in $(p-k)^2$ alone when
  closing the integration contour in the upper half plane. The diagram
  is then given by the propagators of $k$ and $k+\Delta$, evaluated at
  the value of $k^-$ where $p-k$ is on shell.  Applying an analogous
  argument to the integral $\int \d k^-\, \Psi_{BS}(k)$ we find that a
  LCWF can be written as the hadron-parton coupling times one parton
  propagator, evaluated at the value of $k^-$ where the other
  propagator is on shell. As a by-product one finds that the
  plus momentum fractions of the partons w.r.t.\ the hadron are always
  between $0$ and $1$, otherwise the integral is zero. In total we
  find that the diagram for the form factor is given by the product of
  two hadron LCWFs, as stated in the Drell-Yan formula.
\item For $\zeta>x>0$ we can pick up the residue at the pole in $k^2$,
  or alternatively the sum of residues for the poles in $(p-k)^2$ and
  $(k+\Delta)^2$. In the term where $k$ (or $k+\Delta$) is on shell
  both partons in the hadron $p'$ (or $p$) are both off-shell, which
  \emph{cannot} be rewritten in terms of a LCWF. Note that for
  $\zeta>x>0$ the parton that has been struck by the photon has
  negative plus-momentum fraction $x-\zeta$, which does not correspond
  to a parton going into hadron $p'$; a situation that clearly cannot
  be expressed through a LCWF.  This contribution is missing if one
  naively writes down the Drell-Yan formula in a frame where $\zeta
  \neq 0$: here is the origin of the paradox observed
  in~\cite{isg89}.\footnote{\label{foot}In a recent paper~\cite{bhw98}
    this contribution has been rewritten in terms of the LCWF for the
    hadron $p$ containing partons with momenta $k$ and $-(k+\Delta)$
    plus the hadron with momentum $p'$, and the (trivial) LCWF to find
    the hadron with momentum $p'$ in the hadron $p'$. It would be
    interesting to explore this idea in the context of Compton
    scattering, but this shall not be done here.}
\end{enumerate}
We thus see that in order to obtain an overlap representation in terms
of two-parton LCWFs for each hadron we need to go to a reference frame
where $\zeta$, or in other words, $\Delta^+$ is zero. Such a frame was
in fact chosen in the original work by Drell and
Yan.\footnote{\label{other-foot}In a frame with $\zeta = 0$ there can
  be finite contributions of the type discussed in point 3 if the
  integrand in the interval $\zeta>x>0$ becomes singular for
  $\zeta=0$. This happens for the minus component of the parton
  current~\cite{bhw98,burk}, which we shall not use in our
  applications; the plus component of the current does not exhibit
  this phenomenon.}

The argument goes along the same lines if one has more than one
spectator and takes a Bethe-Salpeter wave function with only the
propagator pole in each parton. Let us label the active parton (i.e.\ 
the one hit by the photon) $k_1$ and the spectators $k_2$, \ldots,
$k_{n-1}$, $k_n = p - k_1 - k_2 \ldots k_{n-1}$ and first perform the
integrations over $k_2^-$, \ldots, $k_{n-1}^-$ to put $n-2$ spectators
on shell while not doing anything to the active parton. Then we are
left with one active parton and a cluster of spectators, and the
situation for the integration over $k_1^-$ is as above.

\subsubsection{Compton scattering}
\label{sub-compton}

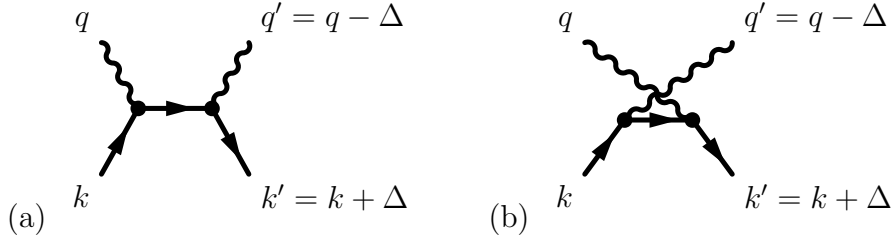
\begin{figure}[hbtp]
\parbox{\textwidth}{\begin{center}
(a)
\unitlength0.7cm
\fmfframe(0.5,1.0)(4.0,1.0){
\begin{fmfgraph*}(3.5,2.5)
\fmfpen{thick}
\fmfleft{Q1,p1}\fmfright{Q2,p2}
\fmf{fermion}{Q1,v1,v2,Q2}
\fmf{photon}{p1,v1} \fmf{photon}{p2,v2}
\fmfv{label=$k$}{Q1}\fmfv{label=$k'=k+\Delta$}{Q2}
\fmfv{label=$q$}{p1}\fmfv{label=$q'=q-\Delta$}{p2}
\fmfdot{v1}\fmfdot{v2}
\end{fmfgraph*}}
%\
(b)
\fmfframe(0.5,1.0)(4.0,1.0){
\begin{fmfgraph*}(3.5,2.5)
\fmfpen{thick}
\fmfleft{Q1,p1}\fmfright{Q2,p2}
\fmf{fermion}{Q1,v1,v2,Q2}
\fmf{phantom,tension=0.7}{p1,v1} \fmf{phantom,tension=0.7}{p2,v2}
\fmffreeze
\fmf{photon}{p1,v2} \fmf{photon}{p2,v1}
\fmfv{label=$k$}{Q1}\fmfv{label=$k'=k+\Delta$}{Q2}
\fmfv{label=$q$}{p1}\fmfv{label=$q'=q-\Delta$}{p2}
\fmfdot{v1}\fmfdot{v2}
\end{fmfgraph*}}
\end{center}}
\caption{\label{handbagfig} (a) $s$-channel and (b) $u$-channel
  diagram for quark-photon or antiquark-photon scattering.}
\end{figure}

We shall now see that the two-photon process of Fig.~\ref{fig1} (b)
involves a new difficulty. Let us take again our toy model of a hadron
with a pointlike coupling to two partons. To leading order in the
electromagnetic coupling the parton-photon vertex is given by the two
diagrams of Fig.~\ref{handbagfig}. Compared to the form factor case we
thus have an extra propagator in the overall process, corresponding to
a squared momentum $\hat{s} = (k+q)^2$ or $\hat{u} = (k - q')^2$, so
that (\ref{k-minus}) is completed by
\begin{equation}
k^- = \frac{(k+q)^2 + (\vk+{\bf q}_\perp)^2}{2 p^+ (x - \zeta + \eta )}
- q^- 
\end{equation}
in the $s$-channel and
\begin{equation}
k^- = \frac{(k-q')^2 + (\vk^{\phantom{.}}
                        -{\bf q}_\perp')^2}{2 p^+ (x - \eta)} + q'^- 
\end{equation}
in the $u$-channel diagram.

\begin{table}[ht]
\begin{center}
\begin{tabular}{|c|c|lcl|} \hline
diagram & region &
\multicolumn{3}{c|}{propagator pole in} \\ \hline $s$-channel
& $1>x>\zeta$      & $p-k$      & or & $k$, $k+\Delta$, $k+q$ \\ 
& $\zeta>x>0$      & $k$, $k+q$ & or & $p-k$, $k+\Delta$ \\
& $0>x>\zeta-\eta$ & $k+q$      & or & $p-k$, $k+\Delta$, $k$ \\
\hline
$u$-channel
& $1>x>\eta$     & $p-k$         & or & $k$, $k+\Delta$, $k-q'$ \\ 
& $\eta>x>\zeta$ & $p-k$, $k-q'$ & or & $k$, $k+\Delta$ \\
& $\zeta>x>0$    & $k$           & or & $p-k$, $k-q'$, $k+\Delta$ \\
\hline
\end{tabular}
\end{center}
\caption{\label{tab:poles}Possibilities to pick up propagator poles in
  the $k^-$-integration for the case $1 > \eta > \zeta > 0$.}
\end{table}
In the case $1 > \eta > \zeta > 0$ one has the possibilities listed in
Tab.~\ref{tab:poles} to pick up propagator poles in the $k^-$-plane.
Proceeding in the same way as in the form factor case we see that only
in the region $1>x>\eta$ we obtain an expression in terms of the
two-parton LCWFs for both hadrons. In all other regions we have
further contributions, either from a parton attached to one hadron but
not the other ($k$ or $k+\Delta$) or not attached to a hadron at all
($k+q$ or $k-q'$).

The situation is analogous in other cases than $1 > \eta > \zeta > 0$,
and also if there is more than one spectator. Notice that in general
we cannot find a frame where $\eta = \zeta = 0$ to solve our problem:
if $\eta = 0$ then $q'^2 = 0$ implies ${\bf q}_\perp' = 0$, and if
also $\zeta = 0$ then $t + Q^2 = - 2 q \cdot q' = 0$ which is a
special kinematical situation.

At this point we look beyond our toy model and remember that we want
to evaluate \emph{soft} overlap contributions in QCD, which involve
the soft parts of the hadron wave functions, not the hard parts that
are generated perturbatively~\cite{bro80}. We will now see that in
certain cases we can obtain an \emph{approximate} expression for the
soft overlap contribution that involves only the LCWFs of the two
hadrons. To this end we first chose a frame with $\zeta = 0$ so as to
eliminate the interval $\zeta >x>0$, as we did for the form factor. It
turns out that with appropriate external kinematics the contributions
from the poles in $\hat{s}$ and $\hat{u}$ go with a highly virtual
parton in at least one of the two hadrons. Since large parton
virtualities are strongly suppressed in the soft parts of the hadron
wave functions (they constitute their hard parts) we can neglect these
pole contributions, restrict $x$ to the interval from 0 to 1 and only
take into account the contribution from the pole in $(p-k)^2$, which
just leads to an expression with two hadron LCWFs as in the form
factor case.

To see when this is the case we write
\begin{eqnarray}
\hat{s} + Q^2 &=& x (s + Q^2 - m^2) + k^2 \nonumber \\ 
 &-&    2 (\vk - x\vp)\cdot {\bf q}_\perp
      - \frac{\eta-\zeta}{x} \left\{ x^2 m^2 - k^2 -
        (\vk - x\vp)\cdot(\vk + x\vp) \right\}  \eqcm
      \nonumber \\
\hat{u} &=& x (u - m^2) + k^2 \nonumber \\ 
 &+&    2 (\vk - x\vp)\cdot {\bf q}_\perp' 
      + \frac{\eta}{x} \left\{ x^2 m^2 - k^2 - 
        (\vk - x\vp)\cdot(\vk + x\vp) \right\}
  \label{s-and-u-hat}
\end{eqnarray}
and make the hypothesis that the soft hadron wave functions are
dominated by intrinsic transverse parton momenta ${\bf k}_{\perp i}$
satisfying ${\bf k}_{\perp i}^2 /x_i^{\phantom{.}}  \lsim \had^2$
(this is implemented in our ansatz for the LCWFs in Sect.~\ref{wave}),
where $\had$ is a hadronic scale in the GeV region, and by parton
virtualities in the range $|k_i^2| \lsim \had^2$. {}From now on we
concentrate on two cases.

\subsubsection*{Large angle Compton scattering (large $s$, $-t$ and
  $-u$)}

We now work in the symmetric frame of Sect.~\ref{symmetric-frame}. Let
us for a moment stick with the case where there is only one spectator
parton; the expressions $\tilde{\bf k}^2_{\perp i}
/\tilde{x}_i^{\phantom{.}}$ and $\hat{\bf k}'^{\,2}_{\perp i}
/\hat{x}'_i$ for this spectator in the initial and final state hadron
(cf.\ Sect.~\ref{hadron-frames}) can then be rewritten in terms of the
active parton momenta $k$ and $k'$. For their sum we obtain
\begin{equation}
\frac{(\vk - x\vp)^2}{1-x} + \frac{(\vk' - x\vp')^2}{1-x} = 
(1-x)\, \vd^2 /2 + \frac{2(\vk + \vd/2)^2}{1-x}  \lsim \had^2 \eqcm
  \label{relative-k-t}
\end{equation}
which implies
\begin{equation}
|1 - x| \lsim \had^2 /(-t) \eqcm \hspace{3em} 
|\vk - x \vp| \lsim \had^2 / \sqrt{-t} \eqpt
  \label{constraints}
\end{equation}
In the case of several spectators the argument goes along the same
lines, now summing $\tilde{\bf k}^2_{\perp i}
/\tilde{x}_i^{\phantom{.}} + \hat{\bf k}'^{\,2}_{\perp i} /\hat{x}'_i$
over all spectators.

We remark in passing that a restriction to intrinsic transverse
momenta ${\bf k}_{\perp i}^2 \lsim \had^2$ instead of ${\bf k}_{\perp
  i}^2 /x_i^{\phantom{.}} \lsim \had^2$ would not be enough to ensure
small parton virtualities in the hadrons: instead of
(\ref{relative-k-t}) we would then only have $(\vk - x\vp)^2 + (\vk' -
x\vp')^2 \lsim \had^2$, which gives $|1 - x| \lsim \had /\sqrt{-t}$
and $|\vk - x \vp| \lsim \had$, and in particular $|\vk + \vd/2| \lsim
\had$. {}From $k^2 - k'^2 = 2 \vd \cdot (\vk + \vd/2)$ we see that
then at least one of the parton virtualities would be of order $\had
\sqrt{-t}$ and not $\had^2$.

With (\ref{s-and-u-hat}), (\ref{constraints}) and
(\ref{symm-frame-2}), (\ref{eta}) we have $s \approx \hat{s}$ and $u
\approx \hat{u}$ up to corrections of order $\had^2 \, (t \pm Q^2)
/t$, provided that both $s$ and $-u$ are large on a hadronic
scale.\footnote{One may admit a two-scale regime $\had^2 \ll -t \ll
  Q^2$ provided that $s$ and $-u$ are also of order $Q^2$.} This
implies that in order for $\hat{s}$ or $\hat{u}$ to have a pole at
least one parton must have a large virtuality or intrinsic transverse
momentum, so that following our above remarks we can neglect these
pole contributions. Note that apart from $-t$ one also needs $-u$
large: when the latter becomes too small the propagator $\hat{u}$ can
easily become soft and it is no longer justified to neglect its pole
contribution (which one may relate to the soft, hadronic part of the
final state photon). Similarly one can see that $s$ must be large,
too.

The physical situation clearly is that of a hard photon-parton
scattering and the soft emission and reabsorption of a parton by the
hadron, similar to the familiar handbag diagram for inclusive deeply
inelastic scattering (DIS) or for DVCS.\footnote{Note however that in
  those cases there are factorisation theorems stating that the
  handbag diagrams are dominant when the hard scale becomes infinitely
  large. In the present case we have a less strong situation of
  factorisation since for infinitely large $-t$ the hard scattering
  mechanism of~\cite{bro80} dominates over the soft overlap or handbag
  contribution.} In the hard scattering one can approximate the parton
momenta $k$, $k'$ as being on shell, collinear with their parent
hadrons and with light cone fractions $x = 1$. This also provides
another point of view on neglecting the $\hat{s}$ and $\hat{u}$ pole
contributions: approximating $k^-$ with the value for which the
partons are on shell in the hard scattering we have a $k^-$-integral
where only the parton lines directly attached to hadrons provide a
$k^-$-dependence. This is just as in the case of the elastic form
factor, and thus we have the same situation for expressing the
amplitude in terms of LCWFs as described in Sect.~\ref{sub-form}.

At this point we can also understand why in the conventional hard
scattering mechanism~\cite{bro80} (and also in the modified one of
Botts, Li and Sterman~\cite{bot89}) one always obtains an expression
involving hadron LCWFs, irrespective of the reference frame used.  The
reason is that in the corresponding diagrams the parton lines from
each hadron are directly attached to a hard scattering subprocess,
where the minus components of their momenta can be approximated with
their values for which the partons are on shell. The corresponding
$k^-$-integration then only concerns the hadron-parton vertex alone
and leads to a LCWF. In the case of soft overlap diagrams the
situation becomes more complicated because spectator parton lines are
"shared" by different hadrons, without undergoing a hard scattering.

\subsubsection*{Deeply virtual Compton scattering (small $-t$, large
  $Q^2$ and $s$)}

In the kinematical region of DVCS, where $-t \sim \had^2$, we can no
longer infer from (\ref{relative-k-t}) that $x$ must be close to one.
Furthermore the factors $(t \pm Q^2) /(2t)$ in (\ref{symm-frame-2}),
(\ref{eta}) are large, and the terms involving ${\bf
  q}_\perp^{\phantom{.}}$, ${\bf q}'_\perp$ and $\eta$ in
(\ref{s-and-u-hat}) can thus be of order of the large
scale,\footnote{A more careful discussion is needed in the case where
  $s \gg Q^2$, which we shall not consider here.} so that $\hat{s}$ or
$\hat{u}$ may be zero even if the partons are near shell and nearly
collinear with their parent hadrons. Our previous argument to neglect
the pole terms in $\hat{s}$ and $\hat{u}$ then no longer works in the
frame we have considered so far.  There exist other frames with $\zeta
= 0$, but one can show that $\eta$ cannot be smaller than in
(\ref{eta}) by solving the minimisation problem for $\eta$ with an
arbitrary axis defining plus components under the constraint $\zeta =
0$.

We know however from the factorisation theorem of
DVCS~\cite{rad97,fact} that in a frame such as the c.m.\ where the
incident and the scattered hadron move fast to the right (and where
$\zeta \neq 0$), the process factorises into a skewed parton
distribution describing the soft coupling between partons and hadrons,
and a hard photon-parton scattering calculated with the minus- and
transverse components of $k$ and $k'$ replaced with zero. This
factorisation is not realised in our symmetric frame of
Sect.~\ref{symmetric-frame}, where the hadron momenta become slow of
order $\sqrt{-t}$ in the DVCS limit.

Using this factorisation in the c.m.\ we can again neglect the pole
contributions from $\hat{s}$ and $\hat{u}$ but have now the problem of
the region $\zeta>x>0$ described in connection with the form factor.
What we will do in this paper is to use LCWFs to calculate the
contribution of the lowest Fock state components to skewed parton
distributions in the region $1>x>\zeta$. We are thus not able to
predict the amplitude of the DVCS process but can give a part of the
nonperturbative input needed to calculate it, which furthermore is
process independent and also occurs e.g.\ in exclusive meson
production at large $Q^2$ and small $t$~\cite{CFS}.

It should be noted that even if we were able to express the full DVCS
amplitude through the overlap of LCWFs we could not hope to evaluate
the amplitude from the lowest Fock states only. In the case of the
elastic form factor, where we do have an overlap formula, we know that
all Fock states become important as one goes to low $-t$ and it seems
reasonable to expect the same for DVCS, where $-t$ is always small by
definition. Similarly the usual parton distributions, where we have an
overlap formula in the full range $0<x<1$, can be well described by
the first few Fock states down to some finite value of $x$, but at
some point higher Fock states will become essential. The same holds a
fortiori for skewed parton distributions as we shall see in
Sect.~\ref{skewed}.

\subsection{Cat's ears diagrams in Compton scattering}

So far we have only considered soft overlap contributions with only
one active parton, which is subsequently hit by the two photons. As
already remarked they have the topology of handbag diagrams, i.e.\ 
they factorise into a parton-photon scattering and a soft subamplitude
with two hadron and two parton lines, which we want to describe in
terms of hadron LCWFs. There are other overlap contributions with
\emph{two} active partons, each coupling to one photon; they have the
topology of so called cat's ears diagrams. One can see that in the
large angle region as well as for DVCS one cannot avoid large
virtualities or intrinsic transverse momenta occurring somewhere in
these diagrams, so that we no longer deal with a soft overlap. Working
with soft hadron wave functions one must then add at least one hard
gluon in the diagrams.

That in DVCS cat's ears diagrams become unimportant in the large-$Q^2$
limit is part of the factorisation theorem for that process. In the
large angle region it is interesting to note that the diagrams where
there is just one hard gluon exchanged between the two active partons
consist of a hard scattering subprocess involving two parton lines
(corresponding to the diagrams for Compton scattering off a meson in
the hard scattering mechanism of~\cite{bro80}), and a number of
spectator partons which as in the soft overlap (handbag) diagrams must
be wee partons. It is reasonable to assume that such ``hybrid''
diagrams give contributions to the amplitude whose order of magnitude
is between the pure soft overlap and the pure hard scattering
contributions: compared with the latter they have less hard gluons
(and thus hard propagators and coupling constants), but in contrast to
the pure soft overlap diagrams they require $N-2$ instead of $N-1$ wee
partons in an $N$-particle Fock state, which is less restrictive for
the hadron wave functions.

\subsection{Soft overlap contributions to other processes}

Having discussed in detail the conditions necessary to express soft
overlap contributions in terms of LCWFs for spacelike elastic form
factors and for Compton scattering we wish to make some remarks on
other processes:

\subsubsection{Meson production $\gamma^\ast p \to M p$}

Let us first see what happens if in the overlap diagrams for Compton
scattering we replace the outgoing photon with a meson $M = \rho$,
$\phi$, $\pi$, $K$, \ldots, and the pointlike photon-quark coupling
with the $q\bar{q}$ Bethe-Salpeter wave function of $M$. In the
discussion of our toy model we have seen that the loop integration
over $k^-$ gives a sum over residues, where each term corresponds to a
simple pole in the $k^-$-plane and can be written as the product of
\emph{two} LCWFs (of the two external particles that "share" the
parton which is on its mass pole). This is not the structure we would
need for an expression in terms of \emph{three} LCWFs, two for the
incident and scattered proton and one for the meson. If and how such a
structure can be obtained requires further investigation which goes
beyond the scope of this paper.

{}From our discussion of Compton scattering it is however clear that
in the region of large $s$, $-t$, $-u$ there is no soft overlap,
because if the partons in the protons are all to be soft then there is
a parton with large virtuality $\hat{s}$ or $\hat{u}$, which now
couples to the meson.

In the region of small $-t$ but large $Q^2$ and $s$ the situation is
different. First we remember that in this case it has been
shown~\cite{CFS} that for longitudinal photon polarisation and in the
large-$Q^2$ limit the process factorises into a soft amplitude
involving the two protons and two partons, the soft transition from a
$q\bar{q}$-pair to the meson, and a hard photon-parton scattering with
at least one hard gluon exchange. A soft overlap contribution
competing with this mechanism is possible when the quark line that
directly goes from the meson to the soft proton amplitude is a wee
parton: then one can take out the gluon from the hard scattering
diagrams without any parton line going far off shell.\footnote{Such
  end point configurations are indeed the reason why factorisation
  cannot be established in the case of transverse photon
  polarisation.} As mentioned above it is not clear whether such
contributions can be written in terms of LCWFs for the protons and the
meson. Likewise it remains to be investigated whether it can be
expressed in terms of the LCWF of the meson and a skewed parton
distribution in the proton, the latter being obtained from the
parton-proton amplitude by an integration over $k^-$ in a similar way
as LCWFs are obtained from Bethe-Salpeter wave functions~\cite{DG}.

\subsubsection{Timelike processes}

Crossing relates the timelike ($\gamma^\ast \to p\bar{p}$) to the
spacelike form factor ($\gamma^\ast p \to p$), and the production of
$p\bar{p}$ in a two-photon collision to Compton scattering; the
diagrams for the timelike processes are obtained from those in
Fig.~\ref{fig1} by a rotation of $90^\circ$ counterclockwise. Using
our toy model one easily sees that like their spacelike counterparts
these processes admit soft overlap contributions. They can however not
be expressed in terms of LCWFs: the parton line shared by the proton
and antiproton cannot correspond to an incoming parton for both $p$
and $\bar{p}$ as it would have to be in LCWFs, except for the the
point where its plus momentum is strictly zero. This holds in any
reference frame so that knowledge of the LCWFs is not sufficient to
evaluate the soft overlap contributions to these
processes.\footnote{Again it might be possible to find an expression
  of the overlap along the lines mentioned in our
  footnote~\protect\ref{foot}, but this will not be pursued here.}

%%%%%%%%%%%%%%%%%%%%%%%%%%%%%%%%%%%%%%%%%%%%%%%%%%%%%%%%%%%%%%%%%%%%%%%%%
\section{Large angle Compton scattering with the handbag}
\label{theory-compt}
%%%%%%%%%%%%%%%%%%%%%%%%%%%%%%%%%%%%%%%%%%%%%%%%%%%%%%%%%%%%%%%%%%%%%%%%%
\subsection{Calculation of the handbag diagrams}
\label{handbag}

The calculation of the handbag diagrams for real or virtual Compton
scattering at large $s$, $-t$ and $-u$ can be done using the methods
that are well known for usual DIS and for DVCS. At some points it
presents however additional complications which we shall now discuss.
For simplicity we work in the frame of Sect.~\ref{symmetric-frame}
where $\zeta = 0$, although our derivation can be done in other frames
as well. Our starting point is the expression of the Compton amplitude
in terms of a soft proton matrix element and a hard parton-photon
scattering:
\begin{eqnarray}
  {\cal A} &=& \sum_a (e e_a)^2 \int \d^4 k\, 
  \theta(k^+) \int {\d^4 z\over (2\pi)^4}\, e^{i\, k\cdot z}
     \left[ \langle p'|\, T\, \overline\psi{}_{a \alpha}(0)\,
     \psi_{a \beta}(z) \,|p\rangle \, 
     H_{\alpha\beta}(k',k)\,  \right. \hspace{5em} \nonumber \\
&& \hspace{14em}
     \left. {}+ \langle p'|\,T\, \overline\psi{}_{a \alpha}(z)\,
     \psi_{a \beta}(0) \,|p\rangle \, 
     H_{\alpha\beta}(-k,-k') \right]  \eqcm
 \label{starting-point}
\end{eqnarray}
where
\begin{equation} \label{hard-scattering}
H_{\alpha\beta}(k',k) = \left(
  \varepsilon'^\ast\cdot\gamma\, 
  \frac{(k+q)\cdot\gamma}{(k+q)^2 + i\epsilon} \
  \varepsilon\cdot\gamma +
  \varepsilon\cdot\gamma\, 
  \frac{(k'-q)\cdot\gamma}{(k'-q)^2 + i\epsilon} \
  \varepsilon'^\ast\cdot\gamma \right)_{\alpha\beta} \nonumber \\
\end{equation}
is the tree level expression for the hard scattering, with
polarisation vectors $\varepsilon$ and $\varepsilon'$ for the incoming
and outgoing photon. The sum is over quark flavours $a$, $e_a$ being
the electric charge of quark $a$ in units of the positron charge~$e$.
The first term in (\ref{starting-point}) corresponds to the case where
the incoming parton $k$ in the hard subprocess is a quark, the second
term corresponds to an incoming antiquark. For ease of writing we do
not display the spin labels for the proton states here and in the
following.

Using that the photon-parton scattering is dominated by a large scale
we now neglect the variation of the transverse and minus components of
$k$ and $k'$ in $H$, where we replace them with momentum vectors that
are on shell and lie in the scattering plane, namely with
\begin{equation} \label{on-shell}
\bar{k}  = \lcvec{k^+}{-\frac{t/4}{2 k^+}}{-\frac{1}{2}\,\vd}
\eqcm \hspace{2em}
\bar{k}' = \lcvec{k^+}{-\frac{t/4}{2 k^+}}{\frac{1}{2}\,\vd}
\eqcm
\end{equation}
respectively. The integrations over $k^-$ and $\vk$ in
(\ref{starting-point}) can then be performed explicitly, leaving us
with an integral $\int\d k^+ \int\d z^-$ and forcing the relative
distance of fields in the matrix elements on the light cone, $z = [\,
0, z^-, {\bf 0}_\perp]$. After this the time ordering of the fields
can also be dropped~\cite{DG}.

At this point one might be tempted to proceed as in standard DIS (or
in DVCS) and decompose $H$ on the Dirac matrices $\gamma^\rho$ and
$\gamma^\rho \gamma_5$. This leads to the Fourier transforms of the
nonlocal matrix elements $\langle p'|\, \overline\psi{}_{a}(0)\,
\gamma^\rho\, \psi_{a}(z) \,|p\rangle$, $\langle p'|\,
\overline\psi{}_{a}(0)\, \gamma^\rho \gamma_5\, \psi_{a}(z)
\,|p\rangle$, and the corresponding ones with the arguments 0 and $z$
interchanged. In DIS or DVCS, where only the \emph{plus} components of
the proton momenta are large, one has that only the plus components of
the currents give a leading contribution in the limit of large $Q^2$.
Now however we have a large scattering angle, and the proton momenta
have large plus, minus and transverse components, so that it does not
follow from kinematic considerations that the plus component of, say,
$\int\d z^-\, e^{i\, k^+ z^-} \langle p'|\, \overline\psi{}_{a}(0)\,
\gamma^\rho\, \psi_{a}(z) \,|p\rangle$ is large compared to its minus
or transverse components and thus dominates in the Compton amplitude.

To show that the plus components indeed dominate also in large angle
scattering we use that the proton-parton amplitudes described by the
soft matrix elements can be written as the amplitude for a proton with
momentum $p$ emitting the active parton with momentum $k$ and a number
of on-shell spectators times the corresponding conjugated amplitude
for momenta $p'$ and $k'$, summed over all spectator configurations;
this just corresponds to inserting a complete set of intermediate
states between the quark and antiquark fields in the matrix elements.
We note that for small $k^2$, $k'^2$ and small intrinsic transverse
parton momenta, ${\bf k}_{\perp i}^2 /x^{\phantom{.}}_i \lsim \had^2$,
one cannot form large kinematical invariants at the hadron-parton
vertices.\footnote{The situation is special for small momentum
  fraction $x$ of the active parton, when Fock states with large $N$
  are important; a case we do not consider here.}

For each of the proton-parton vertices we now go to a frame where the
momentum $\bar{k}$ or $\bar{k}'$ has a zero transverse (and thus also
a zero minus) component, performing a transverse boost as described in
Sect.~\ref{hadron-frames}. Considering for definiteness the case where
the parton coming out of the proton is a quark we write in this frame
\begin{eqnarray}  \label{projection}
\psi(z) &=& 
\frac{1}{2}\,\gamma^-\gamma^+\, \psi(z) +
\frac{1}{2}\,\gamma^+\gamma^-\, \psi(z) \nonumber \\
&=& \frac{1}{2k^+} \sum_{\lambda} 
    \left[ u(\bar{k},\lambda)\, 
    \left(\bar{u}(\bar{k},\lambda) \gamma^+ \psi(z) \right) +
    \gamma^+ u(\bar{k},\lambda)\,
    \left(\bar{u}(\bar{k},\lambda) \psi(z) \right) \right]
\end{eqnarray}
with a sum over helicities $\lambda /2 = \pm 1/2$. We can now argue
that in the matrix element of (\ref{projection}) between the incoming
proton and the spectator system the term with $\bar{u}
\gamma^+\psi(z)$ dominates over the one with $\bar{u} \psi(z)$ because
at the vertex we have a large plus component but no large invariant,
and thus retain only the first term in the decomposition
(\ref{projection}).\footnote{We note that this corresponds to the
  ``good'' component of the Dirac field in the context of light cone
  quantisation~\cite{bro89}.} Now we use that the boost
(\ref{eq:plustrafo}) to the frame where $\bar{k}$ has vanishing
transverse and minus components leaves the plus component of any
vector unchanged so that (\ref{projection}) also holds in the overall
symmetric frame of Sect.~\ref{symmetric-frame}. Repeating our argument
for the antiquark field we arrive at the replacement
\begin{equation} \label{replace}
 \overline\psi{}_{\alpha}(0)\, \psi_{\beta}(z) \to 
 \left( \frac{1}{2k^+} \right)^2 \sum_{\lambda,\lambda'}
 \left( \overline\psi(0) \gamma^+ u(\bar{k}',\lambda') \right)
 \left( \bar{u}(\bar{k},\lambda) \gamma^+ \psi(z) \right) \times
 \bar{u}_{\alpha}(\bar{k}',\lambda') \, u_{\beta}(\bar{k},\lambda)
\end{equation}
and an analogous one involving antiquark spinors for
$\overline\psi{}_{\alpha}(z)\, \psi_{\alpha'}(0)$. In
(\ref{starting-point}) the hard scattering kernels are then multiplied
with the spinors for on-shell (anti)quarks, which guarantees
electromagnetic gauge invariance of our result. Note that the full
expression (\ref{starting-point}) need \emph{not} be gauge invariant
since the handbag diagrams are not the complete set of diagrams for
our process.

To further simplify the hadronic matrix elements we use that the hard
scattering, where of course we neglect quark masses, conserves the
parton helicity: $\lambda' = \lambda$. In a suitable convention for
massless spinors one has $u(\bar{k},\lambda) = - v(\bar{k},-\lambda)$
and $\arg[\bar{u}(\bar{k}',\lambda) \gamma^+ u(\bar{k},\lambda)] = 1$
for any on-shell momenta $\bar{k}$, $\bar{k}'$, so that we can
multiply (\ref{replace}) with
\begin{equation} \label{trick}
1 = \frac{\bar{u}(\bar{k}',\lambda) \gamma^+ 
          u(\bar{k},\lambda)}{2k^+}  \eqpt
\end{equation}
With $u(\bar{k},\lambda)\, \bar{u}(\bar{k},\lambda) =
\bar{k}\cdot\gamma\, (1-\lambda \gamma_5) /2$ and analogous
expressions for $\bar{k}'$ and for antiquark spinors we obtain after a
little algebra
\begin{eqnarray} \label{half-way}
  \lefteqn{ {\cal A} = \sum_a (e e_a)^2 \int \d k^+\,
    \theta(k^+) \int {\d z^-\over 2\pi}\, e^{i\, k^+ z^-}\,
    \frac{1}{2k^+} \sum_{\lambda} } 
\nonumber \\ 
&& \times \left[ \langle p'|\, \overline\psi{}_{a}(0)\, \gamma^+
          \frac{1+\lambda\gamma_5}{2}\,\psi_{a}(z^-) \,
          |p\rangle \;
          \bar{u}(\bar{k}',\lambda) H(\bar{k}',\bar{k})
                 u(\bar{k},\lambda) \, \right.
\nonumber \\ 
&& \hspace{0.2em} \left. {}+
          \langle p'|\, \overline\psi{}_{a}(z^-)\, \gamma^+
          \frac{1-\lambda\gamma_5}{2}\,\psi_{a}(0) \,|p\rangle \;
          \bar{v}(\bar{k},\lambda) H(-\bar{k},-\bar{k}')
                 v(\bar{k}',\lambda) \, \right]  \eqcm
\end{eqnarray}
where we write $\psi(z^-)$ as a shorthand notation for $\psi(z)$ with
$z = [\, 0, z^-, {\bf 0}_\perp]$. We thus find that the plus component
of the nonlocal currents dominates as we have anticipated in our
footnote~\ref{other-foot}, and that the operators in the matrix
elements are in fact the same as those of the leading-twist parton
distributions occurring in DIS or DVCS.

We now must discuss what to take for $k^+$ in the hard scattering. As
shown in Sect.~\ref{sub-compton} the requirement to have no hard
partons directly coupling to the protons forces the active partons $k$
and $k'$ to have small intrinsic transverse momenta in their parent
hadrons and a momentum fraction $x$ close to one when $-t$ is large.
This corresponds to the approximation (\ref{on-shell}) with $k^+ =
p^+$ we will make in the hard scattering factors, i.e.\ the
expressions after the proton matrix elements in (\ref{half-way}). Some
degree of arbitrariness is associated with the global factor
$1/(2k^+)$ in (\ref{half-way}), which has its origin in
(\ref{projection}), (\ref{replace}) and for which we choose to keep
$k^+ = x p^+$. Admittedly there is no clear-cut way to associate it to
either the hard scattering, where we set $x=1$, or the soft matrix
elements, where setting $x=1$ would not even make sense since for $x$
strictly at its end point our proton LCWFs are zero.

Making use of the charge conjugation properties of Dirac matrices and
spinors in order to rewrite the term corresponding to antiquark-photon
scattering we obtain our final expression for the handbag diagrams in
large angle Compton scattering,
\begin{eqnarray} \label{final}
   {\cal A} &=& \frac{1}{4} \sum_{\lambda}
   \bar{u}(\bar{k}',\lambda) H(\bar{k}',\bar{k})
                 u(\bar{k},\lambda) \, 
   \sum_a (e e_a)^2\, \int_0^1 \frac{\d x}{x} \,
   \int {\d z^-\over 2\pi}\, e^{i\, x p^+ z^-} \nonumber \\ 
&& \times \left[ \langle p'|\,
     \overline\psi{}_{a}(0)\, \gamma^+\,\psi_{a}(z^-) - 
     \overline\psi{}_{a}(z^-)\, \gamma^+\,\psi_{a}(0)
     \,|p\rangle \right. \nonumber \\
&& + \left. \lambda\, \langle p'|\,
     \overline\psi{}_{a}(0)\, \gamma^+\gamma_5\,\psi_{a}(z^-) +
     \overline\psi{}_{a}(z^-)\, \gamma^+\gamma_5\,\psi_{a}(0) 
     \,|p\rangle \right]  \eqcm
\end{eqnarray}
with (\ref{hard-scattering}) and with (\ref{on-shell}) for $k^+ =
p^+$. We note that the Fourier transformed matrix elements in
(\ref{final}) are skewed parton distributions at $\zeta = 0$ and large
$-t$, as was already remarked in~\cite{rad98a}.  In (\ref{final}) we
have incorporated their support property $x<1$, cf.~\cite{rad97,DG}.
Following Radyushkin~\cite{rad98a} we introduce a form factor
decomposition
\begin{eqnarray} \label{R-form-factors}
\lefteqn{ \sum_a e_a^2\, \int_0^1 \frac{\d x}{x} \, p^+
   \int {\d z^-\over 2\pi}\, e^{i\, x p^+ z^-}
     \langle p'|\,
     \overline\psi{}_{a}(0)\, \gamma^+\,\psi_{a}(z^-) - 
     \overline\psi{}_{a}(z^-)\, \gamma^+\,\psi_{a}(0) 
     \,| p\rangle } \nonumber \\
&& \hspace{11em} = R_V(t)\, \bar{u}(p')\, \gamma^+ u(p)\, +
   R_T(t)\, \frac{i}{2m}\, \bar{u}(p')\, \sigma^{+\nu}
                  \Delta_\nu \, u(p) \eqcm \nonumber \\
\lefteqn{ \sum_a e_a^2\, \int_0^1 \frac{\d x}{x} \, p^+
   \int {\d z^-\over 2\pi}\, e^{i\, x p^+ z^-}
     \langle p'|\,
     \overline\psi{}_{a}(0)\, \gamma^+\gamma_5\,\psi_{a}(z^-) +
     \overline\psi{}_{a}(z^-)\, \gamma^+\gamma_5\,\psi_{a}(0) 
     \,| p\rangle } \nonumber \\
&& \hspace{11em} = R_A(t)\, \bar{u}(p')\, \gamma^+\gamma_5 u(p)\, +
   R_P(t)\, \frac{\Delta^+}{2m}\, \bar{u}(p')\, \gamma_5\, u(p)
\end{eqnarray}
for the $x$-integrals over these skewed distributions.
$R_V$, $R_T$, $R_A$ and $R_P$ are new form factors 
specific to Compton scattering;
note that $R_P$ does not contribute to the Compton amplitude in our
symmetric frame with $\Delta^+=0$.

One may ask how to improve on the approximation (\ref{on-shell}) with
$k^+ = p^+$ when calculating the hard scattering. There will be
corrections due to the facts that in the hard scattering
\begin{enumerate}
\item $x$ is not strictly one,
\item the intrinsic transverse momenta of the partons $k$, $k'$ are
  nonzero, and
\item the virtualities $k^2$, $k'^2$ are not zero.
\end{enumerate}
The order of magnitude of all these corrections is controlled by the
parameter $\had^2 /(-t)$ as discussed in Sect.~\ref{sub-compton}.
Note that in order to express the amplitude in terms of LCWFs or of
the light cone matrix elements in (\ref{half-way}) it was essential to
neglect the $k^-$-dependence of the hard scattering. The inclusion of
off-shell corrections (point 1) would thus necessitate an extension of
the framework we are using here. We emphasise that the on-shell
condition in the hard scattering is our guarantee to obtain a gauge
invariant result; ``exactly'' evaluating the handbag diagrams would
only have a limited sense since a part of the corrections to
(\ref{on-shell}) will break gauge invariance and be cancelled by other
diagrams. Furthermore points~1, 2 and 3 are kinematically related:
from $k^2 - k'^2 = 2 \vd \cdot (\vk + \vd/2)$ in our symmetric frame
we see that if we insist on taking on-shell partons in the hard
scattering then we must fix $\vk = -\vd/2$ (as we did in
(\ref{on-shell})), which forbids us to evaluate the effect from the
variation of $\vk$ in the hard scattering kernel. We also see that for
$x\neq 1$ the choice $\vk = -\vd/2$ no longer corresponds to zero
intrinsic transverse momenta $\vk+x\vd/2$ and $\vk+(2-x)\, \vd/2$ of
$k$ and $k'$ in their parent hadrons.

Compared with fixing $\vk$ the approximation $x=1$ in the hard
scattering presents the particularity that $x$ is taken at its
kinematical end point; the soft part of the process can only select
$x$ around some value smaller than 1. Moreover, we find that with our
ansatz for the LCWFs (Sect.~\ref{wave}) both the $x$-integrals in
(\ref{R-form-factors}) and the corresponding one for the elastic form
factor are dominated by values of $x$ not very close to 1 for $-t$
between, say, 5 and 20 GeV$^2$, with the peaks of the integrands being
of order 0.45 to 0.75. The reason is that with our wave functions the
end point $x=1$ is rather strongly suppressed in the integrands of
(\ref{nform}), (\ref{rva}) by a third power $(1-x)^3$,
cf.~(\ref{disN}), (\ref{qpowers}), (\ref{disNS}), and that the
suppression of large $\vk^2 /x$ in the LCWFs is only effective for
values clearly larger than 1 GeV$^2$. It turns out that the factor
$1/x$ in (\ref{R-form-factors}) does not significantly shift the
values of $x$ where the integrand has its maximum, but rather
increases the height of the peak.

One might think of only dropping the approximation $x=1$ then, but
allowing $x$ to be different from 1 in the hard scattering would lead
to serious problems: in the case of a real incident photon for
instance one easily calculates that for $\vk = -\vd/2$ and $x=\eta=
(\sqrt{s} - \sqrt{-u}) /(\sqrt{s} + \sqrt{-u})$ one has $\hat{u}=0$.
It would however be mistaken to treat this as a pole in the hard
scattering (\ref{hard-scattering}) which gives an imaginary part to
the scattering amplitude. We must remember from our discussion of the
$k^-$-integration in Sect.~\ref{sub-compton} that we have already
neglected certain terms where $\hat{u}$ has a pole. Retaining others
by allowing $x$ to range from 0 to 1 in the hard scattering is then
inconsistent and would give misleading results. What happens in this
example is that the factorisation into a hard scattering and a soft
proton matrix element breaks down for $x$ not sufficiently large.
Keeping $x=1$ fixed in $\bar{u}(\bar{k}') H(\bar{k}',\bar{k})
u(\bar{k})$ is thus related to our approximation of factorising the
soft overlap contribution to Compton scattering into a hard
parton-photon scattering and a soft proton matrix element.

The fact that in our numerical applications the hadron wave functions
are probed at intermediate rather than very large $x$ means on one
hand that our results are not too sensitive to the precise behaviour
of the LCWFs near $x=1$, and also not to a possible Sudakov
suppression (cf.~\cite{bol96} for comments on these points in the case
of the elastic form factor). On the other hand our approximation
$x=1$ in the hard scattering of the Compton process has only a limited
accuracy for $-t$ not very large.

We finally also neglect the proton mass when relating $\hat{s}$ and
$\hat{u}$ to the external variables. Comparing (\ref{symm-frame-1})
with (\ref{on-shell}) at $x=1$ we see that this means $\bar{k} \approx
p$ and $\bar{k}' \approx p'$ so that we have $\hat{s} \approx s$ and
$\hat{u} \approx u$. Corrections to this will be of relative order
$m^2/(-t)$ and thus of the same size as other terms we do not control.

\subsection{Proton spin}
\label{proton-spin}

We have already remarked that the hard scattering subprocess does not
change the helicity of the active parton (the same holds for the
quark-photon coupling in the elastic form factor). As the helicities
of the spectators do not change either a change in the proton helicity
implies that for at least one of the incident or scattered proton the
parton helicities do not add up to the hadron helicity. In other words
the calculation of proton spin flip amplitudes requires to take into
account LCWFs with nonzero orbital angular momentum $L_3$ of the
partons in a detailed manner;\footnote{In this respect one has the
  same situation as in the hard scattering formalism~\cite{bro81}.}
this will not be attempted in the present work. For the lowest, three
quark Fock state we only take a wave function with zero $L_3$, which
has been constrained by several physical observables in~\cite{bol96},
and do not endeavour to model wave functions with $L_3 \neq 0$. For
higher Fock states, which for sufficiently large $-t$ provide only a
correction to the three-quark contribution in Compton scattering and
the elastic form factor, we will not specify how the orbital angular
momenta between the various partons are explicitly coupled; describing
such detailed effects is not within the scope of this paper.

Due to its finite mass the helicity of a proton depends of course on
the choice of reference frame. Taking the incident proton for
definiteness we can express this dependence in a covariant way using
its spin four-vector $s$. In the hadron frame of
Sect.~\ref{hadron-frames}, where $p$ has zero transverse momentum, the
spin vector for a state of definite helicity is a linear combination
of $p$ and the vector $v' = [\, 0,1, {\bf 0}_\perp]$, which is
unchanged by the boost to the overall symmetric frame. This choice of
spin quantisation axis is natural in our context of LCWFs, which are
defined with respect to the same vector $v'$ through the integration
over the minus components of parton momenta. A corresponding argument
holds for the scattered proton, with the same vector $v'$. We find
that in our symmetric frame with $\zeta=0$  the helicity flip
amplitudes are only due to $R_T$, which we will therefore
not be able to model here, whereas the helicity conserving ones go
with $R_V$ and $R_A$. The same holds for the elastic form factors:
$F_2$ changes helicity and $F_1$ does not, and we will only calculate
$F_1$.

We know from experiment that in the transition $\gamma^\ast p \to p$
proton helicity flip becomes small compared with no flip for large
enough $-t$, so that neglecting the former can be justified as an
approximation. The measured difference between the Dirac form factor
$F_1$ and the magnetic Sachs form factor $G_M = F_1 + F_2$ at a given
$-t$ shows the degree of accuracy of neglecting spin flip
contributions, and it is reasonable to assume that the situation will
be similar for the new form factors (\ref{R-form-factors}).

\subsection{The hard scattering}
\label{hard-scatter}

We now give the hard scattering amplitudes
\begin{equation} \label{hard-amplitudes-def}
  {\cal H}_{\lambda, \, \mu\mu'} = 
  \bar{u}(\bar{k}',\lambda) H(\bar{k}',\bar{k})
                   u(\bar{k},\lambda)
\end{equation}
where $\mu$ and $\mu'$ respectively denote the helicity of the initial
and final state photon. For virtual Compton scattering the initial
photon helicity depends on the reference frame and we choose to define
it in the photon-proton c.m., i.e\ with respect to the $p$-$q$ axis:
our symmetric $\zeta=0$ frame is adapted to discuss the physics of our
reaction mechanism, but $\gamma^\ast$-polarisations defined in the
c.m.\ are well suited for the consideration of azimuthal asymmetries
we shall briefly mention below, apart from being a standard choice
that facilitates comparison with other work. With our approximation
$\bar{k} \approx p$, $\bar{k}' \approx p'$ the photon-proton c.m.\ is
identical to the c.m.\ of the hard subprocess $q(\bar{k})\,
\gamma^\ast(q) \to q(\bar{k}')\, \gamma(q')$. In our phase convention,
where $\arg[\bar{u}(\bar{k}',\lambda) \gamma^+ u(\bar{k},\lambda)] =
1$, we explicitly find
\begin{eqnarray} \label{hard-amplitudes}
{\cal H}_{+,\, ++} &=&   2\, \sqrt{\frac{s}{-u}}\: \frac{s+Q^2}{s}
\eqcm \hspace{3em}
{\cal H}_{+,\, --} \:=\: 2\, \sqrt{\frac{-u}{s}}\: \frac{s}{s+Q^2}
\eqcm \nonumber \\
{\cal H}_{+,\, +-} &=&   2\, \frac{Q^2}{s+Q^2}\: \frac{t}{\sqrt{-s u}}
\eqcm \hspace{2.5em}
{\cal H}_{+,\, -+} \:=\: 0
\eqcm \nonumber \\
{\cal H}_{+,\, 0-} &=&   - 2\: \frac{Q}{s+Q^2}\: \sqrt{-2t}
\eqcm \hspace{2.3em}
{\cal H}_{+,\, 0+} \:=\: 0
\eqcm
\end{eqnarray}
with the kernels for $\lambda=-1$ given by parity invariance as ${\cal
  H}_{\lambda, \, \mu\mu'} = (-1)^{\mu-\mu'} {\cal H}_{-\lambda, \,
  -\mu-\mu'}$.

With (\ref{final}) and (\ref{hard-amplitudes}) we have all necessary
ingredients to calculate the cross section in terms of the form
factors $R_V$, $R_A$ (and $R_T$ which we will neglect in this
work). We present a numerical study of real Compton scattering in
Sect.~\ref{compt}. Virtual Compton scattering is measured in
electroproduction, $e p \to e p \gamma$, where it interferes with the
Bethe-Heitler process, i.e.\ the emission of the final state photon
from the lepton, and its detailed study shall not be attempted here.
The results in the handbag mechanism have however some general
features, both for real and virtual initial photons, which we discuss
now.

The first point is that the photon-proton amplitude comes out as
purely real: the form factors $R_V$, $R_A$, $R_T$ are real due
to time reversal invariance, and the hard scattering kernel does not
have an imaginary part because the corresponding diagrams cannot be
cut with $\hat{s}$ and $\hat{u}$ being far off-shell; such cuts only
arise at the level of $\alpha_s$-corrections to the photon-parton
scattering. In the hard scattering mechanism~\cite{bro80} the
situation is very different: there one has cuts already to leading
order in $\alpha_s$, which lead to nontrivial phases in the scattering
amplitude. This may offer a valuable tool to distinguish
experimentally which reaction mechanism is at work: in $e p \to e p
\gamma$ with longitudinally polarised lepton beams the beam
polarisation asymmetry is proportional to the imaginary parts of the
$\gamma^\ast p \to \gamma p$ helicity amplitudes, with the
Bethe-Heitler amplitude being purely real. In the handbag mechanism
this polarisation asymmetry is then predicted to be small, arising
only at the level of loop corrections, while in the hard scattering
mechanism it can have a substantial value. This was for instance shown
in ~\cite{kro96}, where virtual Compton scattering was studied within
the hard scattering approximation using a quark-diquark wave function
for the proton.

A second remarkable feature of (\ref{hard-amplitudes}) is the
dependence on the photon helicities: transitions between positive and
negative helicities are forbidden for real photons and suppressed by
$Q^2 /(s+Q^2)$ if the photon virtuality is small compared with
$s$.\footnote{Whether this still holds at the level of
  $\alpha_s$-corrections would need further investigation.} This
helicity selection rule could be tested in real Compton scattering
with linearly polarised incident photons: it leads to the absence of a
dependence of the cross section on the azimuth $\Phi$ between the
plane of photon polarisation and the scattering plane; nonzero photon
helicity flip amplitudes will in general give a $\cos
2\Phi$-contribution to the differential cross section. For finite but
not very large $Q^2$ the situation is more complicated, because the
helicity flip amplitude is only suppressed and not zero, and because
the process interferes with Bethe-Heitler.

%%%%%%%%%%%%%%%%%%%%%%%%%%%%%%%%%%%%%%%%%%%%%%%%%%%%%%%%%%%%%%%%%%%%%%%%%
\section{The Fock state wave functions}
\label{wave}
%%%%%%%%%%%%%%%%%%%%%%%%%%%%%%%%%%%%%%%%%%%%%%%%%%%%%%%%%%%%%%%%%%%%%%%%%
The valence Fock state of the nucleon has been investigated in some
detail in Ref.~\cite{bol96}. The explicit form of the corresponding
wave function has been extracted from a fit to the valence quark
distribution functions derived in Ref.~\cite{GRV} and to the Dirac
form factor of the proton assuming dominance of the soft overlap
contribution. This is just the physics we are interested in here;
therefore we take over the results of Ref.~\cite{bol96} as a starting
point.  The wave function proposed in Ref.~\cite{bol96} has also been
shown to work successfully for $J/\psi$ decays into proton-antiproton
pairs, a process that is well under control of perturbative physics in
contrast to, for instance, the form factors in the experimentally
accessible region of momentum transfer. In the subsequent sections we
will test that wave function in further observables, namely in Compton
scattering and in the polarised parton distributions. We will even go
a step further than in Ref.~\cite{bol96} and explore the next two
higher Fock states consisting of four and five partons in order to
determine their gross features. Moreover, we are going to investigate
the global effect of all Fock states in an approximate way.  As has
been shown recently by Radyushkin~\cite{rad98a}, one can then directly
relate the parton distributions controlling deeply inelastic
lepton-nucleon scattering with exclusive observables such as form
factors or the Compton cross section, without assuming an explicit
form of the distribution amplitudes.

For the reader's convenience we will start with a brief description of
the properties of the LCWF for the proton's valence Fock state derived
in~\cite{bol96}. According to Sotiropoulos and Sterman~\cite{sot94}
the valence Fock state of a proton with momentum $p$ and positive
helicity can be written as
\begin{eqnarray}
  |\,P,p,+;\q\q\q \rangle\; &=&
                    \int [{\rm d}x]_3 [{\rm d}^{2}\vk]_3\;
                    \Bigl\{
                    \Psi _{123}\,{\cal M}_{+-+} +
                    \Psi _{213}\,{\cal M}_{-++}     \nonumber \\
 && \hspace{7em}  - \Bigl(\Psi _{132}\, + \,
                    \Psi _{231}\Bigr){\cal M}_{++-}  \Bigr\}  \eqcm
\label{state}
\end{eqnarray}
with plane wave exponentials and the colour wave functions omitted
here and in the following. The integration measures in Eq.\ 
(\ref{state}) are defined by
\begin{equation}
[{\rm d}x]_N \equiv \prod_{i=1}^N {\rm d}x_i\,
     \delta(1-\sum_i x_i) \eqcm  \hspace{3em}
[{\rm d}^2 \vk]_N \equiv \frac{1}{(16\pi^3)^{N-1}}\,\prod_{i=1}^N 
     {\rm d}^2 \vk{}_i \, 
     \delta^{(2)}(\sum_i \vk{}_i) \eqpt
\label{mass}
\end{equation}
The quark $i$ is characterised by its plus momentum $k^+_i = x_i \,
p^+$, its transverse momentum ${\bf k}_{\perp i}$ with respect to the
proton's momentum, and by its helicity $\lambda_i$. A three-quark
state is then given by
\begin{equation}
  \hspace{-1cm}
  {\cal M}_{\lambda_1 \lambda_2 \lambda_3 } =
  \frac{1}{\sqrt{x_1 x_2 x_3}} \,
    |\,\u; x_1 p^+, \vk{}_1, \lambda_1\rangle \,
    |\,\u; x_2 p^+, \vk{}_2, \lambda_2\rangle \,
    |\,\d; x_3 p^+, \vk{}_3, \lambda_3\rangle
  \hspace{-1cm}
\label{quark}
\end{equation}
with a normalisation
\begin{equation}
  \langle \q; x'_i p^+, {\bf k}'_{\perp i},
  \lambda'_i \,|\, 
    \q; x_i p^+, \vk{}_i, \lambda_i\rangle =
     2 x_i p^+ (2\pi)^3 \,
     \delta_{\lambda'_i \lambda^{\phantom{.}}_i} \delta(x'_i p^+
      -x^{\phantom{.}}_i p^+) \,
     \delta({\bf k}'_{\perp i} - 
            {\bf k}^{\phantom{.}}_{\perp i}) \eqpt
\label{normq}
\end{equation}
A neutron state is obtained by the exchange $\u \leftrightarrow \d$.

We only consider the part of the wave function with zero orbital
angular momentum $L_3$ along the 3-axis, so that the quark helicities
sum up to the proton's helicity. As has been demonstrated in
Ref.~\cite{dzi88} Eq.\ (\ref{state}) is the most general ansatz for
the $L_3 = 0$ projection of the three-quark proton wave function:
{}From the permutation symmetry between the two $\u$-quarks and from
the requirement that the three quarks have to be coupled in an isospin
$1/2$ state it follows that there is only one independent scalar wave
function, which for convenience is parametrised as
\begin{equation}
  \Psi _{123}(x_i,{\bf k}_{\perp i})
\equiv
   \Psi(x_1,x_2,x_3;{\bf k}_{\perp 1},
   {\bf k}_{\perp 2},{\bf k}_{\perp 3})
=              \frac{f_3}{8\sqrt{6}}\, \phi_{123}(x_i)\,
               \Omega_3 (x_i,{\bf k}_{\perp i})
\label{Psiansatz}
\end{equation}
with the normalisation conditions
\begin{equation}
  \int [{\rm d}x]_3\, \phi_{123}(x_i) = 1 \eqcm  \hspace{3em}
  \int [{\rm d}^2 \vk]_N \, \Omega_N(x_i,\vk{}_i) = 1 \eqpt
\label{norm}
\end{equation}
$f_3$ plays the role of the nucleon wave function at the origin of
coordinate space and $\phi_{123}(x_i) \equiv \phi(x_1,x_2,x_3)$ is the
nucleon's valence distribution amplitude. Both quantities depend on a
factorisation scale $\mu_F$ and are subject to evolution. Expanding
$\phi_{123}(x_i)$ as
\begin{equation}
  \phi_{123}(x_i,\mu_F) = \phi_{\rm AS}(x_i) \left[ \, 1 +
  B_1(\mu_F)\, \tilde \phi_{123}^1(x_i) + 
  B_2(\mu_F)\, \tilde \phi_{123}^2(x_i) + \ldots\, \right]  \eqcm
\label{da-expansion}
\end{equation}
where $\phi_{\rm AS}(x_i) = 120\,x_1 x_2 x_3$ is the asymptotic \da{}
and $\tilde \phi_{123}^1(x_i)$, $\tilde \phi_{123}^2(x_i)$, etc.\ are
the eigenfunctions of the evolution kernel~\cite{bro80}, one has
\begin{equation}
  \hspace{-17mm}
  f_3(\mu_F)  =  f_3(\mu_0)\,\left(
    \frac{\ln(\mu_0/\LQCD)}{\ln(\mu_F/\LQCD)} \right)^{2/(3\beta_0)}
     \hspace{-0.8cm} , \hspace{1cm}
  B_n(\mu_F)  =  B_n(\mu_0)\,\left(
    \frac{\ln(\mu_0/\LQCD)}{\ln(\mu_F/\LQCD)} \right)^{\tilde
    \gamma_n/\beta_0} \hspace{-8mm} 
  \hspace{5mm}
\label{BnFNevol}
\end{equation}
with $\beta_0 \equiv 11 - 2 n_f /3$, $\tilde\gamma_1=20/9$,
$\tilde\gamma_2=8/3$, etc.  In Ref.~\cite{bol96} it has been shown
that it is sufficient to retain only the first two terms in the
expansion (\ref{da-expansion}). They are taken as $B_1(\mu_0)=3/4$ and
$B_2(\mu_0)=1/4$ at a factorisation scale of $\mu_0 = 1\gev$. At this
scale one then has the simple form
\begin{equation}
\phi_{123}(x_i) = 60\,x_1 x_2 x_3 \,( 1 + 3 x_1 )
\label{da3}
\end{equation}
for the valence \da{}.

When calculating the overlap contributions to the elastic form factor
and to large angle Compton scattering in Sect.~\ref{form} and
\ref{compt} we will use the \da{} at a factorisation scale $\mu_F^2 =
-t$ given by the momentum transfer to the nucleon.\footnote{For the
  higher Fock state LCWFs to be discussed below the evolution will be
  neglected.} The parton distributions in Sect.~\ref{distributions}
and \ref{skewed} will be calculated and compared with the
parametrisations from global fits at our starting scale $\mu_0^2 = 1
\gev^2$.

The transverse momentum dependence of the wave function is contained
in the function $\Omega_N$. A simple symmetric Gaussian
parametrisation,
\begin{equation}
  \Omega_N(x_i,{\bf k}_{\perp i}) =
  \frac{(16\pi^{2}a^{2}_N)^{N-1}}{x_{1}x_{2}\ldots x_{N}}
  \exp
     \left [
            -a_N^{2} \sum_{i=1}^{N}{{\bf k}_{\perp i}^2 \over x_{i}}
     \right ]\eqcm
\label{BLHMOmega}
\end{equation}
suffices and meets various theoretical requirements, see for instance
\cite{BHL,chi95,bro98} and our remark following
Eq.~(\ref{constraints}). This ansatz keeps the model simple and allows
one to carry through the $\vk$-integrations analytically. Note that
the ansatz (\ref{Psiansatz}), (\ref{BLHMOmega}) represents a
\emph{soft} wave function, i.e.\ the full wave function where the
perturbative tail with its power-like decrease is
removed~\cite{bro80}. Integrating $\Omega_N$ in Eq.~(\ref{norm}) to
infinity instead of to a cut-off scale given by the hard scale in a
process introduces only a small negligible error.
  
The values of the normalisation $f_3$ and the transverse
size parameter $a_3$ have been determined in~\cite{bol96} as
\begin{equation} \label{da3-parameters}
f_3 = 6.64 \cdot 10^{-3}\, \gev^2 \eqcm \hspace{3em}
a_3 = 0.75 \,\gev^{-1}
\end{equation}
at the scale of reference $\mu_F=\mu_0$. With these parameters the
valence Fock state wave function has a value of 0.17 for its
probability and a value of 411 MeV for the rms transverse momentum.
The valence Fock state thus appears to be rather compact, with a
radius of only about a half of the charge radius. For further
discussion of the properties of the valence Fock state wave function
see~\cite{bol96}.

With the valence Fock state fully specified we can now turn to the
higher ones. Explicitly we only consider the Fock states with an
additional gluon ($N=4$) and with an additional sea quark-antiquark
pair ($N=5$). Due to parity conservation both require one unit of
orbital angular momentum. One therefore encounters many different
possibilities of coupling the various partons in a nucleon, each
coming with a new wave function. It seems plausible to assume that the
effect of the orbital angular momentum is averaged out in the sum over
all different coupling possibilities.\footnote{In principle there is
  no difficulty in treating all possibilities explicitly. Each of them
  is described by an appropriate covariant spin wave
  function~\cite{hus91,bol94} that is proportional to $K \cdot
  \gamma$, where $K_{\mu}$ is the relative momentum of two clusters of
  partons. These $K \cdot \gamma$-terms, representing the orbital
  angular momentum between the two clusters, give rise to an
  additional factor $ \sim K'_{\mu}\,K^{\mu}$ in the expressions for
  observables like the overlap integral for the nucleon form factor.}

With this proviso in mind we take
\begin{eqnarray}
  |\,P,p,+;\q\q\q\g \rangle\;& =&\; 
            \int  [{\rm d}x]_4  [{\rm d}^{2}{\bf k}_{\perp}]_4\;
  \Psi_{1234}(x_i,\vk{}_i) \nn\\
    &&  \times \left[ {\cal M}_{++-} \,-\, {\cal M}_{+-+} \right] \,
             \frac{1}{\sqrt{x_4}}\,
    |\,\g; x_4 p^+, {\bf k}_{\perp 4}, \lambda_4\rangle \,
\label{Psi4}
\end{eqnarray}
as a representative of all $N=4$ Fock states, with the gluon state
$|\,\g; x_4 p^+, {\bf k_{\perp}}_4, \lambda_4\rangle$ normalised as in
(\ref{normq}) and
\begin{eqnarray}
\Psi_{1234}(x_i,\vk{}_i) &=&\frac{f_4}{8\sqrt{2}}\;
         \phi_{1234}(x_i)\, \Omega_4 (x_i,\vk{}_i) \ .
\label{wf4}
\end{eqnarray}
For the \da{} of this Fock state we take (at the scale
$\mu_0$)
\begin{equation}
\phi_{1234}(x_i)\,=\, 
    \frac{9!}{30}\,x_1 x_2 x_3 x_4^2\;(1+3x_1 ) \eqcm
\label{da4}
\end{equation}
i.e.\ the \da{} has the asymptotic form multiplied by an asymmetry
factor of the same type as in the \da{} for $N=3$. The spin-isospin
coupling of the valence quarks requires a \da{} that is symmetric
under the exchange $2\leftrightarrow 3$. The gluon is supposed to
couple with the orbital angular momentum in a spin zero state. Thus
the ansatz (\ref{Psi4}) satisfies the minimal requirement that the
partons of this Fock state are coupled in a spin-isospin $1/2$ state.

For the $N=5$ Fock state we assume a sea that is colourless, SU(3)
flavour symmetric and coupled to total spin zero.\footnote{ The $N=5$
  Fock state with two gluons in it is discarded since its contribution
  to physical quantities is highly suppressed in the kinematical
  region of interest, cf.\ our remark after (\ref{gpowers}) below. If
  in the following we talk about higher Fock states $(N>5)$, this
  particular Fock state is understood to be included.} The
generalisation to a more complicated sea is straightforward, requiring
flavour-dependent wave functions which may also have additional
asymmetries in their $x_i\,$-dependence, but in order to keep the
model simple we refrain from introducing such wave functions. With our
simple ansatz the valence quarks are in a totally symmetric state in
flavour-spin-momentum-space, just as the valence Fock state itself,
and the valence sector of the $N=5$ Fock state therefore exhibits the
same structure as (\ref{state}). Assuming its wave function to equal
that of the valence Fock state we make the ansatz
\begin{eqnarray} 
\lefteqn{ \hspace{-1em} |\,P,p,+;\q\q\q\, \q\bar{\q} \rangle \; =\; 
          \int  [{\rm d}x]_5  [{\rm d}^{2}{\bf k}_{\perp}]_5 } \nn \\
 && \hspace{-1em} \times\, \Bigl\{
                    \Psi _{12345}\,{\cal M}_{+-+} +
                    \Psi _{21345}\,{\cal M}_{-++}  
      - \Bigl(\Psi _{13245}\, + \,
                    \Psi _{23145}\Bigr){\cal M}_{++-}  \Bigr\}  \nn \\
 && \hspace{-1em} \times\,
    \sum_{q=u,d,s} \, \frac{1}{\sqrt{x_4 x_5}} \Bigl\{
      |\,\q; x_4 p^+, {\bf k}_{\perp 4}, +\rangle \,
      |\,\bar{\q}; x_5 p^+, {\bf k}_{\perp 5}, -\rangle \nn \\
 && \hspace{5.2em}
  {}- |\,\q; x_4 p^+, {\bf k}_{\perp 4}, -\rangle \,
      |\,\bar{\q}; x_5 p^+, {\bf k}_{\perp 5}, +\rangle \Bigr\}
\label{wav5}
\end{eqnarray}
with
\begin{equation}
\Psi_{12345}(x_i,\vk{}_i) = \frac{f_5}{48}\; 
                 \phi_{12345}(x_i)\, \Omega_5(x_i,\vk{}_i)
\end{equation}
for the wave function and
\begin{equation}
  \phi_{12345}(x_i)\,=\, 
  \frac{10!}{16}\,x_1 x_2 x_3 x_4 x_5\;(1+3x_1)
\label{da5}
\end{equation}
for the \da{} at scale $\mu_0$. The symmetrisation between the sea and
valence quarks required by the Pauli principle is ignored here. We
argue that it cannot play a major role because of the fairly large
spatial separations between sea and valence quarks: the sea quarks
have to build up the full charge radius of the nucleon while the
valence quarks form a compact core.

Admittedly our parametrisations of the higher Fock states are
oversimplified. For the physical processes and in the kinematical
region of interest here they give however only small contributions
compared with the valence Fock state, and to investigate these
corrections we deem our ansatz to be sufficiently accurate.

We finally give an integral which will appear in our overlap formulae,
namely
\begin{equation} 
I_{N}(x_i,\zeta,\vd^2) = \int [\d^2 \vk]_N \; 
                    \Omega_N(x'_i,{\vk'}_i) \, \Omega_N(x_i,\vk{}_i) 
\label{gint1}
\end{equation} 
where the relation between the primed and unprimed variables is given
by (\ref{breve-args}) and for $\zeta=0$ also by (\ref{tilde-args}). An
appropriate tilde, hat or breve upon the variables $x_i,\vk{}_i$ is
understood depending on the case. For our Gaussian $\vk$-dependence
the integral (\ref{gint1}) evaluates to
\begin{equation} 
I_{N}(x_i,\zeta,\vd^2)\,=\, \frac{\rho_N} {x_1'\ldots x_N'}  \;
                           \Upsilon_N (x_j,\zeta; \vd^2)  \eqcm
\label{gint2}
\end{equation} 
where
\begin{equation}
\Upsilon_N (x_j,\zeta; \vd^2)=
\left(\frac{2}{2-\zeta}\right)^{N-2}
      \frac{2 (x_j-\zeta)}{(x_j-\zeta)+x_j(1-\zeta)^2}\;
      \exp\left[\frac{-\,a_N^2\,\vd^2\,(1-x_j)}
                     {(x_j-\zeta)+x_j(1-\zeta)^2}\right] 
\label{eq:SKgauss}
\end{equation} 
and
\begin{equation} 
\rho_N = (8\pi^2 a_N^2)^{N-1} \eqpt
\end{equation}
Notice that $I_N$ is a function of all momentum fractions $x_i$
whereas $\Upsilon_N$ only depends on the fraction $x_j$ of the active
quark.

Turning to a more generic notation we have that each Fock state is
described by a sum of terms, each with its own momentum space wave
function $\Psi_{N\beta}$, where $\beta$ labels different spin-flavour
combinations of the partons. On the basis of this notation the Fock
state probabilities are given by $P_N \equiv \sum_\beta \int [\d x]_N
[\d^2 {\bf k}_\perp]_N \, |\Psi_{N\beta}(x_i,\vk{}_i)|^2$. For our
parametrisations we have
\begin{equation}
P_3=\frac{435}{224}\,\rho_3\,f_3^2 \eqcm \qquad  
P_4=\frac{27972}{275}\,\rho_4\,f_4^2 \eqcm \qquad
P_5=\frac{685125}{352}\,\rho_5\,f_5^2 \eqcm 
\label{prob}
\end{equation}
and with (\ref{da3-parameters}) we obtain $P_3=0.17$ as already
mentioned above.

%%%%%%%%%%%%%%%%%%%%%%%%%%%%%%%%%%%%%%%%%%%%%%%%%%%%%%%%%%%%%%%%%%%%%%%%%%%%
\section{Parton distributions}
\label{distributions}
%%%%%%%%%%%%%%%%%%%%%%%%%%%%%%%%%%%%%%%%%%%%%%%%%%%%%%%%%%%%%%%%%%%%%%%%%%%%
As shown by Brodsky and Lepage~\cite{bro80} the contribution of an
$N$-particle Fock state to the distribution function for a parton of
type $a$ in the proton is generically given by
\begin{equation}  
\q^{(N)}_a(x)\,=\, \sum_j\; \sum_\beta 
\int [\d x]_N [\d^2 {\bf k}_\perp]_N \,
   \delta(x - x_j)\; |\Psi_{N\beta}(x_i,\vk{}_i)|^2
\label{disf}
\end{equation}
where the sum $j$ runs over all partons of type $a$. Summation over
all Fock states leads to the full distribution functions
\begin{equation}
  \q_a(x)\,=\, \sum_N\, \q^{(N)}_a(x) \eqpt
\label{dis}
\end{equation}
Note that in our notation $\q_a$ stands for the distributions of
quarks, antiquarks or gluons. {}From the wave functions defined in
Sect.~\ref{wave} and with the help of (\ref{gint2}) for $\zeta=0$ and
$\vd=0$ one easily finds the individual contributions to the
distribution functions from the $N=3,4,5$ Fock states as a function of
the parton momentum fraction $x$:
\begin{equation}
\q_a^{(N)}(x) = b_a^{(N)}\,P_N\, 
                     x^{n_a} \,(1-x)^{m_a(N)} \,
                     \;\left[ 1 + c_a^{(N)}\, (1-x)
                            + d_a^{(N)}\, (1-x)^2 \right]  \eqcm
\label{disN}
\end{equation}
where the coefficients $b_a^{(N)}$, $c_a^{(N)}$ and $d_a^{(N)}$ are
compiled in Tab.~\ref{tabc}. 

As usual we define a valence quark distribution by $\q_v^{(N)}(x)
\equiv \q^{(N)}(x) - \bar \q^{(N)}(x)$ for $\q = \u, \d$.  The sea is
flavour symmetric in our simple model , hence
\begin{equation}  \label{sea-symmetric}
\ubar^{(5)}(x)=\dbar{}^{(5)}(x) = \sbar^{(5)}(x) = \s^{(5)}(x)  \eqpt
\end{equation}
With our particular ansatz (\ref{wav5}), (\ref{da5}) we also have 
\begin{equation}  \label{sea-special}
\dbar{}^{(5)}(x) = \d_v^{(5)}(x)/3  \eqpt
\end{equation}

\renewcommand{\arraystretch}{1.5}
\begin{table*}
  \begin{center}
  \begin{tabular}{|c||c|c|c||c|} \hline
  $\;\q_a^{(N)}$ & $b_a^{(N)}$ 
& $c_a^{(N)}$ & $d_a^{(N)}$ 
& $m_a(N)$ \\ \hline\hline
  $\u_v^{(3)}$ & $14\cdot\frac{140}{29}$ 
& $ -\frac{6}{7}$ & $\frac{12}{35}$ 
& 3 \\
  $\d_v^{(3)}$ & $\phantom{14\cdot}\;\frac{140}{29}$ 
& $\phantom{-} 3$ & $\frac{12}{5}$  
& 3 \\
  $\u_v^{(4)}$ & $17\cdot\frac{990}{37}$ 
& $-\frac{45}{34} $&$ \frac{39}{68}$ 
& 7 \\
  $\d_v^{(4)}$ & $\phantom{14\cdot}\;\frac{990}{37}$ 
& $\phantom{-}\frac{3}{2}$ & $\frac{3}{4}$ 
& 7 \\
  $\g_{\phantom{v}}^{(4)}$ & $\phantom{1}7\cdot\frac{990}{37}$ 
& $\phantom{-} 2 $&$ \frac{9}{7}$ 
& 5 \\
  $\u_v^{(5)}$ & $14\cdot\frac{792}{29}$ 
& $-\frac{15}{14} $&$ \frac{5}{14}$ 
& 7 \\
  $\d_v^{(5)}$ & $\phantom{14\cdot}\;\frac{792}{29}$ 
& $\phantom{-}\frac{3}{2}$ & $\frac{2}{3}$
& 7 \\ \hline
  \end{tabular}
  \end{center}
  \caption[]{\label{tabc}Coefficients for the Fock state contributions
    to the parton distribution functions according to
    Eq.~(\protect\ref{disN}). The powers $m_a(N)$ of
    (\protect\ref{qpowers}) and (\protect\ref{gpowers}) are also
    listed.}
\end{table*}
\renewcommand{\arraystretch}{1.0}

One observes that all contributions appear in the form $x^n \,
(1-x)^m$ times a polynomial in $(1-x)$ which is generated by the
asymmetries in the \da{}s, i.e.\ their departure from the asymptotic
form. This holds for polynomial \da{}s in general. The leading power
$m_a(N)$ of $(1-x)$ in $\q^{(N)}_a(x)$ is generated by the
corresponding asymptotic \da{}; for quark distributions one has
\begin{equation}
n_q=1 \eqcm \qquad\qquad  m_q(N)= 2N + 2l_g -3 \eqcm
\label{qpowers}
\end{equation}
and for the gluon distribution 
\begin{equation}
n_g=3 \eqcm \qquad\qquad  m_g(N)= 2N + 2l_g -5 \eqcm 
\label{gpowers}
\end{equation}
where $l_g$ is the number of gluons in the $N$-particle Fock state. We
see that higher Fock states generate higher powers $m_a(N)$. Summing
over all Fock states the leading powers of $(1-x)$ for valence quark,
gluon and sea quark distributions come out as 3, 5 and 7,
respectively.  For the contributions from the $N=5$ Fock state with
three quarks and two gluons the leading powers are very high,
$m_q=11$, $m_g=9$, which is why we do not consider it here.

Our results for the valence parton distributions respect the usual
counting rule behaviour~\cite{BHL}. In other cases our results for the
leading powers of $1-x$ differ from those obtained from perturbative
QCD arguments by Brodsky, Burkardt and Schmidt~\cite{bro95}. This is
not a contradiction since we are dealing with soft physics
contributions.  The perturbative results of Ref.~\cite{bro95} manifest
themselves only in a region $1-\epsilon\leq x \leq 1$ where the
perturbative QCD contribution dominates over the soft contribution. To
estimate $\epsilon$ we remark that the overlap formulae (\ref{disf}),
(\ref{eformn}) for parton distributions and elastic form factors are
exact if one takes the \emph{full} wave functions instead of their
soft parts considered in this work \cite{bro80,BHL}. Using the
relations (\ref{nform}) or (\ref{radf}) between both types of
quantities we obtain $\epsilon \sim 1 /(-a_3^2\, \bar{t}\,)$, where
$\bar{t}$ is the momentum transfer in $F_{1\,p}(t)$ at which the
perturbative components of the wave functions start to dominate over
the soft ones. For the wave function we consider here, $-\bar{t}$ is
of the order of $500~\gev^2$~\cite{bol96}.

\begin{figure}[hbtp]
\parbox{\textwidth}{\begin{center}
\psfig{file=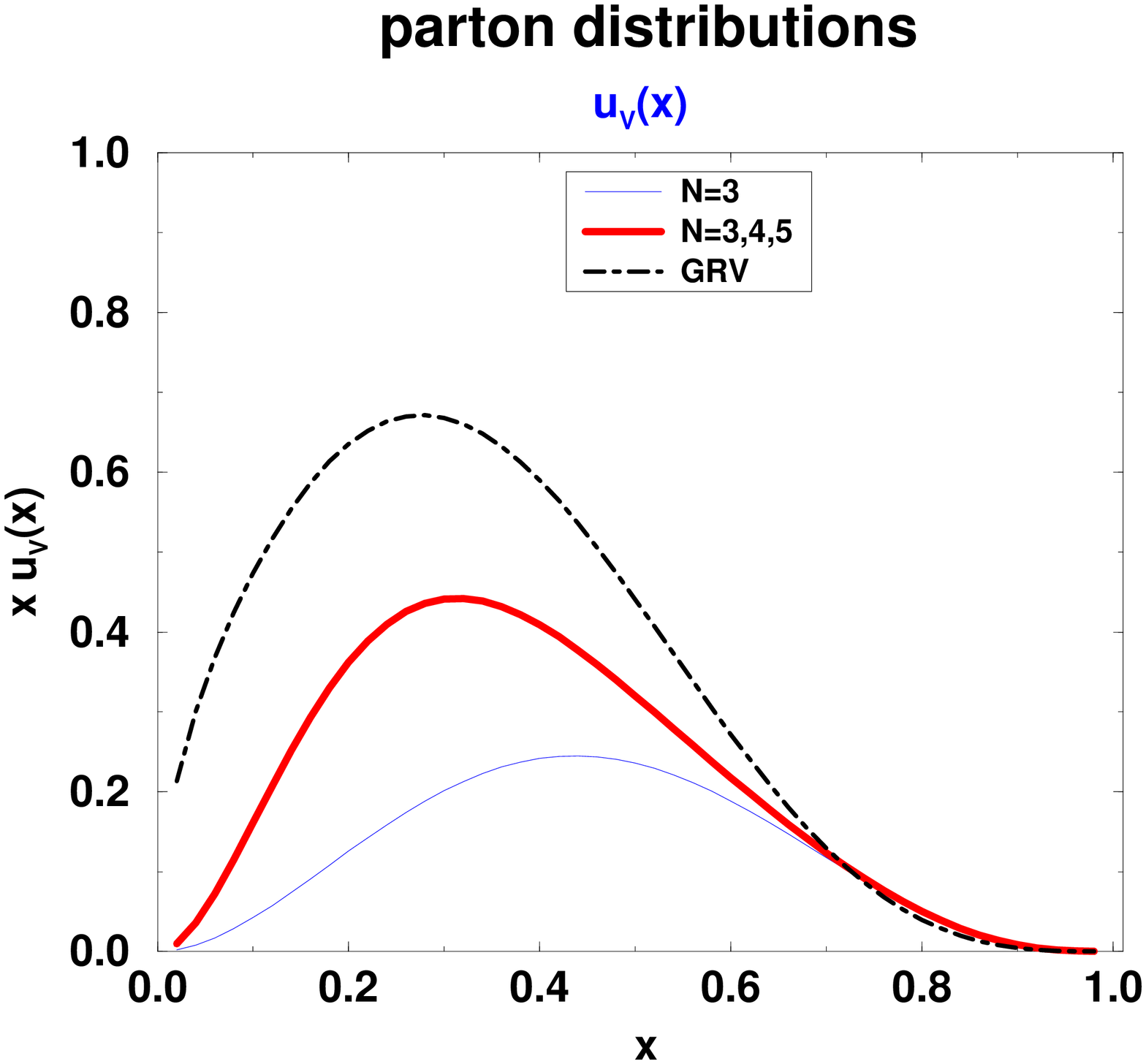, width=6.5cm, angle=-90} \
\psfig{file=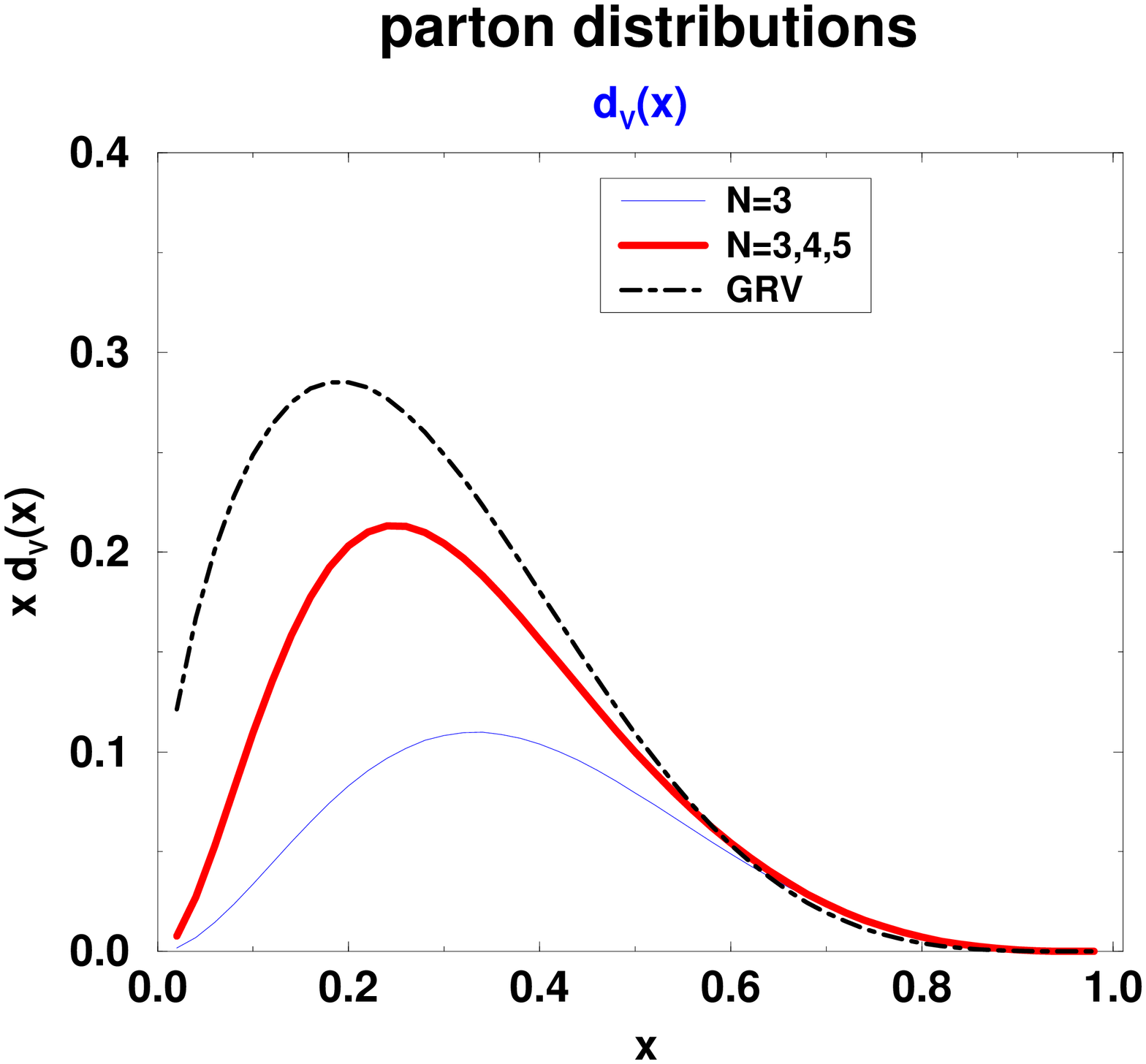, width=6.5cm, angle=-90} \\[1em]
\psfig{file=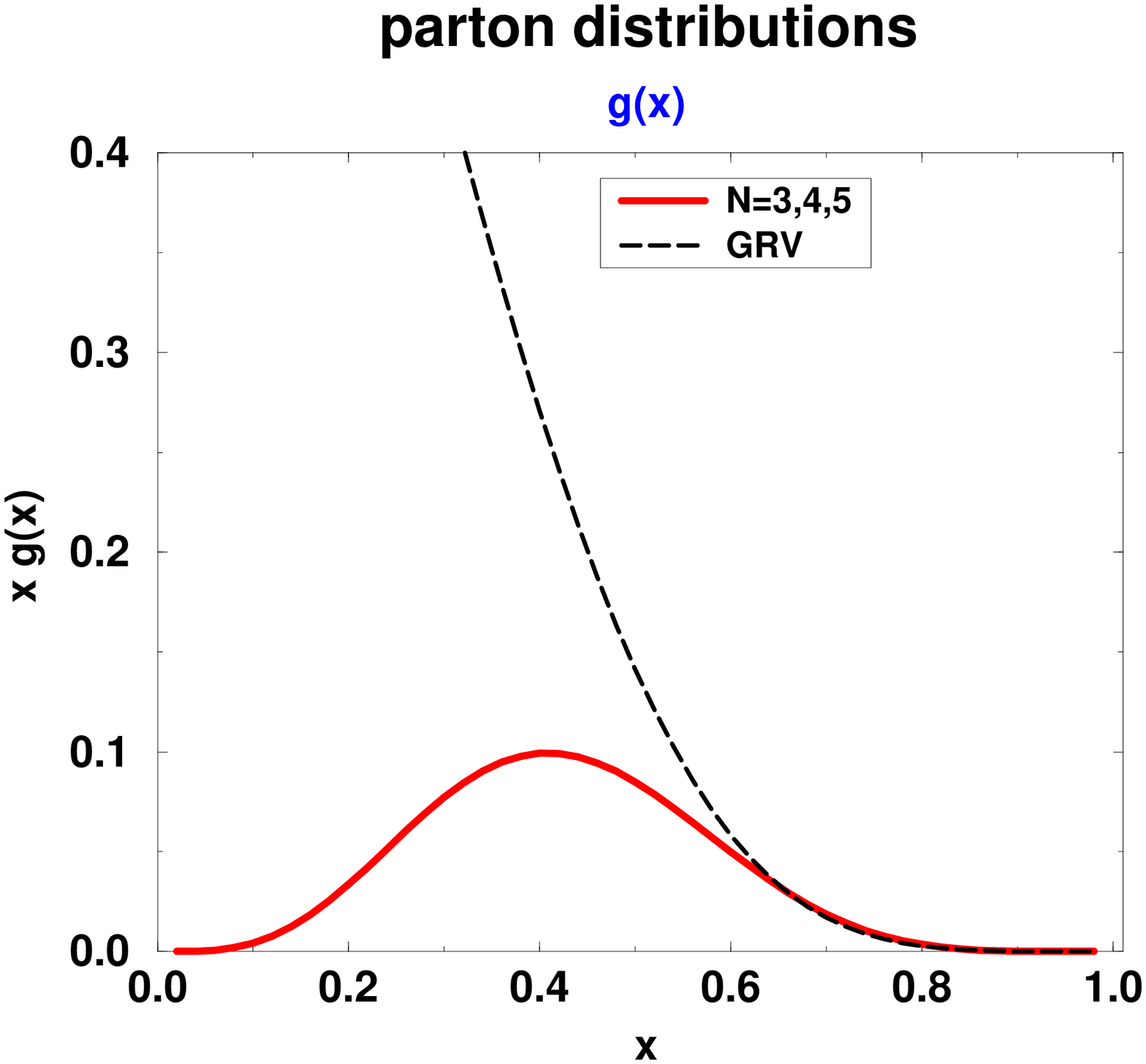, width=6.5cm, angle=-90} \
\psfig{file=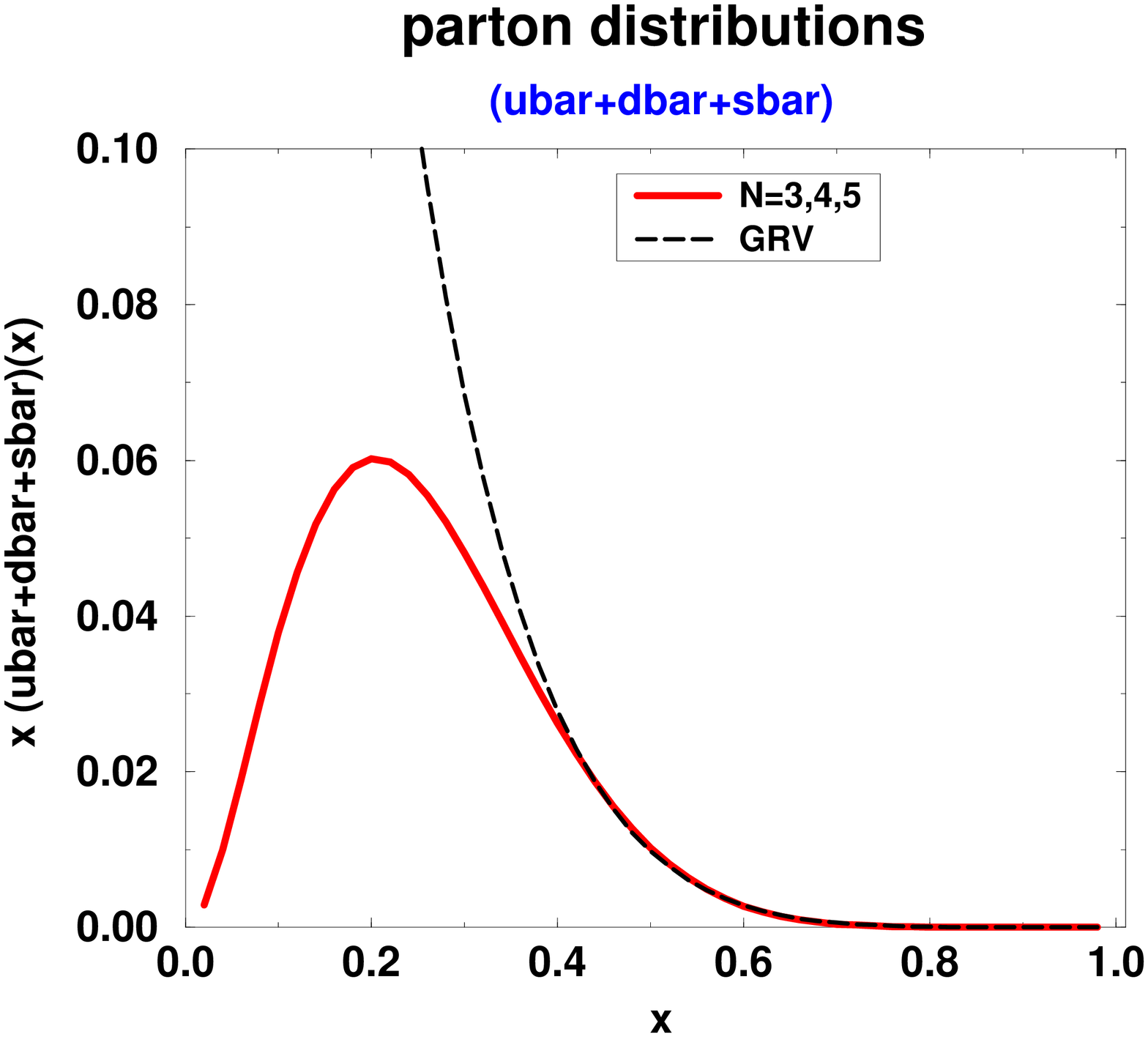, width=6.5cm,angle=-90}
\end{center}}
\caption{\label{partonfig}Parton distributions obtained from the
  $N=3,4,5$ Fock states ($P_3=0.17$, $P_4=P_5=0.1$). The model results
  are compared to the 1995 GRV LO parametrisation~\protect\cite{GRV}
  at a factorisation scale of 1~GeV. For the sea distributions we sum
  over the three flavours.}
\end{figure}

If for simplicity we take the transverse size parameters for the
$N=3,4,5$ Fock states to be equal,
\begin{equation}
a_5 = a_4 = a_3  \eqcm
\end{equation}
then only one parameter remains free for each of the new Fock state
wave functions, namely its probability (or the constants $f_N$, cf.\ 
Eq.\ (\ref{prob})). We fix these two parameters by fitting our gluon
and antiquark distributions (\ref{disN}) to the Gl\"uck-Reya-Vogt
(GRV) parametrisation~\cite{GRV} at large $x$. A best fit is obtained
for the values
\begin{equation}
  P_4=P_5=0.1 \eqcm \hspace{3em} f_4=1.06\times 10^{-4}\,\gev^3 \eqcm
  \hspace{2em} f_5=3.64\times 10^{-6}\,\gev^4 \eqpt
\end{equation}
The results of the fit are shown in Fig.~\ref{partonfig} and compared
to the GRV parametrisation.\footnote{ At large $x$ the 1998 GRV
  parametrisation is rather close to the 1995 version. We compare here
  with the LO parametrisation of the 1995 version.} The agreement with
the distribution functions given in Ref.~\cite{MRST} is of similar
quality at large $x$.  All distribution functions of the proton are
reproduced quite well down to $x$ of about $0.5$, for the sea quark
distribution even to lower values. We see how the first three Fock
states control the large-$x$ ($x\gsim 0.5$) behaviour of the
distribution functions; certainly the situation could be improved by
including even higher Fock states. We emphasise that the asymmetries
in the \da{}s play an important role: they push up $\u_v$ and diminish
$\d_v$ at the same time, thus producing a ratio $\u_v/\d_v$ of about
five at large $x$ while totally symmetric \da{}s yield a ratio of only
two, in sharp conflict with the GRV parametrisation. We also note that
our ratio $\d_v/\u_v$ tends to 1/14 in the limit $x\to 1$ and differs
from the SU(6) result of 1/5 \cite{tho97}.

The spin dependent parton distributions allow another interesting test
of our approach. These distributions measure the difference between
the distributions of type-$a$ partons with positive and negative
helicity. In analogy to the unpolarised distribution discussed above
we find within our model
\begin{equation}
\Delta \q_a^{(N)}(x) = \Delta b_a^{(N)}\,P_N\, 
                     x^{n_a} \,(1-x)^{m_a(N)} \,
                     \;\left[ 1 + \Delta c_a^{(N)}\, (1-x)
                            + \Delta d_a^{(N)}\, (1-x)^2 \right]
\label{disNS}
\end{equation}
with the coefficients listed in Tab.~\ref{tabcd}. The powers $n_a$ and
$m_a(N)$ are the same as the ones for unpolarised distributions, given
by~(\ref{qpowers}). As a consequence of our simple assumptions that
the gluons and sea quark pairs are unpolarised we have $\Delta
\g^{(4)}(x) = \Delta \qbar^{(5)}(x) =0$. Note also that $\Delta
\q_a^{(N)}(x) \propto \q_a^{(N)}(x)$ at large $x$. While the constants
of proportionality are close to unity for the valence $\u$-quark
distributions, they are negative or even zero for valence $\d$-quarks.

\renewcommand{\arraystretch}{1.5}
\begin{table*}
  \begin{center}
  \begin{tabular}{|c||c|c|c||c|} \hline
  $\Delta\q_a^{(N)}$ & $\Delta b_a^{(N)}$
& $\Delta c_a^{(N)}$ & $\Delta d_a^{(N)}$ 
& $m_a(N)$ \\ \hline\hline
  $\Delta\u_v^{(3)}$ &$ 40\cdot\frac{140}{87}$ 
& $-\frac{21}{20}$ & $\frac{9}{40}$ 
& 3 \\
  $\Delta\d_v^{(3)}$ &$  -\phantom{\cdot}\;\,\frac{140}{87}$ 
& $\phantom{-} 3 $&$ \frac{9}{5}$ 
& 3 \\
  $\Delta\u_v^{(4)}$ & $16\cdot\frac{990}{37}$ 
& $-\frac{3}{2}$ & $\frac{9}{16}$ 
& 7 \\
  $\Delta\d_v^{(4)}$ & $\phantom{40\cdot}\;\, 0$ 
& $\phantom{-}0$  & 0  
& 7 \\
  $\Delta\u_v^{(5)}$ & $40\cdot\frac{264}{29}$ 
& $-\frac{6}{5} $&$ \frac{27}{80}$ 
& 7 \\
  $\Delta\d_v^{(5)}$ & $-\phantom{\cdot}\;\,\frac{264}{29}$ 
& $\phantom{-}\frac{3}{2}$ & $\frac{1}{2}$ 
& 7 \\ \hline
  \end{tabular}
  \end{center}
  \caption[]{\label{tabcd}Coefficients for the Fock state
    contributions to the spin-dependent parton distribution functions
  according to Eq.~(\protect\ref{disNS}). The powers $m_a(N)$ from
  (\protect\ref{qpowers}) are also given. } 
\end{table*}
\renewcommand{\arraystretch}{1.0}

\begin{figure}[hbtp]
\parbox{\textwidth}{\begin{center}
\psfig{file=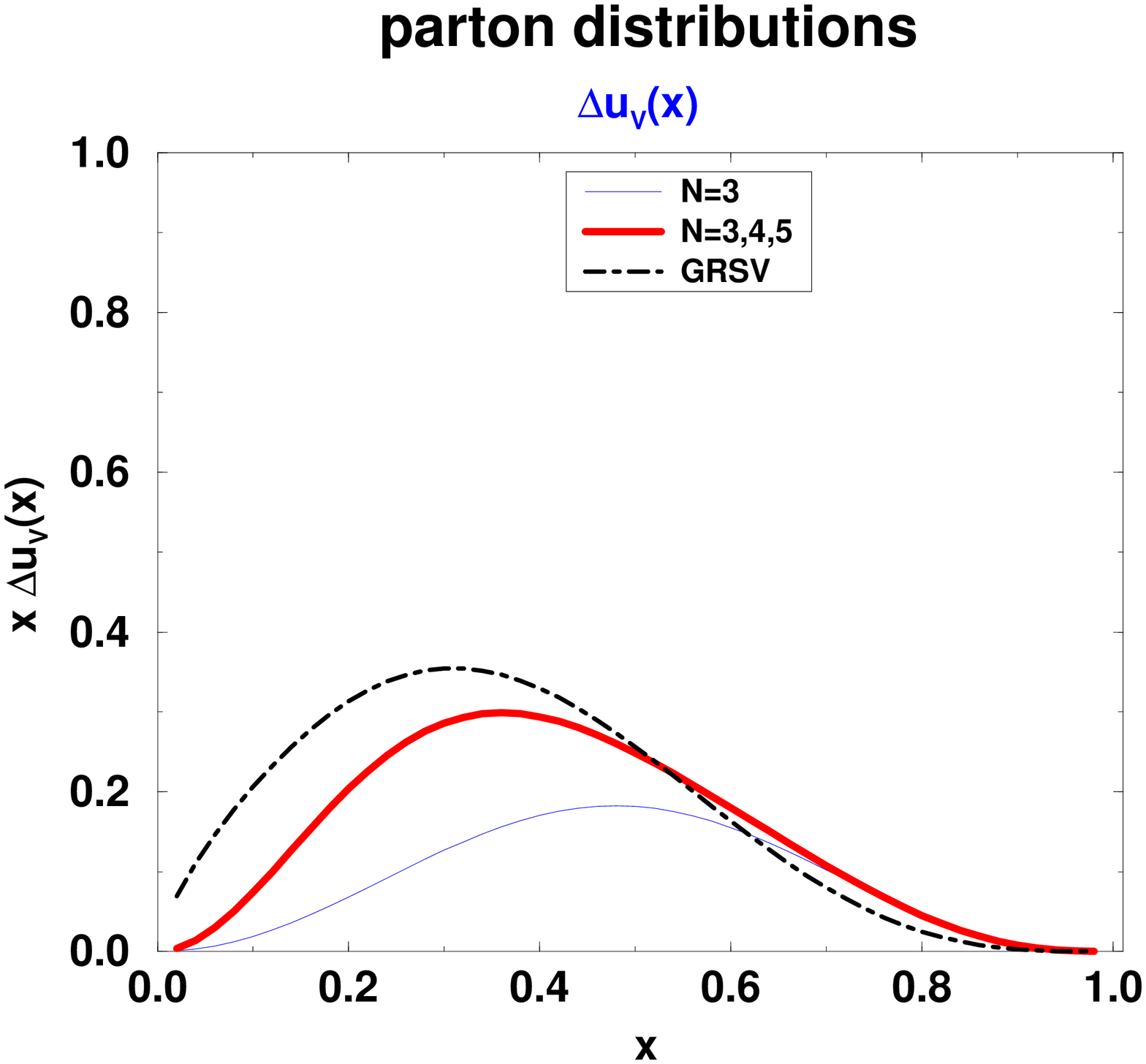, width=6.5cm, angle=-90} \ 
\psfig{file=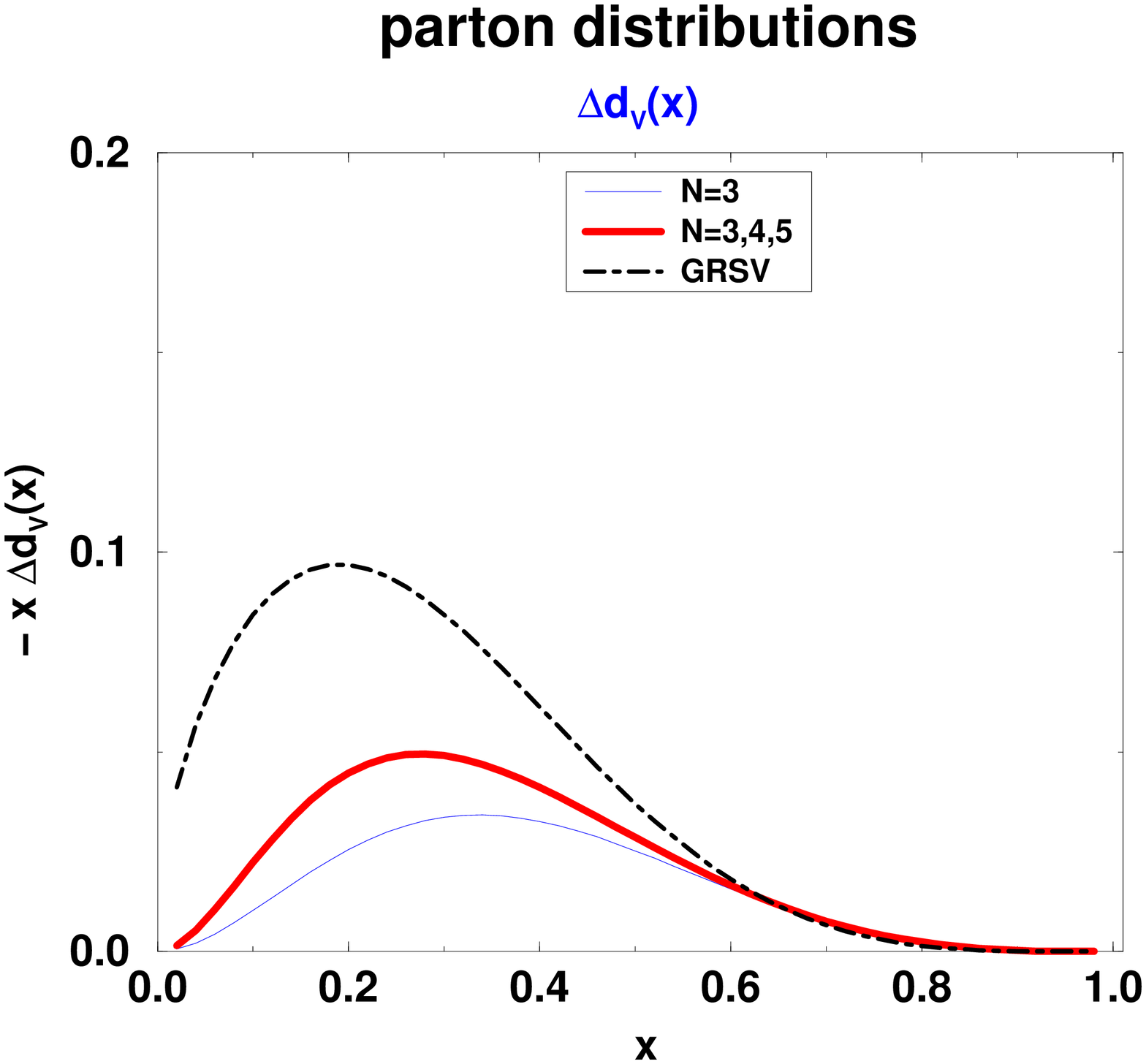, width=6.5cm, angle=-90}
\end{center}}
\caption{\label{udfig}Spin-dependent valence quark distributions
  $\Delta \u_v$ and $\Delta \d_v$. The model results are compared to
  the parametrisation of Ref.~\protect\cite{glu96}.}
\end{figure}

In Fig.~\ref{udfig} we compare our predictions with the
parametrisation proposed in Ref.~\cite{glu96}. As we see, surprisingly
good agreement is obtained in our simple model. There is also fair
agreement with the polarised parton distributions determined
in~\cite{lea98} at large $x$. The relative strength of $\Delta \u_v$
and $\Delta \d_v$ in that region reflects the spin structure of the
valence Fock state and the asymmetry in its \da{}.

%%%%%%%%%%%%%%%%%%%%%%%%%%%%%%%%%%%%%%%%%%%%%%%%%%%%%%%%%%%%%%%%%%%
\section{Form factors}
\label{form}
%%%%%%%%%%%%%%%%%%%%%%%%%%%%%%%%%%%%%%%%%%%%%%%%%%%%%%%%%%%%%%%%%%%
According to Drell and Yan~\cite{DY} the Dirac form factor can be
represented as the overlap of LCWFs as
\begin{equation} 
 F_1(t) \,=\, \sum_N\; F_1^{(N)} (t)
\label{eform}
\end{equation}
with individual Fock state contributions 
\begin{equation}
F_1^{(N)} (t) \,=\, 
      \sum_a\;e_a\, \sum_j \sum_\beta \int\, 
      [\d \tilde{x}]_N [\d^2 \tilde{\bf k}_\perp]_N\,
      \Psi^*_{N\beta}(\hat{x}'_i,{\hat{\bf k}'_{\perp i}})\,
      \Psi_{N\beta}(\tilde{x}_i,{\tilde{\bf k}_{\perp i}})\, \eqcm
\label{eformn}
\end{equation}
where $j$ runs over all partons of type $a$. We use our symmetric
frame to evaluate the overlap, the primed and unprimed arguments in
(\ref{eformn}) are therefore related by (\ref{tilde-args}) and we have
$\vd^2 = -t$.  Performing the $\vk$-integrals for the $N=3,4,5$ Fock
states with the help of Eq.~(\ref{gint2}), we arrive at
\begin{eqnarray}
F_{1\,p}^{(N)}(t) &=& \int \d x \, \exp{\left[ \frac12 \, a^2_N \, t
                 \, \frac{1-x}{x} \right]} 
    \, \left\{ e_u \,\u_v^{(N)}(x)  + e_d \,\d_v^{(N)}(x) \right\}
                                                \eqcm \nn \\[0.2em]
F_{1\,n}^{(N)}(t) &=& \int \d x \, \exp{\left[ \frac12 \, a^2_N \, t
                 \, \frac{1-x}{x} \right]} 
    \, \left\{ e_u \,\d_v^{(N)}(x) 
             + e_d \,\u_v^{(N)}(x)  \right\} \eqpt
\label{nform}
\end{eqnarray}
for the proton and neutron form factors. The appearance of the parton
distributions here is a consequence of the fact that the integrand in
their overlap representation (\ref{disf}) is obtained from the one in
(\ref{eformn}) by setting $\vd =0$. Thus the $\vk$-integrals only
differ by the exponential factor of (\ref{eq:SKgauss}) at $\zeta=0$,
which arises from the Gaussian $\vk$-dependence of our wave functions.

\begin{figure}[hbtp]
\parbox{\textwidth}{\begin{center}
\psfig{file=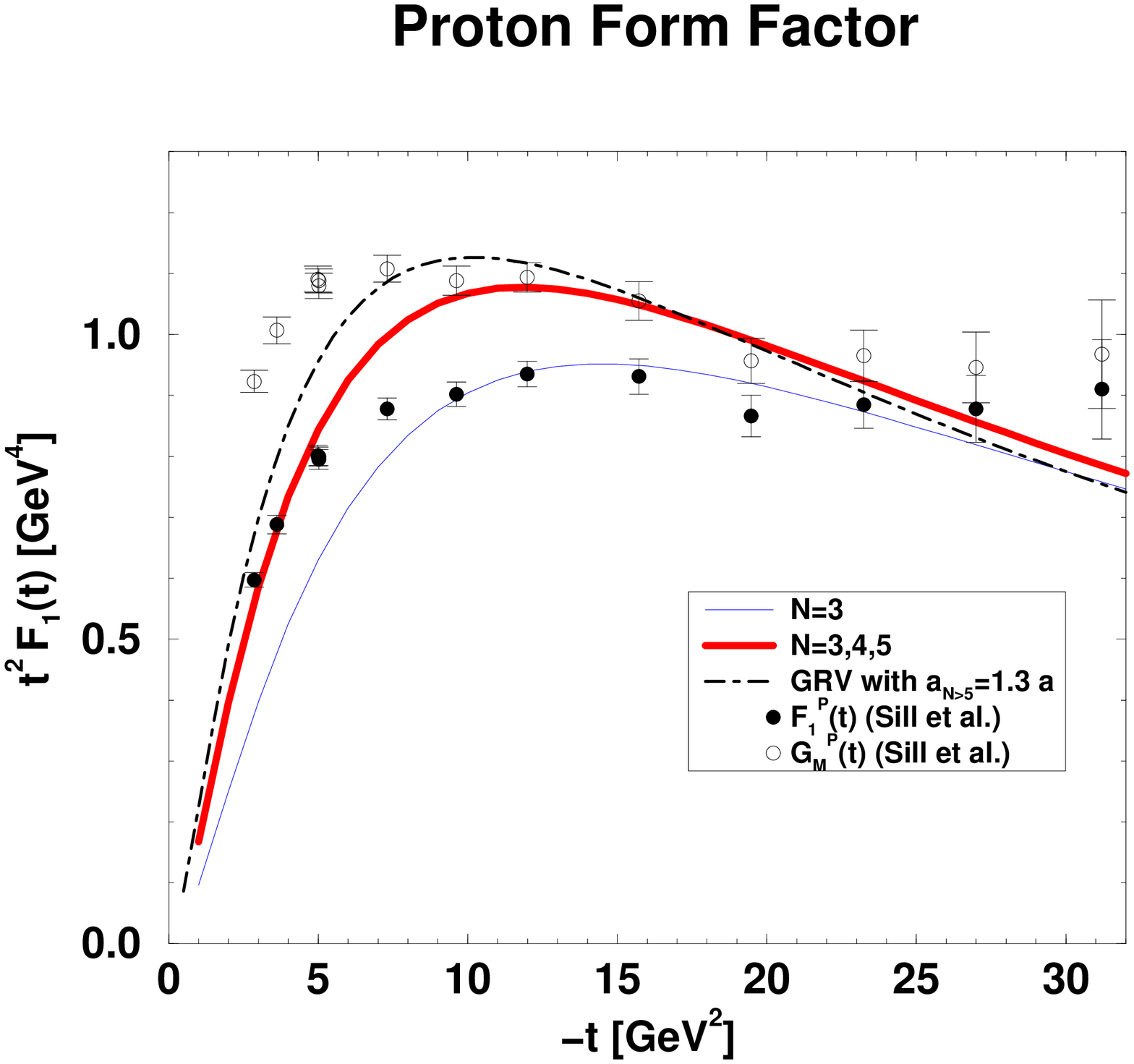, width=6.5cm, angle=-90} \
\psfig{file=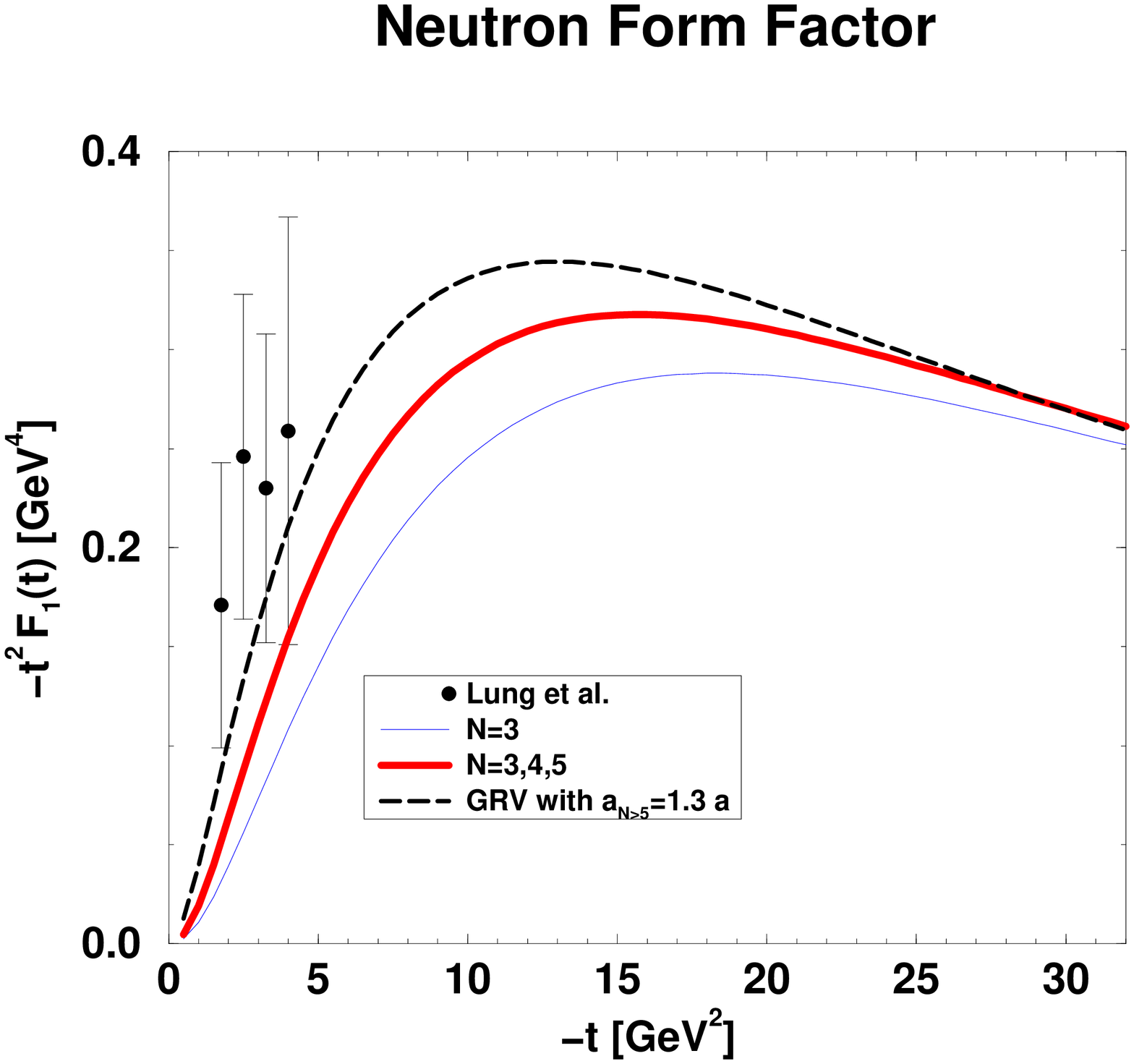, width=6.5cm, angle=-90}
\end{center}}
\caption{\label{fffig}Electromagnetic form factor of the proton and
  neutron using the model parton distributions for the valence Fock
  state only, the $N=3,4,5$ Fock states, and all Fock states on the
  basis of the GRV parametrisation at the factorisation scale 1~GeV
  \protect\cite{GRV}, cf.\ (\protect\ref{diseff}). Data for $F_1$ and
  $G_M$ are taken from~\protect\cite{sil93,lun93}.}
\end{figure}

It is now suggestive to assume that the $\vk$-dependence of all Fock
state wave functions is given by the Gaussian (\ref{BLHMOmega}) and to
approximate all $a_N$ with a common transverse size parameter $a$.
Summing over $N$ in (\ref{eform}) then leads to a representation of
form factors in terms of the valence quark distribution functions:
\begin{eqnarray}
F_{1\,p}(t) &\simeq& \int \d x \, \exp{\left[ \frac12 \, a^2 \, t \,
                             \frac{1-x}{x} \right]} 
                   \, \left\{ e_u \,\u_v (x) 
                    + e_d \,\d_v (x)\right\} \eqcm \nn\\[0.2em]
F_{1\,n}(t) &\simeq& \int \d x \, \exp{\left[ \frac12 \, a^2 \, t \,
                             \frac{1-x}{x} \right]} 
                   \, \left\{ e_u \,\d_v (x) 
                    + e_d \,\u_v (x) \right\} \eqcm
\label{radf}
\end{eqnarray}
a formula recently proposed by Radyushkin~\cite{rad98a}. Remarkably,
inclusive observables are related to exclusive ones. The chief
advantage of this formula is its independence from any explicit form
of the \da{}s. Of course a common value for the transverse size
parameter for all Fock states is unrealistic: as we saw before the
valence Fock state is rather compact corresponding to about a half of
the charge radius. Consequently the higher Fock states have to develop
the full radius. For the purpose of evaluating the form factors from
Eqs.~(\ref{eform}) and (\ref{nform}) we take $a_3=a_4=a_5$ as before
and put as a simple ansatz $a_N=1.3\, a_3$ for $N>5$, where the factor
1.3 is adjusted to the data for $F_{1\,p}$. A substantially larger
factor would strongly suppress the higher Fock state contributions, a
smaller one would lead to large contributions exceeding the form
factor data.\footnote{We note at his point that in contrast to our
  ansatz a transverse size parameter $a = 0.84$~GeV$^{-1}$ common to
  all Fock states was used in~\cite{rad98a}.} Then we set
\begin{equation}
    \sum_{N>5}\, q_a^{(N)}(x)=\, q_a (x) -
                         \sum_{N=3,4,5}\, q_a^{(N)}(x)  \eqcm
\label{diseff}
\end{equation} 
where $\q_a$ is taken from the GRV parametrisation~\cite{GRV} and the
three lowest Fock state contributions from our model. In this way we
account for the sum of all Fock states in a phenomenological way. The
results obtained in this manner are confronted to the data
\cite{sil93,lun93} in Fig.~\ref{fffig}.

{}For large values of the momentum transfer our simple model agrees
very well with the data, i.e.\ the dimensional counting behaviour is
well mimicked by soft physics. Below about $10\,\gev^2$ the model is
not perfect, deviations of the order of 20$\%$, i.e.\ of the order of
$m^2/(-t)$, are to be noticed. Such corrections are to be expected in
our model, where proton spin-flip effects and orbital angular momentum
in the wave functions are not taken into account as we discussed in
see Sect.~\ref{proton-spin}. In reality spin flip effects are not very
small as is indicated by the difference between the Dirac and magnetic
form factors, $F_1$ and $G_M$, see Fig.~\ref{fffig}. In view of these
approximations we are satisfied with our results even in the range
$5\,\gev^2 <-t<10\,\gev^2$. We observe from Fig.~\ref{fffig} the
dominance of the valence Fock state contributions. For $-t>10\,\gev^2$
all other Fock states contribute less than 20$\%$; each individual
Fock state provides only a small correction to the form factor. This
can be regarded as a justification of the rough treatment of the $N=4$
and 5 Fock states introduced in Sect.~\ref{wave}.  We also remark that
the parameters $f_3$ and $a_3$ of (\ref{da3-parameters}) used in this
work have been obtained in~\cite{bol96} by requiring that the data for
$F_{1\, p}$ be saturated by the soft overlap of the valence Fock state
only. Given the uncertainties just discussed and our simplified
treatment of the higher Fock states we think however that readjusting
these parameters is not necessary here. As for the neutron form
factor, we mentioned in Sect.~\ref{distributions} that totally
symmetric wave functions lead to the relation
$\u_v^{(N)}(x)=2\d_v^{(N)}(x)$, which according to Eq.~(\ref{nform})
would lead to a vanishing contribution to $F_{1\,n}$. Hence the
asymmetries in the LCWFs generate the neutron form factor.

For wave functions of the type we are considering here the leading
powers $m_q(N)$ of $(1-x)$ in the valence distributions
$\q_v^{(N)}(x)$ correspond to leading powers $m_q(N)+1$ of $1/t$ in
the asymptotic behaviour of the corresponding Fock state contribution
to $F_1(t)$.\footnote{ The Drell-Yan result~\cite{DY}, for which the
  power $m_q$ of $(1-x)$ in the valence quark distribution functions
  corresponds to a power $(m_q+1)/2$ of $1/t$ in the form factor, is
  only obtained for wave functions factorising in $x$ and $\vk$ (i.e.\ 
  for $\Omega$ not depending on $x_i$).}  Hence for sufficiently large
$-t$ the valence Fock state dominates the form factor with only small
corrections from the next Fock states.  It is important to realise
that this asymptotic behaviour of the overlap contributions does not
set in before $-t\simeq 100 \,\gev^2$ since the expansion of the
integrals appearing in (\ref{nform}) into a power series in $1/t$
converges very slowly.  We remark that the dominance of the soft
overlap contribution is consistent with the strength of the
perturbative contribution to the proton form factor, which drops as
$1/t^2$. As reported in Ref.~\cite{bol96} the perturbative
contribution evaluated from our valence Fock state wave function can
be neglected for experimentally accessible momentum transfers.  For
$-t$ larger than about $500$~GeV$^2$, however, the perturbative
contribution will dominate since our overlap contribution
asymptotically behaves as $1/t^4$.

In analogy to the electromagnetic case we can also calculate the
charged current axial form factor of the nucleon. The various
contributions are now weighted by the quark helicities and isospin,
leading to
\begin{eqnarray}
F_A &=& \sum_N \int \d x \, \exp{\left[ \frac12 \, a_N^2 \, t \,
   \frac{1-x}{x} \right]} \cr
&& \quad \times  
\, \left\{ \Delta \u_v^{(N)}(x) + 2 \, \Delta \ubar^{(N)}(x)  
         - \Delta \d_v^{(N)}(x) - 2 \, \Delta \dbar{}^{(N)}(x) 
   \right\} \eqcm
\end{eqnarray}
Evaluating the axial form factor along the lines described for the
electromagnetic case we find fair agreement with the dipole
parametrisation of the admittedly low-$t$ neutrino data~\cite{kit83}.

%%%%%%%%%%%%%%%%%%%%%%%%%%%%%%%%%%%%%%%%%%%%%%%%%%%%%%%%%%%%%%%%%%%
\section{Large angle Compton scattering}
\label{compt}
\begin{figure}[hbtp]
\parbox{\textwidth}{\begin{center}
\psfig{file=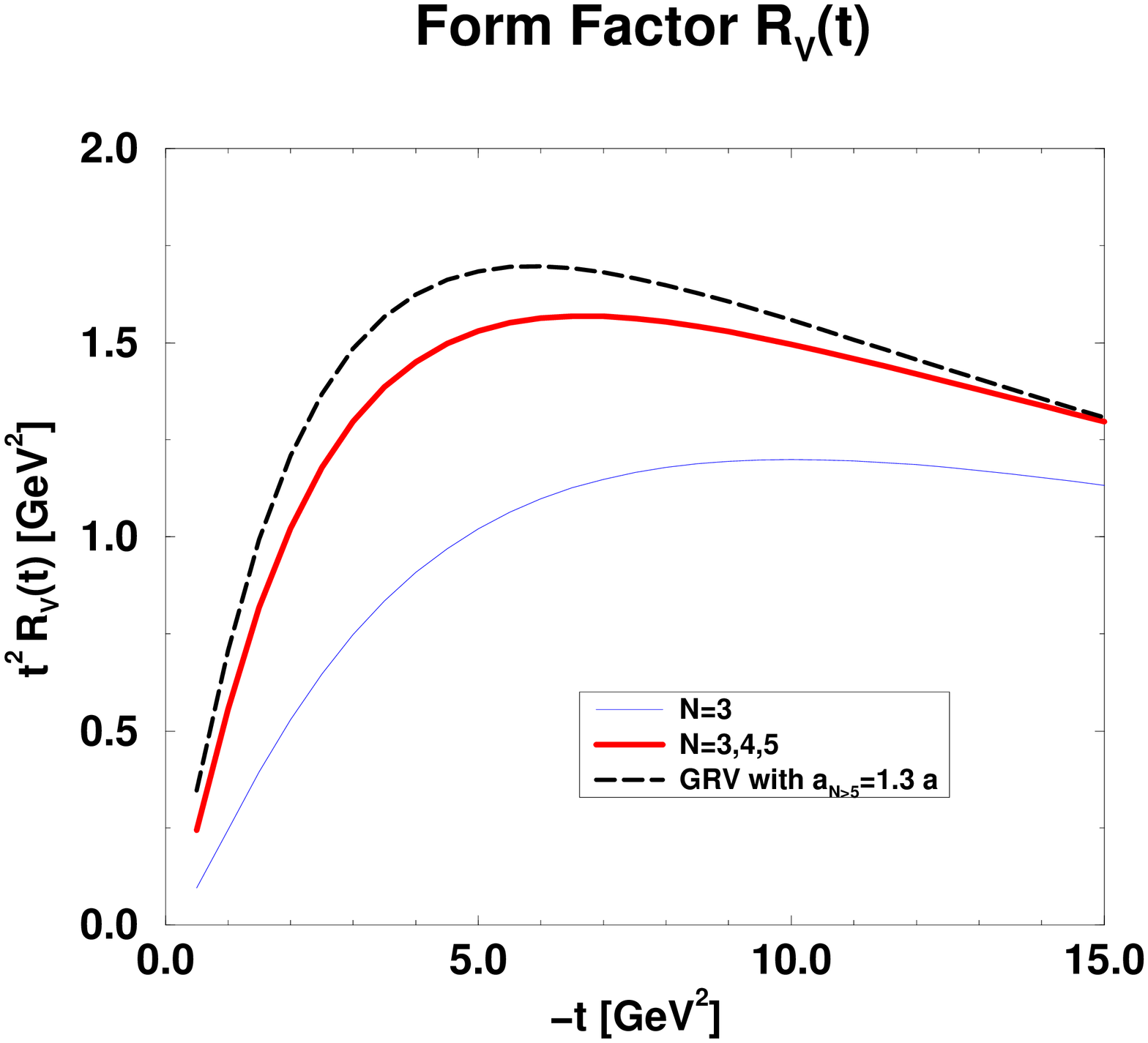, width=6.5cm, angle=-90} \ 
\psfig{file=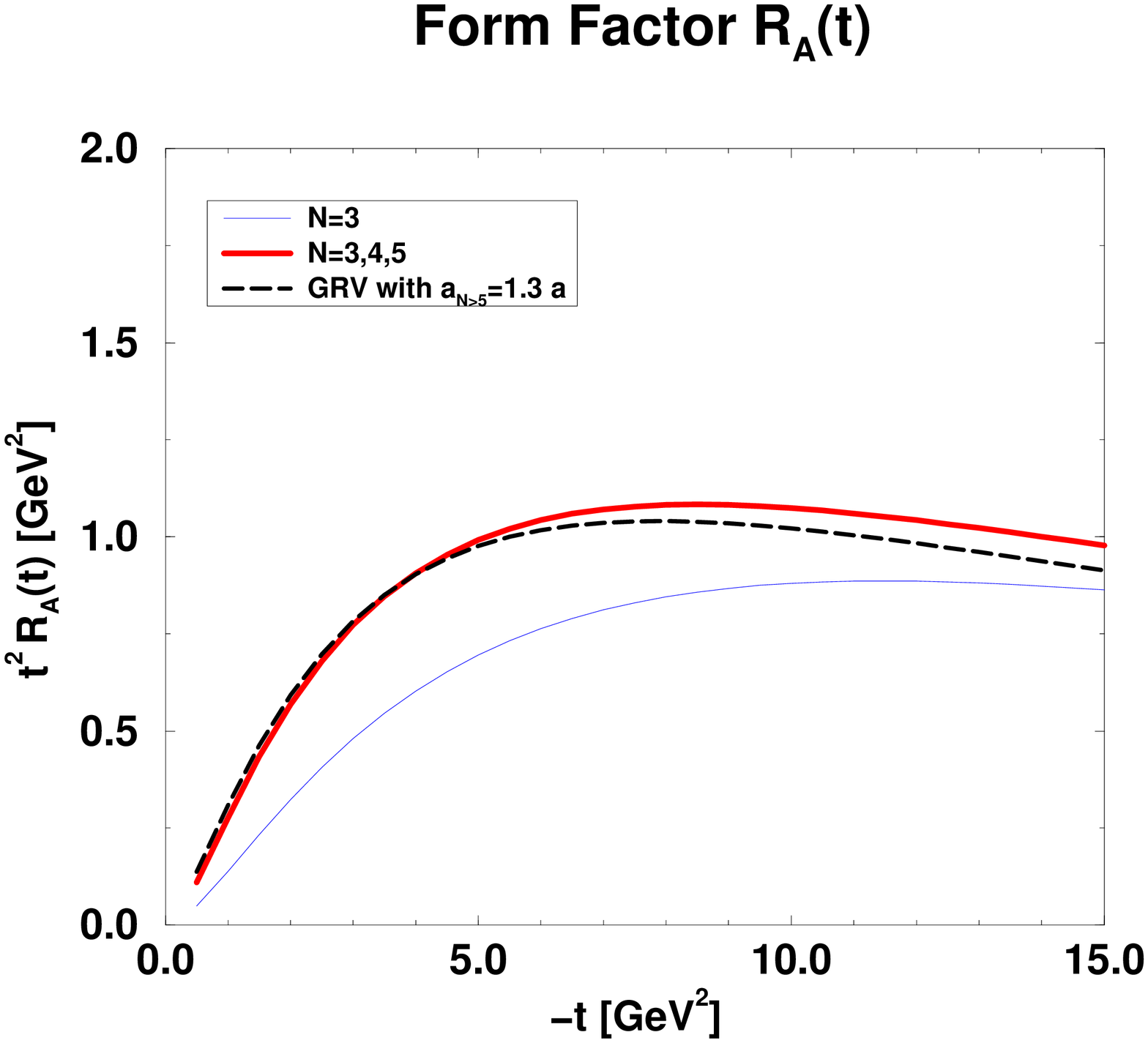, width=6.5cm, angle=-90}
\end{center}}
\caption{\label{Rifig}The form factors $R_V(t)$ and $R_A (t)$,
  evaluated with our model for LCWFs.}
\end{figure}

%%%%%%%%%%%%%%%%%%%%%%%%%%%%%%%%%%%%%%%%%%%%%%%%%%%%%%%%%%%%%%%%%%%
Using our expressions (\ref{final}), (\ref{R-form-factors}) for the
handbag amplitude and neglecting the contribution from the proton spin
flip form factor $R_T$, we obtain the cross section for real
Compton scattering with unpolarised photons and protons as
\begin{equation} \label{compton}
\frac{\d\sigma}{\d t} =  \frac{2\pi\alpha_{\it em}^2}{s^2} \,
           \left[-\frac{u}{s} - \frac{s}{u}\right] \,
           \left\{ \frac12 \, \left(R_V^2(t) + R_A^2(t)\right) - 
           \frac{us}{s^2+u^2}\, \left(R_V^2(t) - R_A^2(t)\right)
                                                     \right\}\,.
\end{equation}
As explained in Sect.~\ref{sub-compton} we can also calculate the
Compton amplitude as an overlap of LCWFs in the symmetric frame of
Sect.~\ref{symmetric-frame}. Using the same approximations as in the
handbag calculation, Sect.~\ref{handbag}, and comparing with
(\ref{final}), (\ref{R-form-factors}) we obtain the analogues of the
Drell-Yan formula (\ref{eform}), (\ref{eformn}) for our form factors
$R_V$ and $R_A$. With our Gaussian ansatz (\ref{BLHMOmega}) for the
$\vk$-dependence of the LCWFs and the integral (\ref{gint1}) the form
factors $R_V$ and $R_A$ can then be written as
\begin{eqnarray}
R_V(t) &=& \sum_N\; \int \frac{\d x}{x} 
       \exp{\left[ \frac12 \, a^2_N \, t \, \frac{1-x}{x} \right]}
       \; \left\{\, e_u^2 \;
                [ \u_v^{(N)}(x) +  2 \, \ubar^{(N)}(x)] 
       \right. \nn\\[0.2em]
 &&  \hspace{6em} \left. {}+ e_d^2 \;
            [ \d_v^{(N)}(x) +  2 \, \dbar{}^{(N)}(x)]
            + e_s^2 \; 2 \, \sbar{}^{(N)}(x)  
     \right\} \eqcm
\nonumber \\  
R_A(t) &=& \sum_N\; \int \frac{\d x}{x} 
       \exp{\left[ \frac12 \, a^2_N \, t \, \frac{1-x}{x} \right]} 
       \; \left\{\, e_u^2 \; 
                [\Delta \u_v^{(N)}(x) + 2\,  \Delta \ubar^{(N)}(x) ] 
       \right. \nn\\[0.2em]
 &&  \hspace{6em} \left. {}+ e_d^2 \;
            [\Delta \d_v^{(N)}(x) + 2 \, \Delta \dbar{}^{(N)}(x)] 
            + e_s^2 \; 2\, \Delta \sbar{}^{(N)}(x)
     \right\} \eqcm
\label{rva}
\end{eqnarray}
in close analogy to the expressions (\ref{nform}) for the Dirac form
factor $F_1$. Our numerical predictions for $R_V$ and $R_A$ are shown
in Fig.~\ref{Rifig}.

If one makes the assumption that $R_A=R_V$ as was done
in~\cite{rad98a} then one obtains the suggestive result that the cross
section for Compton scattering on the proton is just given by the
familiar Klein-Nishina expression for Compton scattering on a free
quark times the square of the form factor $R_V(t)$, which describes
the target structure. In our model the ratio of $R_A$ and $R_V$ is
however rather far from 1 for the values of $-t$ we consider, cf.\ 
Fig.~\ref{Rifig}. {}From (\ref{rva}) one sees that $R_A \approx R_V$
would require all quarks and antiquarks to be completely polarised
along the proton spin, i.e.\ $\q_a^{(N)}(x) \approx
\Delta\q_a^{(N)}(x)$ for all $N$ and $a$, in the range of $x$
dominating the integrals. For $\u$-quarks this holds indeed if $x$ is
close to 1, but not for the intermediate $x$ that are important at our
values of $-t$, while for $\d$-quarks the unpolarised and polarised
quark distributions even have opposite sign. We therefore keep both
terms $R_V^2(t) + R_A^2(t)$ and $R_V^2(t) - R_A^2(t)$ in
(\ref{compton}); they reflect the fact that the proton target has a
nontrivial quark spin structure. Using measurements at different
values of $s$ and $t$ and a Rosenbluth-type separation it will in
principle be possible to isolate the new form factors $|R_V(t)|$ and
$|R_A(t)|$ from sufficiently accurate experimental data, and to
compare them with our predictions.

\begin{figure}[hbtp]
\parbox{\textwidth}{\begin{center}
\psfig{file=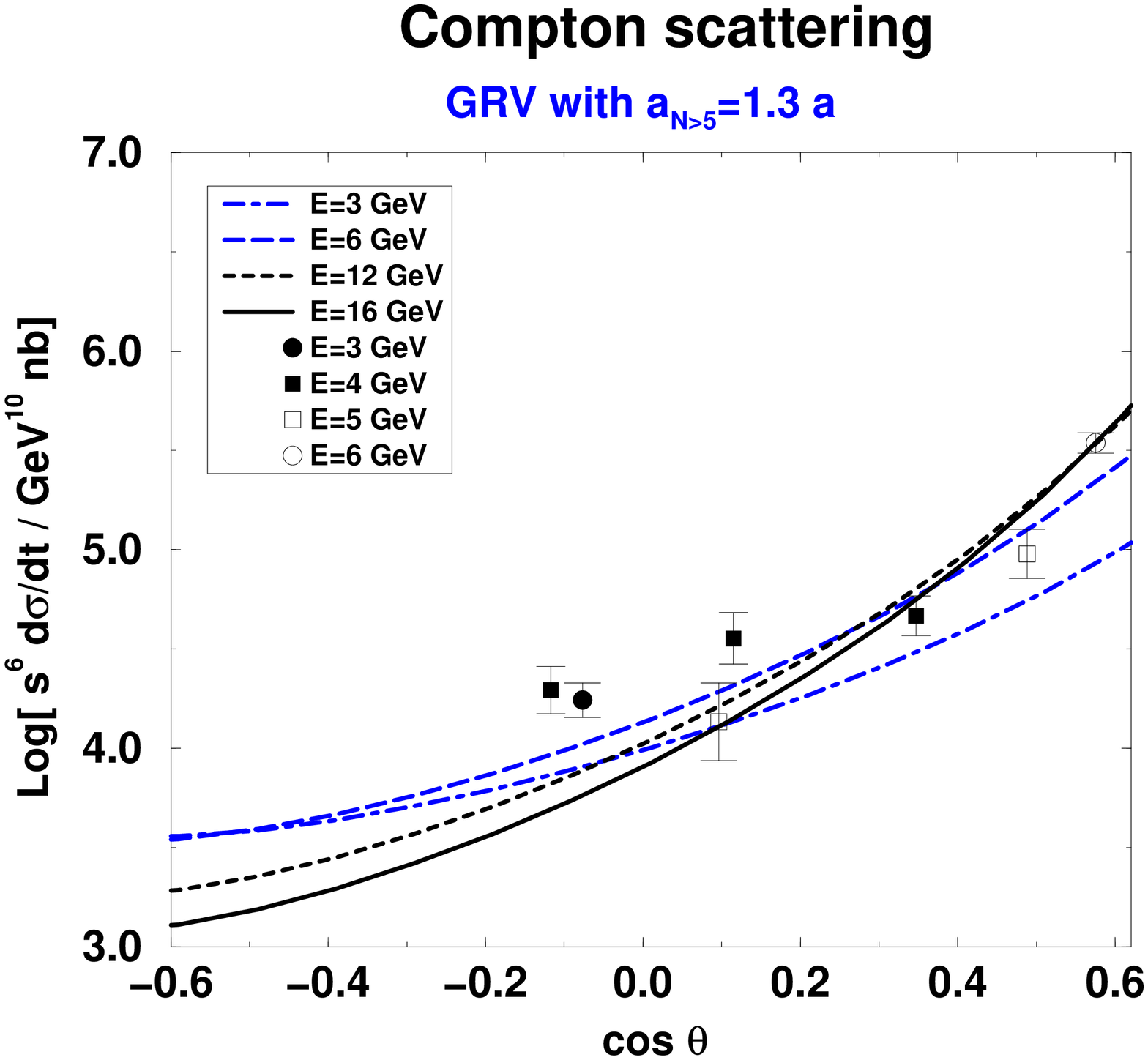, width=6.5cm, angle=-90} \
\psfig{file=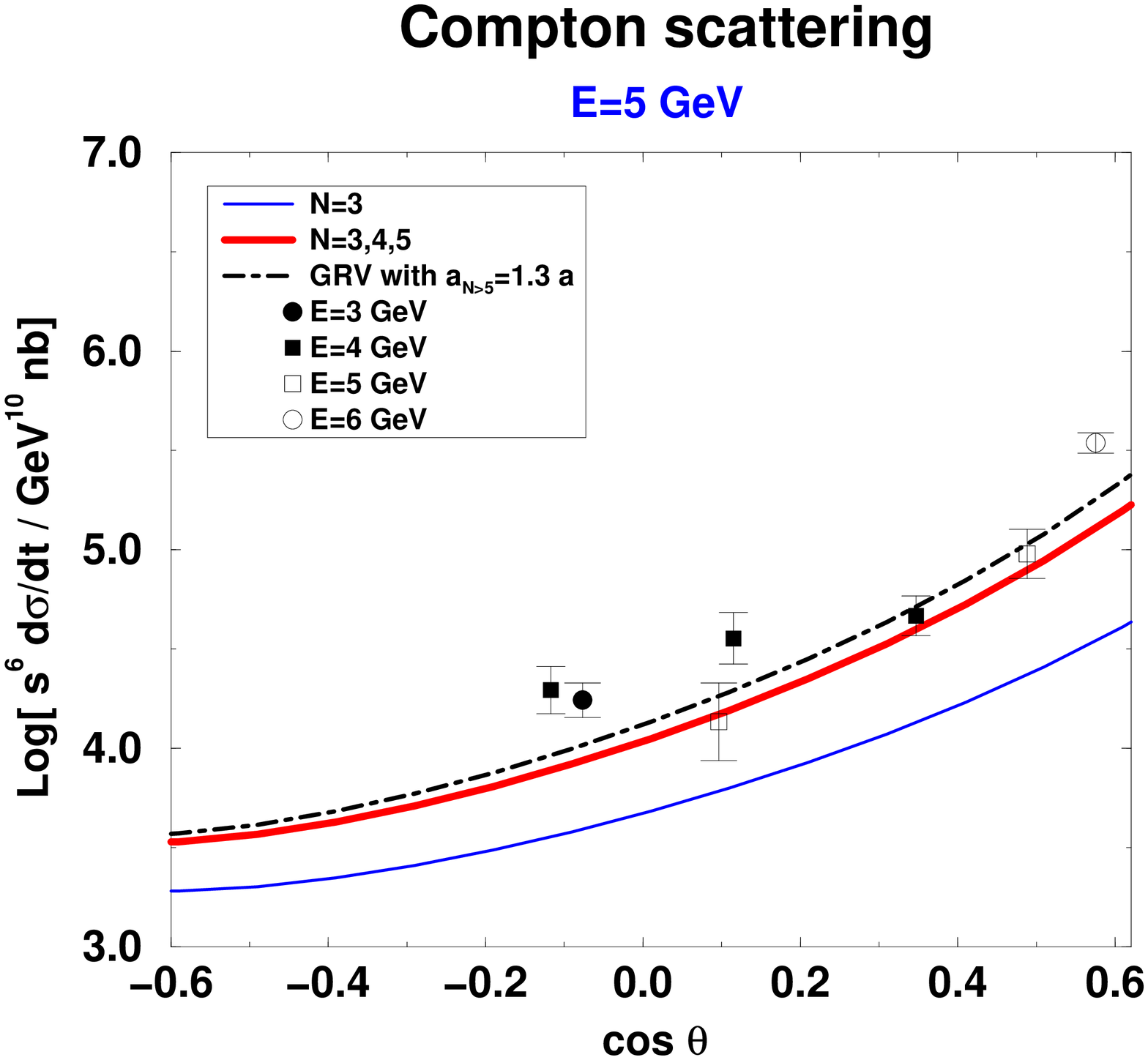, width=6.5cm, angle=-90}
\end{center}}
\caption{\label{dsigfig}The Compton cross section scaled by $s^6$
  versus $\cos{\theta}$, where $\theta$ is the scattering angle in the
  c.m. Data, for $-t,\,-u \geq 2.5\,\gev^2$ only, are taken from
  Ref.~\protect\cite{shu79}. Left: Model predictions obtained from the
  GRV parametrisation~\protect\cite{GRV} for various photon energies
  in the laboratory frame. Right: Model predictions decomposed into
  separate Fock state contributions at a photon energy of 5~GeV.  }
\end{figure}

In Fig.~\ref{dsigfig} we show our results for the Compton cross
section. Given the quality of the data, and the small energies and low
values of $-t$ and $-u$ at which they are available, our predictions
compare fairly well with experiment. As a minimum condition for our
approximations discussed in Sect.~\ref{handbag} to be applicable we
only take into account data points satisfying $-t,\, -u \geq
2.5\,\gev^2$. Better data and data at larger energies are definitely
required for a severe check of the new approach and its confrontation
with the hard scattering mechanism.  For comparison we also show
predictions for the Compton cross section at a photon energy of 12
\gev{} that may be reached at an upgraded JLab facility \cite{nat98}.
At such an energy and at c.m.\ scattering angles around $90^\circ$ the
kinematical conditions for the approach presented here would be
satisfied. Still higher energies, perhaps accessible at
ELFE~\cite{ELFE}, would be even better.

Dimensional counting~\cite{far73} predicts that the Compton cross
section scaled by $s^6$ only depends on the ratio $t/s$ or,
equivalently, on the scattering angle $\theta$ in the photon-proton
c.m. {}From Fig.~\ref{dsigfig} one observes that the soft
contributions do not exhibit this counting rule behaviour, although
they are close to it.  $s^6$-scaling holds in our approach as long as
$R_V$ and $R_A$ behave as $1/t^2$. As one can observe from
Fig.~\ref{Rifig} this is approximately true for $-t$ in the range from
about 5 to 15 GeV$^2$.  For energies between, say, 3 and 6 GeV in the
laboratory frame such $t$-values are only reached in the backward
hemisphere. In this case the energy dependence of the scaled Compton
cross section is hence much milder in the backward than in the forward
hemisphere (see Fig.~\ref{dsigfig}). For energies as large as for
instance 12 to 15 GeV the situation is reversed. The $t$-values are so
large in the backward hemisphere that $R_V$ and $R_A$ do not behave as
$1/t^2$ any more but gradually turn into the soft physics asymptotics
$1/t^4$.  Consequently the scaled Compton cross section exhibits a
stronger energy dependence in the backward hemisphere than in the
forward one.  For very high energies the soft physics contribution to
the large angle Compton cross section scales as $s^{-10}$.  We note
that Radyushkin's result~\cite{rad98a} that all the curves for
different energies intersect each other at $\cos{\theta}=-0.6$ does
not hold in general. This result may depend on specific assumptions
made in~\cite{rad98a} and holds at best in a rather limited region of
energy.  It is an goal of utmost importance to test the energy
dependence of the Compton cross section experimentally in the relevant
kinematical region $s, -t, -u \gg m^2$.  The present data are neither
accurate enough nor really satisfy the kinematical requirements.

The Compton cross section has also been calculated within perturbative
QCD~\cite{kron91} and within a diquark model~\cite{kro96} that
combines perturbative elements with additional soft physics
(correlations in the proton wave function modelled as diquarks). Both
models can also account for the data although, as we said before, the
quality of the data is insufficient for a severe test of the models.
The diquark model does not lead to the dimensional counting behaviour
either; it turns out that the energy dependence of the scaled cross
section in the forward and backward hemisphere predicted by that model
is opposite to the one of the approach proposed here and shown in
Fig.~\ref{dsigfig}.

In the leading twist hard scattering calculation of \cite{kron91}
proton distribution amplitudes are employed which are strongly
concentrated in the end point regions, and thus differ drastically
from the one determined in Ref.~\cite{bol96} and used here (cf.~Eq.\ 
(\ref{da3})). For such distribution amplitudes the perturbative
analysis of Compton scattering, quite like that of the nucleon form
factor~\cite{isg89}, may be afflicted by large contributions from the
soft end point regions, where perturbative QCD is not applicable as we
mentioned in the introduction.\footnote{ $R_V(t)$ and $R_A(t)$
  evaluated from a wave function composed of the distribution
  amplitude proposed in Ref.~\cite{COZ} and the Gaussian
  (\ref{BLHMOmega}) exhibit approximate $1/t^2$ scaling behaviour in a
  much larger $t$-region than found from the distribution amplitude
  (\ref{da3}). Also the maximum values of $R_V(t)$ and $R_A(t)$ are
  larger by a factor 5 to 8. This parallels the behaviour of the
  electromagnetic form factors, see Ref.~\cite{bol96}. As a
  consequence the Compton cross section does not show approximate
  $s^{-6}$ scaling behaviour for photon energies between, say, 3 and
  15 GeV.} We emphasise that in the perturbative approach the
dimensional counting rule behaviour of the Compton cross section is
modified by powers of $\log{s}$ arising from running of the the strong
coupling constant $\alpha_s$ ($d\sigma/dt \propto \alpha_s^4$) and
from the evolution of the proton wave function. These effects have not
been taken into account as yet. It remains to be seen how much these
$\log$s will change the results quoted in Ref.~\cite{kron91}. One may
also expect that the inclusion of transverse momentum effects and
Sudakov suppressions in the perturbative analysis leads to a similarly
strong reduction of the Compton amplitude as was found for the proton
form factor~\cite{ber}.  In view of this it seems premature to us to
claim a success of the purely perturbative approach in Compton
scattering.

%%%%%%%%%%%%%%%%%%%%%%%%%%%%%%%%%%%%%%%%%%%%%%%%%%%%%%%%%%%%%%%%%%%
\section{Skewed parton distributions}
\label{skewed}
%%%%%%%%%%%%%%%%%%%%%%%%%%%%%%%%%%%%%%%%%%%%%%%%%%%%%%%%%%%%%%%%%%%
In this section we are going to investigate skewed parton
distributions (SPDs)~\cite{rad97,ji98}. These distributions are the
non-perturbative input for Compton scattering in the deep virtual
region of small $-t$ but large $Q^2$ and $s$. Factorisation of the
process into hard and soft physics~\cite{rad97,fact} assures that,
like the usual parton distributions, the SPDs are universal in the
sense that they occur in different hard processes, e.g.\ in hard meson
production.

As explained in Sect.~\ref{sub-compton} we will restrict our
investigation of SPDs to the kinematical region where they describe
how a parton with momentum $k$ is taken out from the proton with
momentum $p$ and, having undergone a hard scattering, inserted back
with momentum $k+\Delta$ as a parton inside the scattered proton with
momentum $p+\Delta$ (see Fig.~\ref{fig1}(b)).  Due to this restriction
we are unable to calculate the full amplitude of the deep virtual
Compton scattering process, which includes the region $0 < x <\zeta$
where we do not have a simple representation of SPDs as an overlap of
LCWFs. In a restricted kinematical region, however, we are able to
calculate the process independent SPDs, which are of interest in their
own. We will also be able to check whether they behave correctly in
the formal forward limit $\Delta = 0$, and whether they satisfy bounds
coming from positivity requirements~\cite{MaRy,pir98,rad48a}.

To date essentially nothing is known experimentally about skewed
distributions. However, various model estimates of the SPDs have been
made recently: for instance a bag model calculation~\cite{jms97}, a
chiral quark-soliton model~\cite{pppbgw98}, and a scalar toy
model~\cite{rad97}. A number of simple ans\"atze has also been
proposed~\cite{rad48a,ansatz}. In particular the question whether there is a
strong dependence on the skewedness parameter $\zeta$ is being
debated.

The spin independent skewed distributions are defined by\footnote{For
  convenience we do not display the link-operator needed to render the
  definition gauge invariant, assuming the use of a light-cone gauge
  combined with an appropriate choice for the integration path which
  reduces the link-operator to unity.}
\begin{equation}
p^+\,\int \frac{dz^-}{2\pi}\;
e^{ixp^+z^-}\;
\langle p'|\overline\psi_a(0)\,\gamma^+\,\psi_a(z^-)|p\rangle =
\widetilde{{\cal F}}_\zeta^{\,a}(x;t)\;
\bar u(p')\,\gamma^+\,u(p) + 
\mbox{``$\widetilde{{\cal K}}$-term''}  \eqcm
\label{eq:SKdef}
\end{equation}
where here and in the following $a$ denotes a quark flavour,
antiquarks being explicitly labelled by $\overline{a}$.  The
$\widetilde{{\cal K}}$-term in (\ref{eq:SKdef}) goes with the tensor
current of the proton and is related to proton helicity flip. Like the
Pauli form factor $F_2$ and our form factor $R_T$ we cannot
evaluate it in our model as explained in Sect.~\ref{proton-spin}. In
the definition (\ref{eq:SKdef}) we follow the conventions of
Radyushkin for {\em nonforward distributions}, cf.\ Eqs.~(9.1) and
(9.2) of~\cite{rad97}. The kinematical variables $x$ and $\zeta$ turn
out to be most convenient for calculating the overlap of LCWFs. The
relation to Ji's original definition of {\em off-forward
  distributions}, where a different choice of variables is made, can
be found in Ref.~\cite{rad97}, Eqs.~(9.6) and (9.7), and in
Ref.~\cite{ji98} Eqs.~(24) and (25).

The matrix element in (\ref{eq:SKdef}) is nonzero in the range
$-1+\zeta <x <1$, cf.~\cite{rad97,DG,ji98}. Re-interpreting a quark
with negative momentum fraction as an antiquark with positive fraction
one finds that $-1+\zeta <x< 0$ describes the emission and absorption
of an antiquark, just as $\zeta <x< 1$ does for a quark, while in the
region $0 <x< \zeta$ the proton $p$ emits a quark-antiquark pair and
is left as a proton with momentum $p+\Delta$.

The definition (\ref{eq:SKdef}) reveals the close relationship of SPDs
with the usual quark distributions and with the Dirac form factor.
Indeed one finds the reduction formulas
\begin{equation} \label{dist-reduction}
\widetilde{{\cal F}}_{\zeta=0}^{\,a}(x;t=0)=\q_a(x)
\end{equation}
and
\begin{equation} \label{form-sum-rule}
\sum_a e_a \int_{-1+\zeta}^1 
\widetilde{{\cal F}}_\zeta^{\,a}(x;t)\;\d x = F_1(t)
\eqpt
\end{equation}
Eq.~(\ref{dist-reduction}) can be explicitly checked in our results,
while we cannot evaluate the moments in (\ref{form-sum-rule}), which
contain the region $0 <x< \zeta$ we do not model here, except in the
case $\zeta=0$.

We now turn to the derivation of an overlap formula for the SPDs. In
close analogy to the steps that lead to Eq.~(\ref{half-way}) the
amplitude for DVCS can be written in terms of proton matrix elements
as
\begin{eqnarray} \label{dvcs-distributions}
{\cal A} &=& \sum_a (e e_a)^2 \int_{-1+\zeta}^{1} 
   \frac{\d x}{2 \sqrt{|x x'|}}\, \sum_{\lambda}\,
   \int {\d z^-\over 2\pi}\, e^{i\, x p^+ z^-}\, 
   \langle p'|\, \overline\psi{}_{a}(0)\, \gamma^+
          \frac{1+\lambda\gamma_5}{2}\,\psi_{a}(z^-) \, |p\rangle
\nonumber \\
&& \times 
   \left[ \theta(\zeta <x< 1)\,
          \bar{u}(\bar{k}',\lambda) H(\bar{k}',\bar{k})
                 u(\bar{k},\lambda) \,
\right.
\nonumber \\ 
&& \hspace{0em} \left.
      {}- \theta(0 <x< \zeta)\,
          \bar{v}(-\bar{k}',-\lambda) H(\bar{k}',\bar{k})
                 u(\bar{k},\lambda) \, \right.
\nonumber \\ 
&& \hspace{0em} \left.
      {}+ \theta(-1+\zeta <x< 0)\,
          \bar{v}(-\bar{k}',-\lambda) H(\bar{k}',\bar{k})
                 v(-\bar{k},-\lambda) \, \right]  \eqcm
\end{eqnarray}
with the conventions for spinors given before Eq.~(\ref{trick}). The
different kinematical regions mentioned above can easily be
recognised. The hard scattering is now approximated as collinear,
neglecting $-t$ and $m^2$ compared with $Q^2$ and setting $\bar k=[x
p^+, 0,{\bf 0}_\perp]$, $\bar k'=[x' p^+,0,{\bf 0}_\perp]$. On the
other hand, direct calculation of the overlap diagrams starting from
the Fock state decomposition of the proton (cf.\ Sect.~\ref{wave})
gives the contribution of the region $\zeta<x<1$ to the amplitude as
\begin{eqnarray} \label{dvcs-wave-functions}
{\cal A}' &=& \sum_a (e e_a)^2 \,
\int_\zeta^1 \d x\,  \sum_N
\sum_j \sum_\beta \;\int [\d x]_N [\d^2 {\bf k}_\perp]_N \; 
       \delta(x - x_j) \frac{1}{\sqrt{x_j x_j'}} \,
       \left(1-\zeta\right)^{1-{N \over 2}}
\nonumber \\
&&  \times\,
    \Psi^*_{N\beta}(\breve x'_i,\breve{\bf k}'_{\perp i})\, 
    \Psi_{N\beta}(x_i,{\bf k}_{\perp i}) \;
    \bar{u}(\bar{k}',\lambda) H(\bar{k}',\bar{k})
                 u(\bar{k},\lambda)  \eqcm
\end{eqnarray}
where $j$ runs over quarks of flavour $a$. Note that the label $\beta$
includes a dependence on the parton spin $\lambda$. The arguments of
the outgoing wave function $\breve x'_i$ and $\breve{\bf k}'_{\perp
  i}$ are related to $x_i$ and ${\bf k}_{\perp i}$ by
(\ref{breve-args}). {}From the comparison of
(\ref{dvcs-wave-functions}) with (\ref{dvcs-distributions}) and the
definition (\ref{eq:SKdef}) we obtain the overlap formula for
spin-independent SPDs in the region $\zeta<x<1$:
\begin{equation}
\widetilde{{\cal F}}_\zeta^{\,a\,(N)}(x;t)\,=\,
\left(1-\zeta\right)^{{1-N} \over 2}\, 
\sum_j\sum_\beta \;\int [\d x]_N [\d^2 {\bf k}_\perp]_N \; 
\delta(x - x_j)\; 
\Psi^*_{N\beta}(\breve x'_i,\breve{\bf k}'_{\perp i})\, 
\Psi_{N\beta}(x_i,{\bf k}_{\perp i})
\label{eq:SKoverlap}
\end{equation} 
with $j$ again running over all quarks of flavour $a$. Comparing with
(\ref{disf}) we see that the boundary condition (\ref{dist-reduction})
is correctly implemented in our approach. As for the sum rule
(\ref{form-sum-rule}) we find with (\ref{tilde-args}) and
(\ref{breve-args}) that in the case $\zeta=0$ the overlap expression
(\ref{eq:SKoverlap}) and the corresponding contribution from
antiquarks reproduce the Drell-Yan formula (\ref{eform}).

We notice that for $\zeta<x<1$ the r.h.s\ of (\ref{eq:SKoverlap}) has
the structure of a scalar product in the Hilbert space of wave
functions $\Psi_{N\beta}(x_i,{\bf k}_{\perp i})$. Writing down the
Cauchy-Schwarz inequality for (\ref{eq:SKoverlap}) and using the
reduction formula (\ref{dist-reduction}) we find
\begin{equation}  \label{PST}
\left|\, \widetilde{{\cal F}}_\zeta^{\,a\,(N)}(x;t) \,\right| \leq
\frac{1}{ \sqrt{1-\zeta} }\;
\sqrt{\q_a^{(N)}(x)\; 
      \q_a^{(N)}\left( {\textstyle \frac{x-\zeta}{1-\zeta}} \right)}
\end{equation} 
for the contribution of each Fock state. Notice that at the points
$x=\zeta$ and $x=1$ both sides of (\ref{PST}) are zero because in the
corresponding overlap integrals there are wave functions taken at
their end points. Summing (\ref{eq:SKoverlap}) over all Fock states
one obtains the analogue of (\ref{PST}) for the complete
distributions; it is precisely the positivity constraint on SPDs
derived by Pire, Soffer and Teryaev~\cite{pir98}, which is thus
satisfied by the overlap formula (\ref{eq:SKoverlap}).\footnote{Notice
  that it is satisfied for all $t$ in the physical region $t \le -
  \zeta^2 m^2/(1-\zeta)$, cf.~(\protect\ref{t-min}), with the upper
  bound being $t$-independent.}

To discuss the emission and reabsorption of an antiquark it is useful
to define $\widetilde{{\cal F}}_\zeta^{\,\overline{a}}(x;t)$ by the
r.h.s.\ of (\ref{eq:SKdef}) with the field operators replaced with the
charge conjugated ones. One easily finds the relation
$\widetilde{{\cal F}}_\zeta^{\,\overline{a}}(x;t) = - \widetilde{{\cal
    F}}_\zeta^{\,a}(\zeta-x;t)$. In the region $\zeta <x< 1$ the
distribution $\widetilde{{\cal F}}_\zeta^{\,\overline{a}}(x;t)$
describes the emission of an antiquark with momentum fraction $x$ and
its reabsorption with fraction $x'=x-\zeta$; along the same lines as
above one obtains its overlap representation as the r.h.s.\ of
(\ref{eq:SKoverlap}) with $j$ running over antiquarks instead of
quarks. One then has of course the analogues of the reduction formula
(\ref{dist-reduction}) and the bound (\ref{PST}) for $\widetilde{{\cal
    F}}_\zeta^{\,\overline{a}}(x;t)$ and the usual antiquark
distributions.

Inserting our $N=3,4,5$ Fock state wave functions of Sect.~\ref{wave}
in (\ref{eq:SKoverlap}) we obtain for the skewed $\u$ and $\d$ valence
distributions
\begin{eqnarray}
\lefteqn{
  \widetilde{{\cal F}}_\zeta^{\,a\,(N)}(x;t)
- \widetilde{{\cal F}}_\zeta^{\,\overline{a}\,(N)}(x;t) =
b_a^{(N)}\;P_N\;
(1-\zeta)^{-\frac{N+1}{2} -l_g} \;
\Upsilon_N\left(x,\zeta;  -t\,(1-\zeta) -\zeta^2 m^2 \right)}\nn\\
&& \times\,
x^{n_a}(1-x)^{m_a(N)}\;
\left[(1-\zeta)
      +c_a^{(N)}\,\left(1-\frac{\zeta}{2}\right)(1-x)
      +d_a^{(N)}\,(1-x)^2\right] \eqcm
\label{nfpds}
\end{eqnarray} 
where we remember the expression (\ref{eq:SKgauss}) 
\begin{displaymath}
\Upsilon_N (x,\zeta; \vd^2)=
\left(\frac{2}{2-\zeta}\right)^{N-2}
      \frac{2(x-\zeta)}{(x-\zeta)+x (1-\zeta)^2}\;
      \exp\left[\frac{-\,a_N^2\,\vd^2\,(1-x)}
                     {(x-\zeta)+x(1-\zeta)^2}\right]
\end{displaymath}
given at the end of Sect.~\ref{wave} and make use of the relation
(\ref{t-through-zeta}) between $\vd^2$, $t$ and $\zeta$. The exponents
$n_a$, $m_a(N)$ and the coefficients $b_a^{(N)}$, $c_a^{(N)}$ and
$d_a^{(N)}$ are the same as for the valence quark distributions
discussed in Sect.~\ref{distributions}, cf.\ Eq.~(\ref{qpowers}) and
Tab.~\ref{tabc}, and $l_g$ is the number of gluons in the
corresponding Fock state. For $\zeta=0$ our result (\ref{nfpds})
simplifies to
\begin{equation}
  \widetilde{{\cal F}}_{\zeta=0}^{\,a\,(N)}(x;t)
- \widetilde{{\cal F}}_{\zeta=0}^{\,\overline{a}\,(N)}(x;t) =
\left( \q_a^{(N)}(x) -  \qbar_a^{(N)}(x) \right) \,
\exp{\left[ \frac12 \, a^2_N \, t
                 \, \frac{1-x}{x} \right]}  \eqcm
\end{equation}
which is the origin of our simple representations (\ref{nform}) and
(\ref{rva}) of form factors. Finally we find that with our wave
functions the skewed antiquark and $\s$-quark distributions are
related with the $\d$ valence distribution by the analogues of
Eqs.~(\ref{sea-symmetric}) and (\ref{sea-special}).

\begin{figure}[hbtp]
\parbox{\textwidth}{\begin{center}
\psfig{file=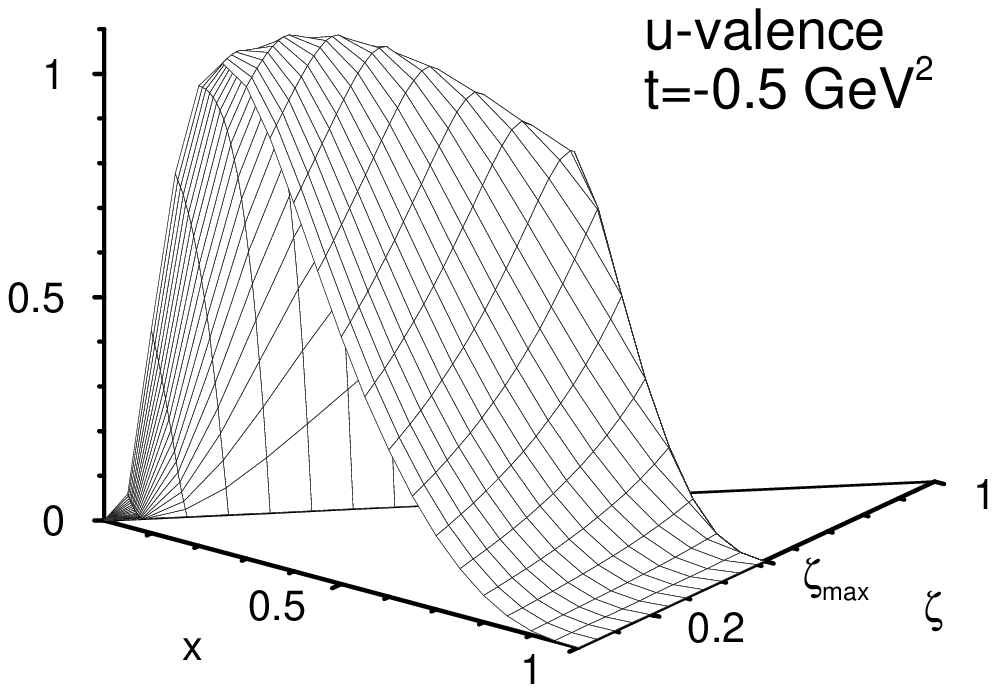, width=7cm} \qquad
\psfig{file=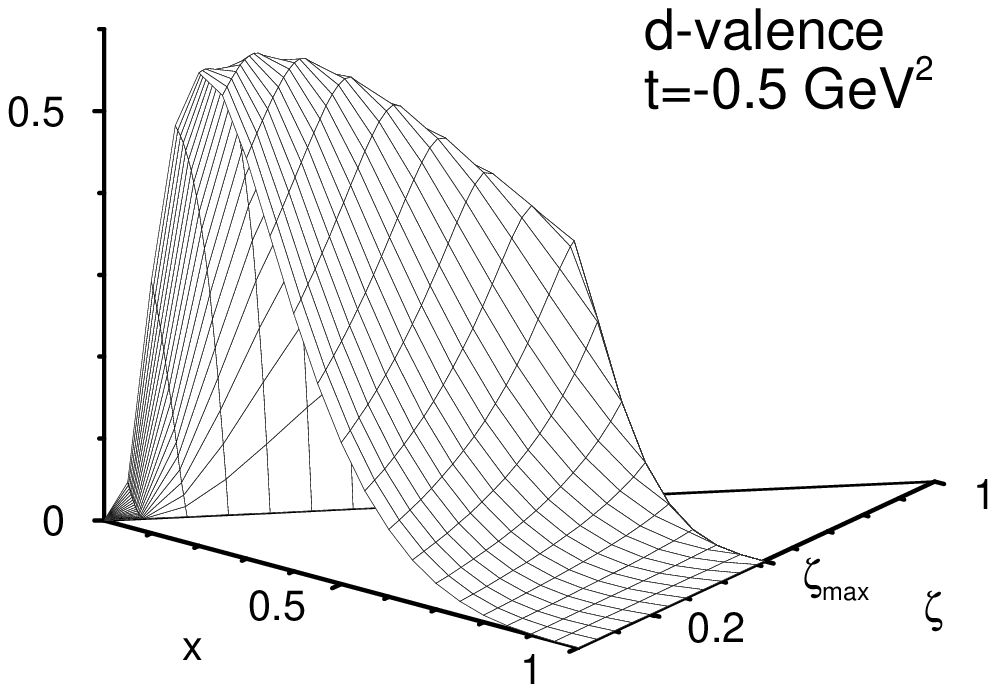, width=7cm}\\
\psfig{file=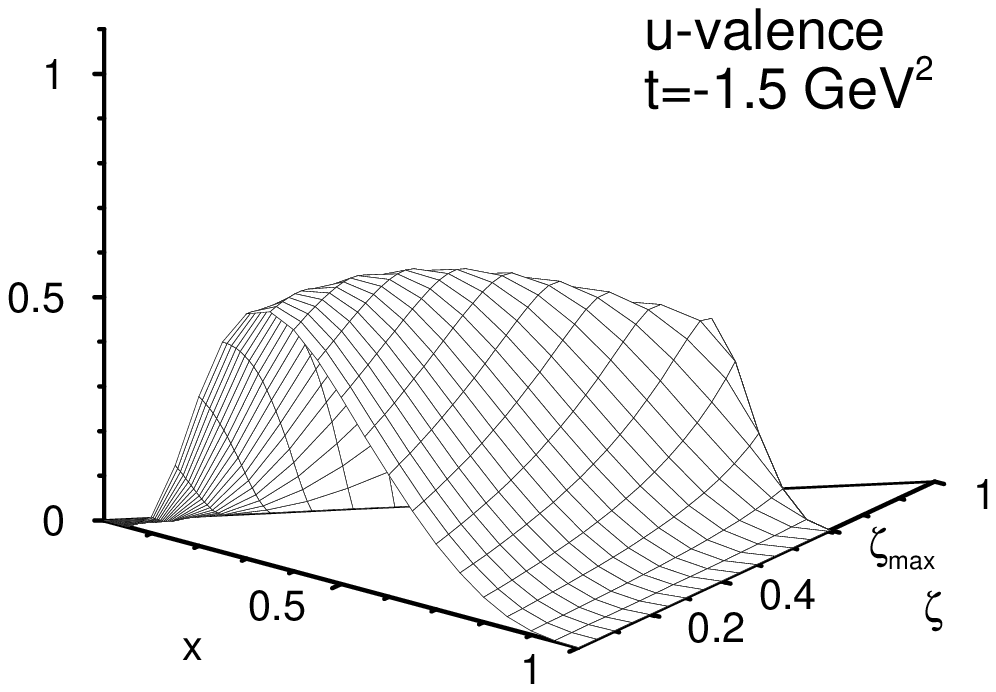, width=7cm} \qquad
\psfig{file=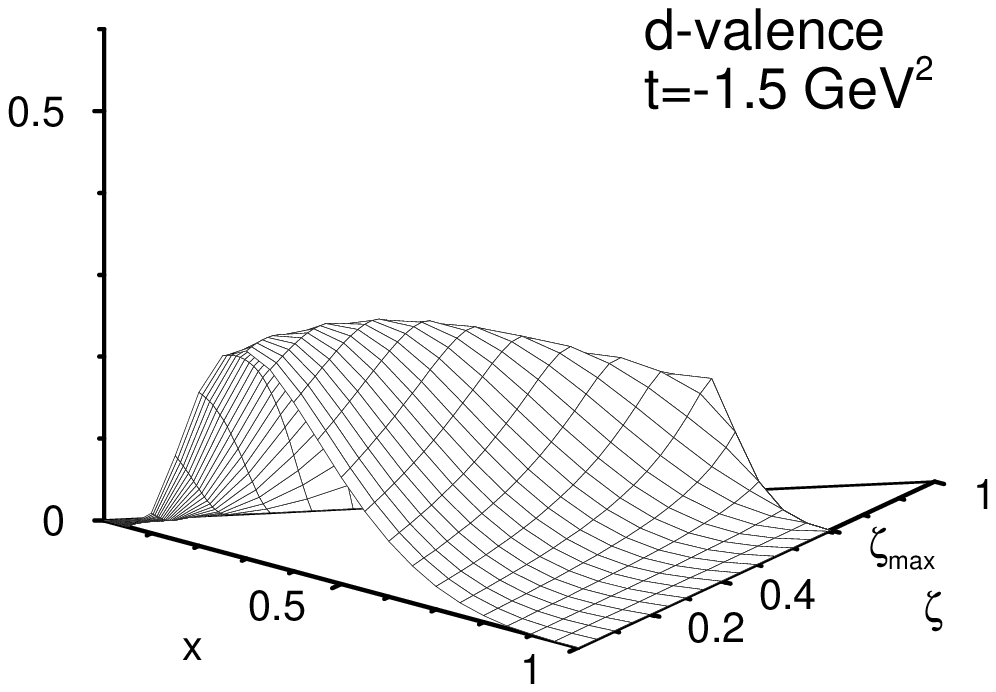, width=7cm}
\end{center}}
\caption{\label{skewedfig}Skewed parton distributions  for $\u$ and
  $\d$ valence quarks in the region $\zeta<x<1$, obtained from the
  $N=3,4,5$ Fock states ($P_3=0.17$, $P_4=P_5=0.1$). The values of
  $\zeta_{\it max}$ are 0.52 at $t=-0.5\,{\rm GeV}^2$ and 0.71 at
  $t=-1.5\,{\rm GeV}^2$.}
\end{figure}

In Fig.~\ref{skewedfig} we display our results (\ref{nfpds}) summed
over the $N=3,4,5$ Fock states for fixed $t$-values of $-0.5\,{\rm
  GeV}^2$ and $-1.5\,{\rm GeV}^2$. We remember from the end of
Sect.~\ref{general-kinematics} that $-t \ge \zeta^2 m^2/(1-\zeta)$. At
fixed $t$ this imposes $\zeta \le \zeta_{\it max}$ with
\begin{equation}  \label{zeta-max}
\zeta_{\it max}=\frac{\sqrt{t\,(t-4 m^2)} + t}{2m^2} \eqpt
\end{equation}
We remark at this point that the $t$-dependence of the SPDs residing
in the factor $\Upsilon_N$ does not factorise in our approach but
mixes with the dependence on $x$ and $\zeta$ in the exponent of
(\ref{eq:SKgauss}); note that the transverse momenta $\breve{\bf
  k}'_{\perp i}$ in the overlap formula (\ref{eq:SKoverlap})
implicitly depend on $x$, $\zeta$ and $\vd$ through
(\ref{breve-args}). A significant dependence on the skewedness
parameter $\zeta$ shows up in our results; a fact which is not
surprising since $\zeta$ determines the momentum fraction of the
active parton in the light-cone wave function of the outgoing nucleon.

\begin{figure}[hbtp]
\parbox{\textwidth}
{\begin{center}
\psfig{file=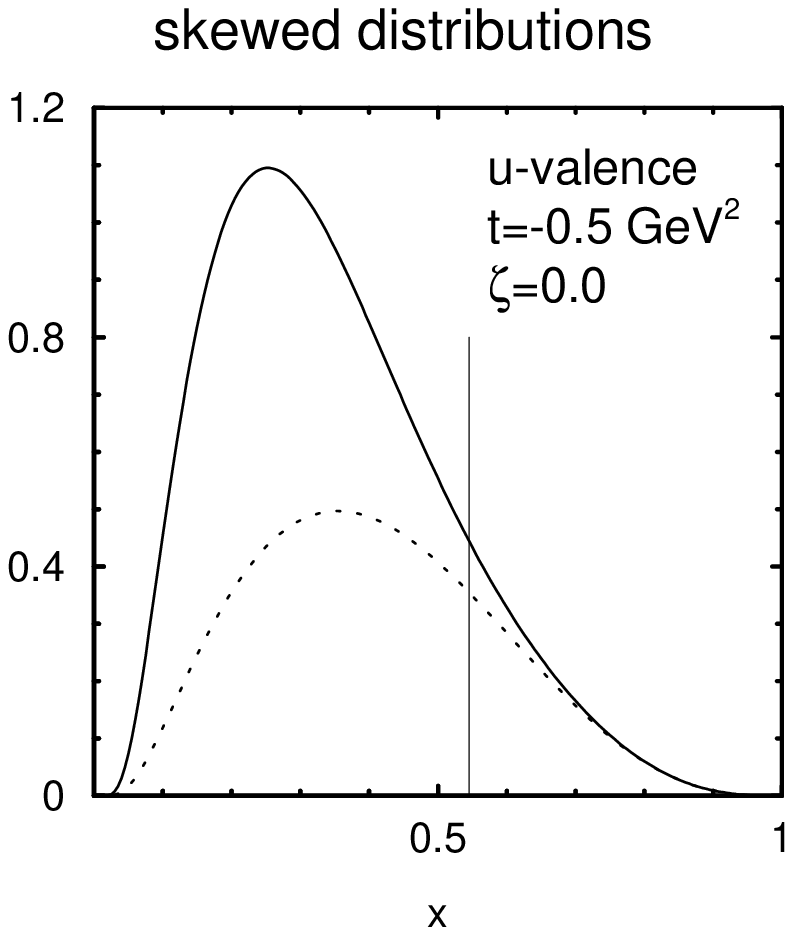, width=5.9cm} \qquad
\psfig{file=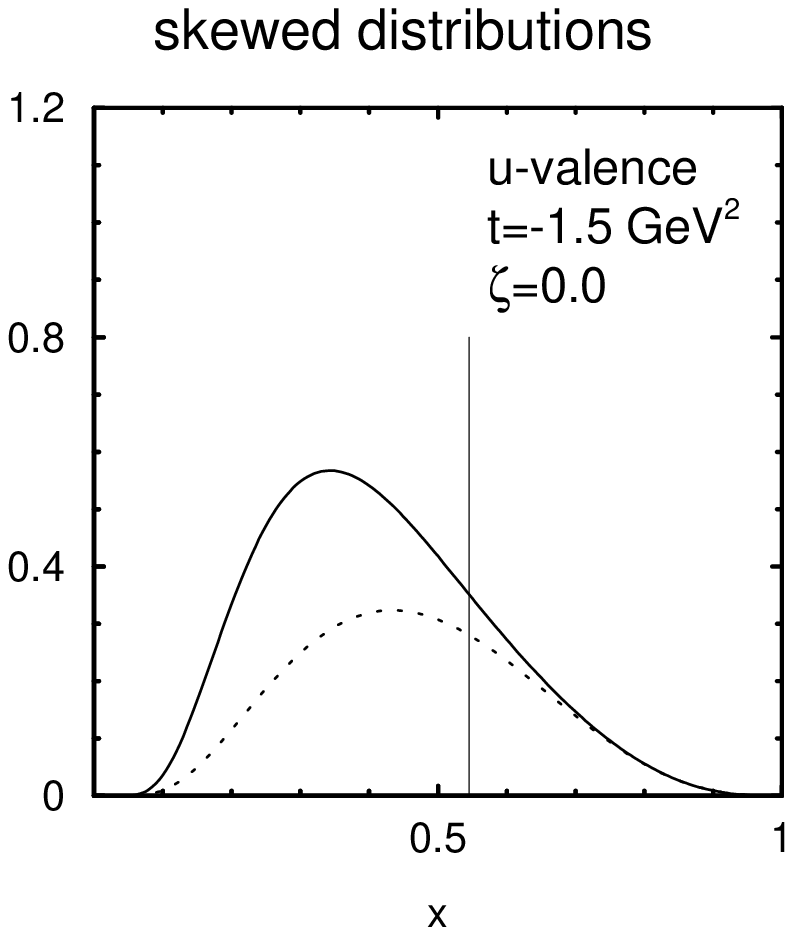, width=5.9cm} \\
\psfig{file=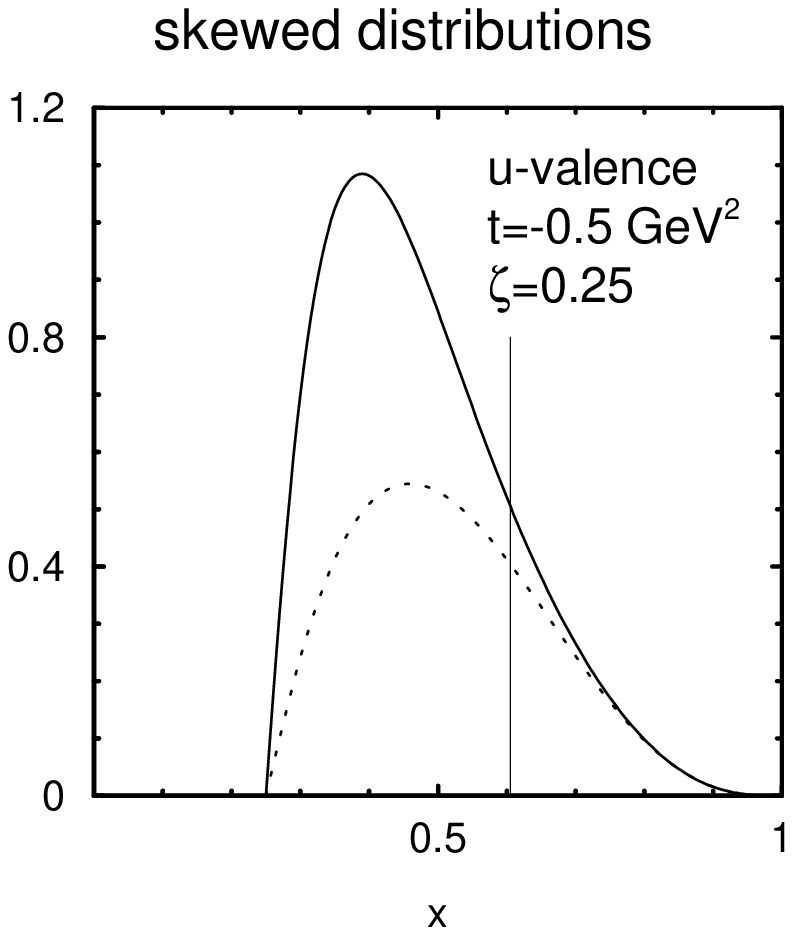, width=5.9cm} \qquad
\psfig{file=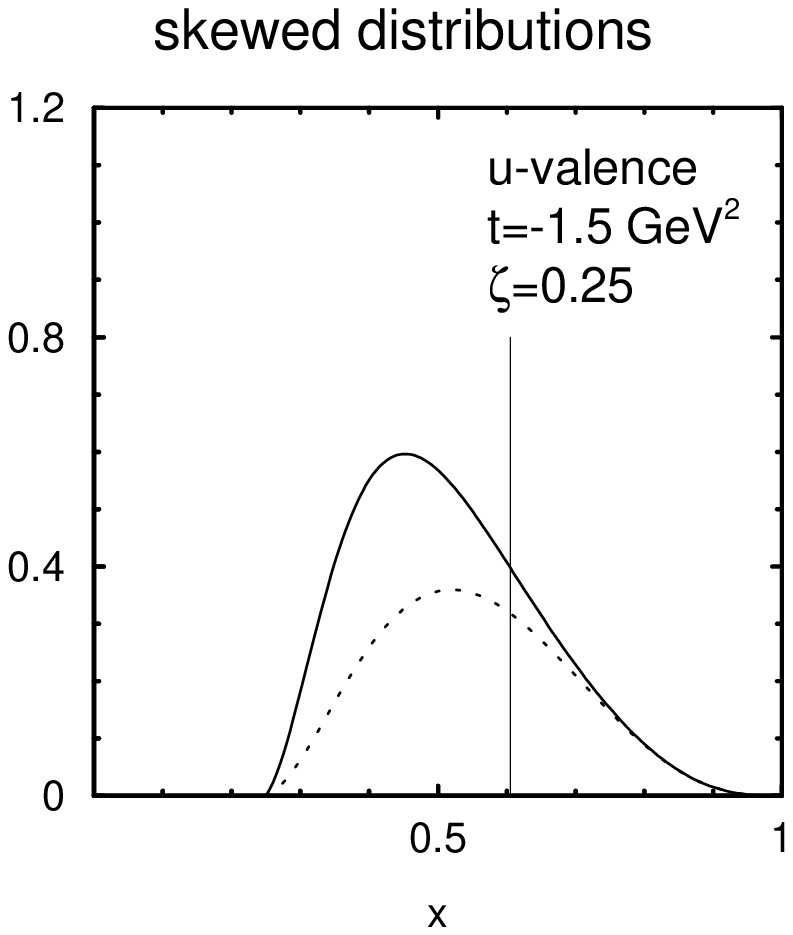, width=5.9cm} \\
\psfig{file=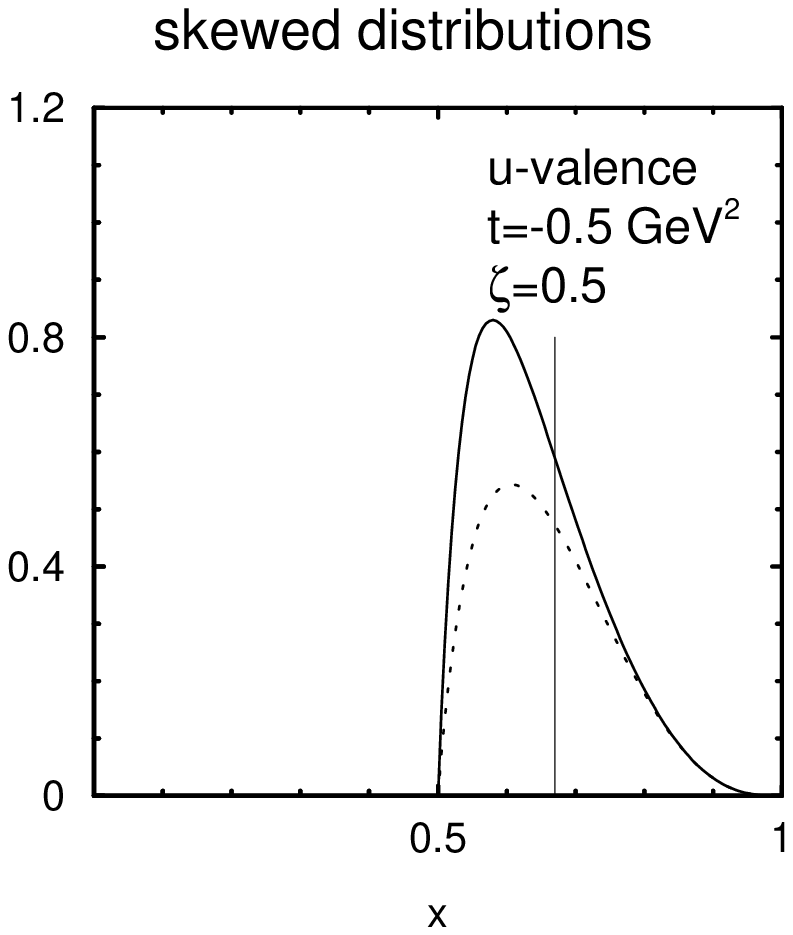, width=5.9cm} \qquad
\psfig{file=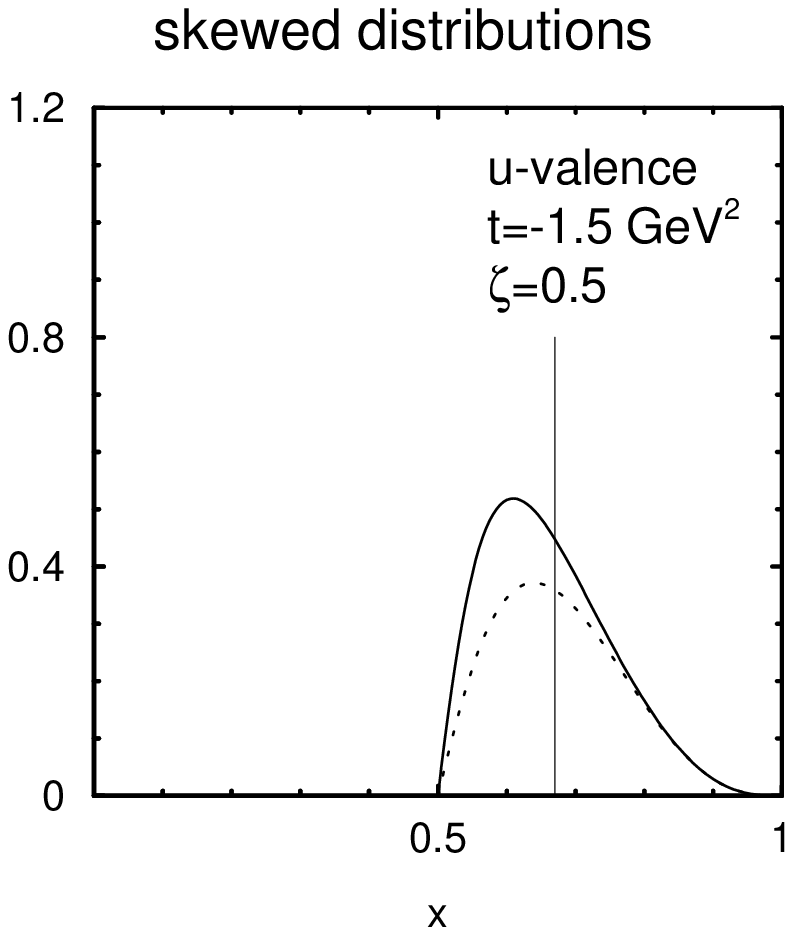, width=5.9cm}
\end{center}}
\caption{\label{fzfig}Skewed valence $\u$-quark parton distributions
  in the region $\zeta<x<1$ as a function of $x$ at fixed $\zeta$ and
  $t$. Full lines show the sum of contributions from the $N=3,4,5$
  Fock states and dashed ones the contribution from the $N=3$ Fock
  state alone. The vertical lines in the plots indicate the value of
  $x$ where the $N=3$ contribution is 80\% of the one summed over
  $N=3,4,5$.}
\end{figure}

In Fig.~\ref{fzfig} we plot the skewed $\u$ valence distributions at
fixed $t$ and $\zeta$ as a function of $x$, comparing the contribution
from the $N=3$ Fock state with the result summed over $N=3,4,5$. As
for the usual parton distributions we see how higher Fock states
become more and more important as $x$ decreases. We notice that the
values of $x$ where this happens increase somewhat with $\zeta$; this
can be understood from the fact that at a given $x$ the momentum
fraction of the parton going back into the proton decreases with
$\zeta$. The area under a curve in Fig.~\ref{fzfig} gives the
$\u$-quark contribution of the regions $-1+\zeta <x< 0$ and $\zeta <x<
1$ to the form factor sum rule~(\ref{form-sum-rule}). We can see that
higher Fock states become less important as $-t$ increases, in
agreement with what we have found for $F_1(t)$ in Sect.~\ref{form}.

For the usual parton distributions we know that both $\q_v(x)$ and
$\qbar(x)$ become singular for $x \to 0$, which cannot be obtained
from any finite number of Fock state contributions, all of which
vanish at $x=0$. The question what the situation is for $x \to \zeta$
in skewed distributions, when the momentum fraction $x'$ becomes zero
while $x$ remains finite, cannot be answered in the framework of this
paper. We therefore do not claim that our results for the contribution
of the first tree Fock states describe the full distribution as $x$
comes close to $\zeta$.

The definition of spin dependent SPDs is obtained from
(\ref{eq:SKdef}) by the replacements $\gamma^+\to\gamma^+\gamma_5$ and
$\widetilde{{\cal F}}^{\,a}_\zeta\to\widetilde{{\cal G}}^{\,a}_\zeta$;
for antiquarks one has $\widetilde{{\cal G}}^{\, \overline{a}
  }_\zeta(x;t) = \widetilde{{\cal G}}^{\,a}_\zeta(\zeta-x;t)$. The
analogue of the $\widetilde{{\cal K}}$-term goes now with the
pseudoscalar current of the proton and is again related to proton spin
flip. {}From the appropriate overlap formulae we find the spin
dependent skewed valence distributions
\begin{eqnarray}
\lefteqn{
  \widetilde{{\cal G}}_\zeta^{\,a\,(N)}(x;t) 
- \widetilde{{\cal G}}_\zeta^{\,\overline{a}\,(N)}(x;t) =
\Delta b_a^{(N)}\;P_N\;
(1-\zeta)^{-\frac{N+1}{2} -l_g} \; 
\Upsilon_N\left(x,\zeta; -t\,(1-\zeta) -\zeta^2 m^2\right)}\nn\\
&& \times\,
x^{n_a}(1-x)^{m_a(N)}\;
\left[(1-\zeta)
      +\Delta c_a^{(N)}\,\left(1-\frac{\zeta}{2}\right)(1-x)
      +\Delta d_a^{(N)}\,(1-x)^2\right] \eqcm
\end{eqnarray} 
where the coefficients $\Delta b_a^{(N)}$, $\Delta c_a^{(N)}$, and
$\Delta d_a^{(N)}$ are the same as the ones for the spin dependent
valence distributions listed in Tab.~\ref{tabcd}. Evidently the spin
dependent skewed distributions reduce correctly to the usual ones in
the limit $\zeta\to 0$ and $t\to 0$.

%%%%%%%%%%%%%%%%%%%%%%%%%%%%%%%%%%%%%%%%%%%%%%%%%%%%%%%%%%%%%%%%%%%
\section{Summary}
\label{sum}
%%%%%%%%%%%%%%%%%%%%%%%%%%%%%%%%%%%%%%%%%%%%%%%%%%%%%%%%%%%%%%%%%%%
In the present paper we have linked ordinary and skewed parton
distributions to soft overlap contributions to elastic form factors
and to large angle Compton scattering using nucleon light-cone wave
functions.

We have investigated how and under which conditions overlap
contributions to exclusive processes can be expressed in terms of
LCWFs. For large angle Compton scattering, at large values of the
Mandelstam invariants $s$, $-t$ and $-u$, we can calculate the soft
overlap contribution using its factorisation into handbag diagrams,
i.e.\ into soft parton emission and reabsorption by the nucleon and a
hard parton-photon scattering. In the case of deeply virtual Compton
scattering, with large $Q^2$ and $s$ but small $-t$, where we cannot
express the amplitude as an overlap of soft LCWFs, we have calculated
the skewed parton distributions in the large-$x$ region.

For the LCWF of the three-quark nucleon Fock state we have taken over
the parametrisation of~\cite{bol96}, which involves only two
parameters adjusted to data. For the Fock states with an additional
gluon or quark-antiquark pair we have taken a very simple ansatz,
introducing only two more parameters, which are fitted to the gluon
and sea quark parton distributions from the GRV analysis~\cite{GRV}.
The values of all four parameters come out in a range compatible with
their physical meaning of Fock state probabilities or a transverse
size parameter. In the overlap contributions to Compton scattering and
the form factor we also estimate the net effect of all higher Fock
states, using as input the difference between the GRV parton
distributions and those calculated from the three lowest Fock states
only.

The phenomenology we can do with our ansatz is very rich: we reproduce
well the unpolarised and polarised parton distributions down to $x$
around 0.5, as well as the data for the nucleon Dirac form factors and
for real Compton scattering at large c.m.\ angles. The inclusion of
higher Fock states in the soft overlap contributions confirms that as
$-t$ increases the lowest Fock states become increasingly dominant and
gives an impression of the accuracy one can hope for by only taking
into account the three quark state. The LCWF of~\cite{bol96} was
constructed so as to saturate the elastic form factor data. The fact
that with the same wave function one obtains a reasonable description
of Compton scattering supports the hypothesis that there is no
sizeable perturbative contribution to either process in the range of
momentum transfers where data exist; soft physics seems to dominate as
was occasionally suggested in the literature~\cite{rad91,isg89,JKR}.

We stress that from the apparent agreement of exclusive data with
dimensional counting rules the dominance of perturbative QCD
contributions cannot be deduced. Soft physics, as for instance the
overlap-type contributions which we propose, provides broad maxima in
scaled observables such as $t^2 F_1(t)$ and the scaled Compton cross
section $s^6\, \d\sigma /\d t$, and thus mimics dimensional counting
rule behaviour in a certain range of $t$.

Compared with the elastic form factors large angle Compton scattering
has a second independent kinematical variable and thus provides an
additional handle to experimentally test how well dimensional counting
rules are satisfied. We further suggest that the imaginary part of the
scattering amplitude, which is accessible in virtual Compton
scattering with a polarised lepton beam, offers a sensible tool to
investigate which dynamical mechanism is at work: in the handbag
mechanism imaginary parts are only generated through loop corrections
to the photon-parton subprocess, whereas in the hard scattering
mechanism real and imaginary parts generically are of the same order
of magnitude. Spin observables may also be sensitive probes of the
underlying physics, given the particular helicity structure of the
photon-parton scattering in the handbag diagrams. In any case we see a
strong motivation to have further and more accurate Compton data at
sufficiently high values of energy and momentum transfer.

%%%%%%%%%%%%%%%%%%%%%%%%%%%%%%%%%%%%%%%%%%%%%%%%%%%%%%%%%%%%%%%%%%%
\section*{Acknowledgments}
%%%%%%%%%%%%%%%%%%%%%%%%%%%%%%%%%%%%%%%%%%%%%%%%%%%%%%%%%%%%%%%%%%%
We wish to acknowledge discussions with 
S.\ Brodsky, A.\ Radyushkin, N.\ Stefanis and E.~Reya.

This work has been partially funded through the European TMR Contract
No.~FMRX--CT96--0008: Hadronic Physics with High Energy
Electromagnetic Probes. T.~F.\ is supported by Deutsche
Forschungsgemeinschaft.

%%%%%%%%%%%%%%%%%%%%%%%%%%%%%%%%%%%%%%%%%%%%%%%%%%%%%%%%%%%%%%%%%%%

\end{fmffile}

\end{document}